\documentclass[english, data-science]{agh-wi}

\titlePL{Zastosowanie generatywnych sieci neuronowych do szybkich\\symulacji eksperymentu ALICE (CERN)}
\titleEN{Applying generative neural networks for fast simulations\\of the ALICE (CERN) experiment}
\author{Maksymilian Wojnar}
\supervisor{Professor Witold Dzwinel}

\usepackage{polski}
\usepackage[
    style=numeric,
    sorting=none,
    language=autobib,
    autolang=other,
    urldate=iso,
    seconds=true,
    backref=false,
    isbn=true,
    url=false,
    maxbibnames=3,
    backend=biber
]{biblatex}
\usepackage{amsmath}
\usepackage{amssymb}
\usepackage[polish, intoc]{nomencl}
\usepackage{graphicx}
\usepackage{xcolor}
\usepackage{tabularx}
\usepackage{multirow}
\usepackage{makecell}
\usepackage{longtable}
\usepackage[ruled,linesnumbered]{algorithm2e}
\usepackage{hyperref}
\usepackage{csquotes}
\usepackage{subcaption}
\usepackage{siunitx}
\usepackage{multicol}
\usepackage{acro}[=v2]
\usepackage[all]{nowidow}
\DeclareAcronym{ML}{short=ML, long=machine learning}
\DeclareAcronym{GPU}{short=GPU, long=graphics processing unit}
\DeclareAcronym{CNN}{short=CNN, long=convolutional neural network}
\DeclareAcronym{ViT}{short=ViT, long=vision transformer}
\DeclareAcronym{MLP}{short=MLP, long=multi-layer perceptron}
\DeclareAcronym{RBM}{short=RBM, long=restricted Boltzmann machine}
\DeclareAcronym{VAE}{short=VAE, long=variational autoencoder}
\DeclareAcronym{VDVAE}{short=VDVAE, long=Very Deep VAE}
\DeclareAcronym{KL}{short=KL, long=Kullback-Leibler}
\DeclareAcronym{SAE}{short=SAE, long=Sinkhorn autoencoder}
\DeclareAcronym{GAN}{short=GAN, long=generative adversarial network}
\DeclareAcronym{WGAN}{short=WGAN, long=Wasserstein GAN}
\DeclareAcronym{MaskGIT}{short=MaskGIT, long=Masked Generative Image Transformer}
\DeclareAcronym{NF}{short=NF, long=normalizing flow}
\DeclareAcronym{VQVAE}{short=VQ-VAE, long=Vector-Quantized Variational Autoencoder}
\DeclareAcronym{ViTVQGAN}{short=ViT-VQ-GAN, long=vision transformer-based VQ-GAN}
\DeclareAcronym{DDPM}{short=DDPM, long=Denoising Diffusion Probabilistic Model}
\DeclareAcronym{DDIM}{short=DDIM, long=Denoising Diffusion Implicit Model}
\DeclareAcronym{SDXL}{short=SDXL, long=Stable Diffusion XL}
\DeclareAcronym{LHC}{short=LHC, long=The Large Hadron Collider}
\DeclareAcronym{ALICE}{short=ALICE, long=A Large Ion Collider Experiment}
\DeclareAcronym{ZDC}{short=ZDC, long=Zero Degree Calorimeter}
\DeclareAcronym{PINN}{short=PINN, long=Physics-informed neural network}
\DeclareAcronym{LAGAN}{short=LAGAN, long=Location-Aware Generative Adversarial Network}
\DeclareAcronym{MMD}{short=MMD, long=maximum mean discrepancy}
\DeclareAcronym{e2eSAE}{short=e2e SAE, long=end-to-end Sinkhorn Autoencoder}
\DeclareAcronym{ELBO}{short=ELBO, long=Evidence Lower Bound}
\DeclareAcronym{BN}{short=BN, long=batch normalization}
\DeclareAcronym{LN}{short=LN, long=layer normalization}
\DeclareAcronym{ReLU}{short=ReLU, long=rectified linear unit}
\DeclareAcronym{GELU}{short=GELU, long=Gaussian Error Linear Unit}
\DeclareAcronym{MSE}{short=MSE, long=mean square error}
\DeclareAcronym{RMSE}{short=RMSE, long=root mean square error}
\DeclareAcronym{MAE}{short=MAE, long=mean absolute error}
\DeclareAcronym{HPC}{short=HPC, long=high-performance computing}
\DeclareAcronym{VQ}{short=VQ, long=vector quantization}
\DeclareAcronym{tSNE}{short=t-SNE, long=t-distributed stochastic neighbor embedding}
\DeclareAcronym{PCA}{short=PCA, long=principal component analysis}
\DeclareAcronym{RBF}{short=RBF, long=radial basis function}
\DeclareAcronym{UMAP}{short=UMAP, long=Uniform Manifold Approximation and Projection}
\DeclareAcronym{MDS}{short=MDS, long=multidimensional scaling}
\DeclareAcronym{EMA}{short=EMA, long=exponential moving average}
\DeclareAcronym{TPE}{short=TPE, long=tree-structured Parzen Estimator}

\definecolor{shadecolor}{gray}{0.9}
\makenomenclature

\begin{document}

\frontmatter
\maketitle

\cleardoublepage
\thispagestyle{empty}
\vspace*{\fill}
\begin{flushright}
    \em
    \begin{minipage}{0.75\textwidth}
        I would like to sincerely thank my supervisor, Professor Witold Dzwinel, for his guidance. I am thankful for his recommendations, especially for proposing the topic of this thesis. I am convinced that his expertise greatly improved the quality of this work. \\
        
        I express my gratitude to everyone who assisted me with the topic and provided valuable feedback on my work, particularly Emilia Majerz. Furthermore, I am grateful to Professor Jacek Otwinowski from the Institute of Nuclear Physics PAS in Krakow and the researchers from AGH University of Krakow and the University of Warsaw for their valuable discussions and suggestions during our meetings. \\
        
        I gratefully acknowledge Polish high-performance computing infrastructure PLGrid (HPC Centers: ACK Cyfronet AGH) for providing computer facilities and support within computational grant no. PLG/2023/016410. \\
        
        I am thankful for the support from a grant funded by the Polish Ministry of Science and Higher Education (Agreement Nr. 2023/WK/07).

    \end{minipage}
\end{flushright}

\begin{abstractEN}
    This thesis investigates the application of state-of-the-art advances in generative neural networks for fast simulation of the Zero Degree Calorimeter (ZDC) neutron detector in the ALICE experiment at CERN. Traditional simulation methods using the GEANT Monte Carlo toolkit, while accurate, are computationally demanding. With increasing computational needs at CERN, efficient simulation techniques are essential. The thesis provides a comprehensive literature review on the application of neural networks in computer vision, fast simulations using machine learning, and generative neural networks in high-energy physics. The theory of the analyzed models is also discussed, along with technical aspects and the challenges associated with a practical implementation. The experiments evaluate various neural network architectures, including convolutional neural networks, vision transformers, and MLP-Mixers, as well as generative frameworks such as autoencoders, generative adversarial networks, vector quantization models, and diffusion models. Key contributions include the implementation and evaluation of these models, a significant improvement in the Wasserstein metric compared to existing methods with a low generation time of 5 milliseconds per sample, and the formulation of a list of recommendations for developing models for fast ZDC simulation. Open-source code and detailed hyperparameter settings are provided for reproducibility. Additionally, the thesis outlines future research directions to further enhance simulation fidelity and efficiency.
\end{abstractEN}

\tableofcontents

\mainmatter

\chapter{Introduction}

\Ac{HPC} has become an essential component of scientific research, standing alongside theory and experiments as a core pillar of science. However, the use of numerical simulations often requires vast computational resources. To address these challenges, researchers have turned to surrogate models, a simplified versions of complex simulations that approximate system behavior balancing accuracy and computational efficiency~\cite{surrogate}. The recent advancements in \ac{ML} have significantly enhanced this approach. Neural networks, with their ability to approximate complex systems, have proven successful in enabling faster and more efficient physical simulations without the need for direct modeling of experiments.

Generative neural networks have revolutionized fields such as computer vision and natural language processing by creating high-quality samples that are nearly indistinguishable from real data. While commonly used for text and image generation, these models also show great potential to generate samples from specific processes, especially physical events, thus facilitating fast simulation tasks.

\section{Motivation and scope of work}

\begin{figure}[h!]
    \centering
    \includegraphics[width=\textwidth]{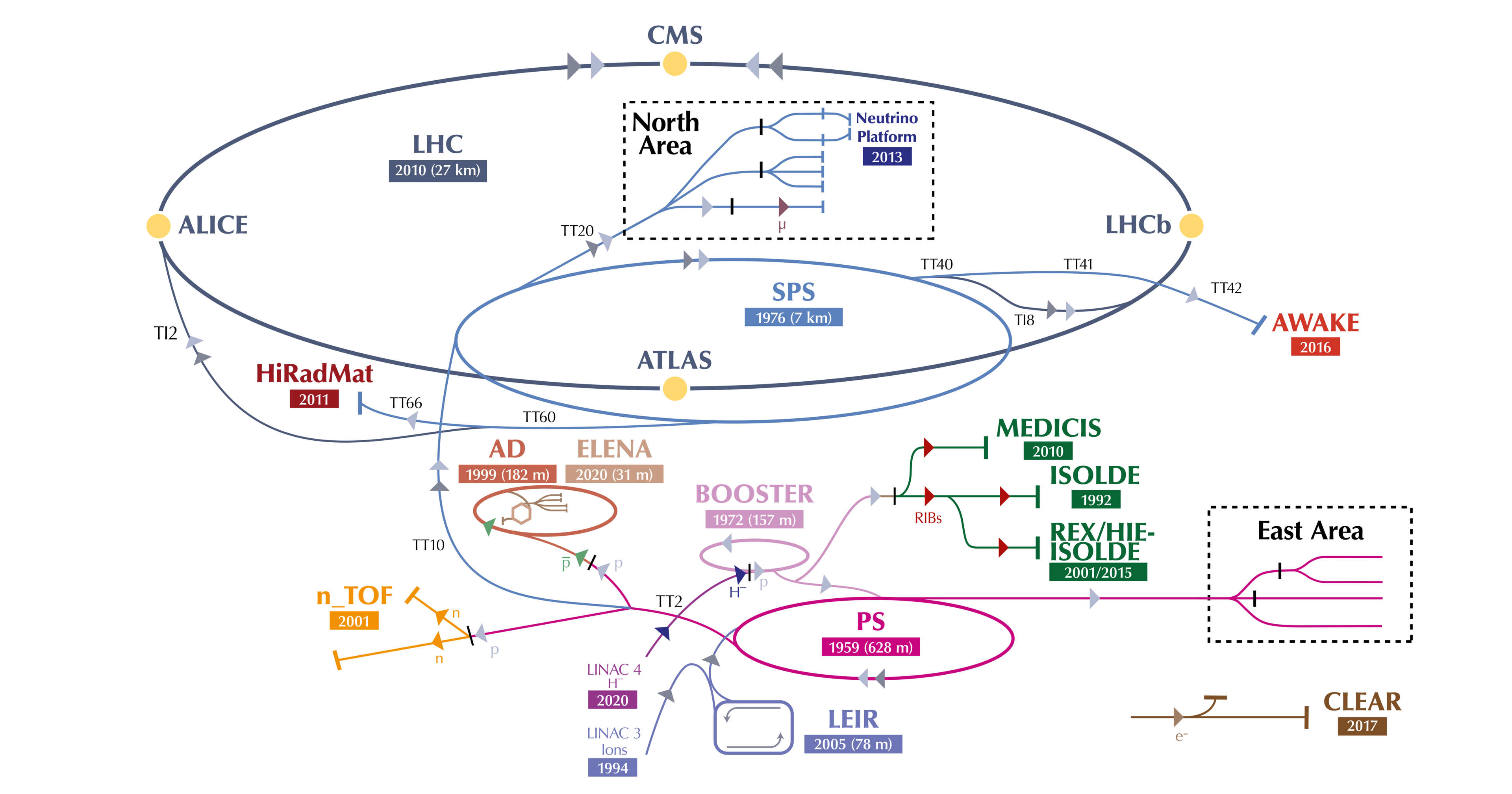}
    \caption{Scheme of the CERN complex in 2022~\cite{lhc_photo}.}
    \label{fig:lhc}
\end{figure}

The European Organization for Nuclear Research (fr. \textit{Conseil Européen pour la Recherche Nucléaire}, CERN) is a core scientific organization operating the largest particle physics laboratory in the world~\cite{fundamental_research}. 
\ac{LHC} at CERN is the most extensive and powerful particle accelerator. Its primary objectives are the measurement of particle properties, the exploration of new particles postulated by supersymmetric theories, and the investigation of various unresolved questions in the area of particle physics~\cite{cern_brochure}. There are four special intersection points at the \ac{LHC} designed for particle collisions, as shown in Figure~\ref{fig:lhc}. The \ac{LHC} is located at the top of the image as a large ellipse. Yellow dots denote the crossing points for key experiments, namely ALICE, CMS, LHCb, and ATLAS.

\ac{ALICE}~\cite{alice} at the \ac{LHC} is specifically designed for the investigation of heavy-ion collisions. \Ac{ALICE} investigates the behavior of strongly interacting matter under extreme conditions, resulting in the creation of a unique state of nuclear matter (quark-gluon plasma) at very high temperatures and densities. As of 2017, \ac{ALICE} employed a comprehensive set of 18 detector systems~\cite{alice_detectors}. %
These detectors serve multiple purposes, providing essential information about the particles involved in collisions, such as the mass, velocity, or electrical charge of the particles.

\begin{figure}[h!]
    \centering
    \begin{subfigure}[]{0.32\textwidth}
        \includegraphics[width=\linewidth]{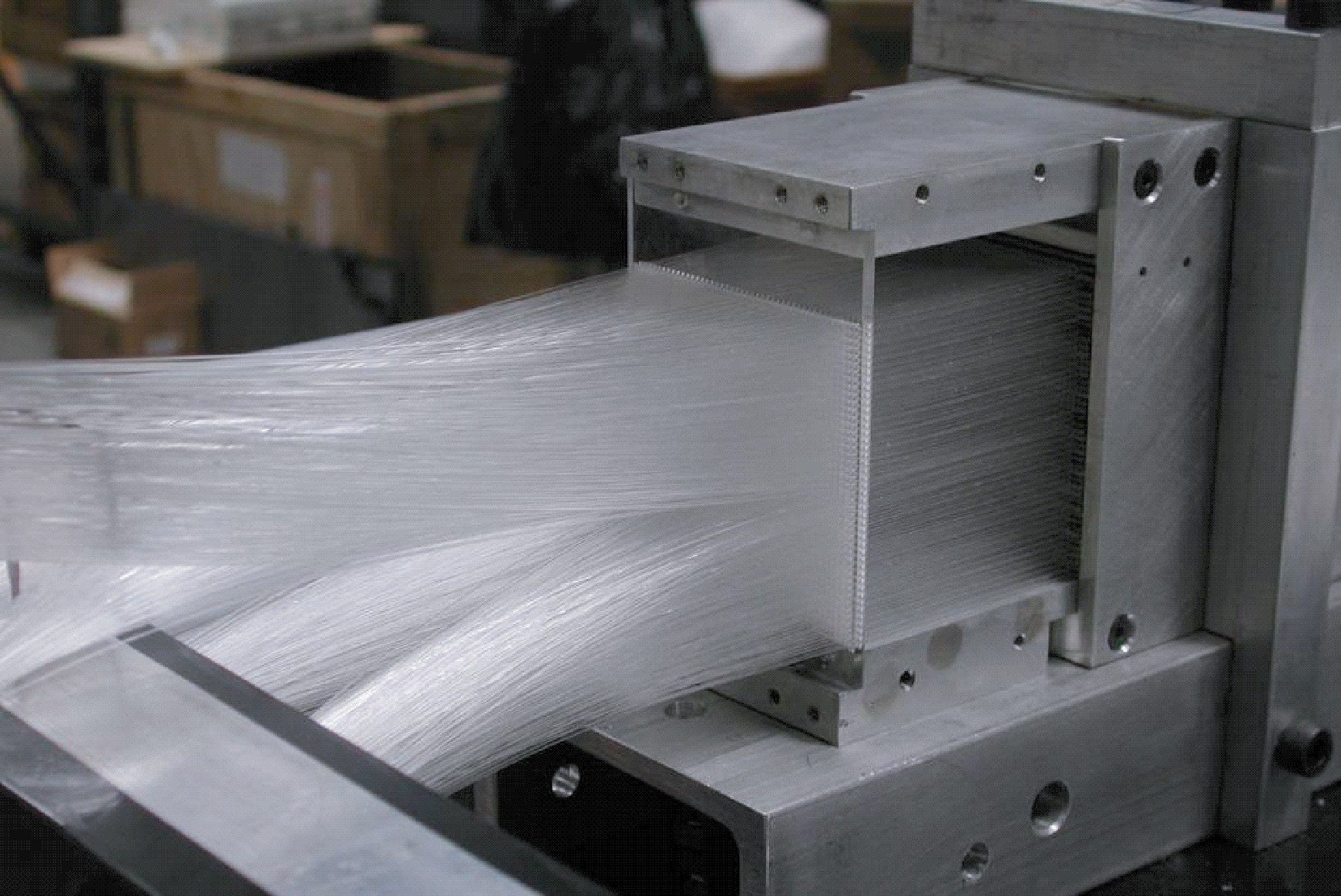}
        \caption{The quartz fibers divided into channels.}
        \label{fig:zdc_1}
    \end{subfigure}
    \hfill
    \begin{subfigure}[]{0.3\textwidth}
        \includegraphics[width=\linewidth]{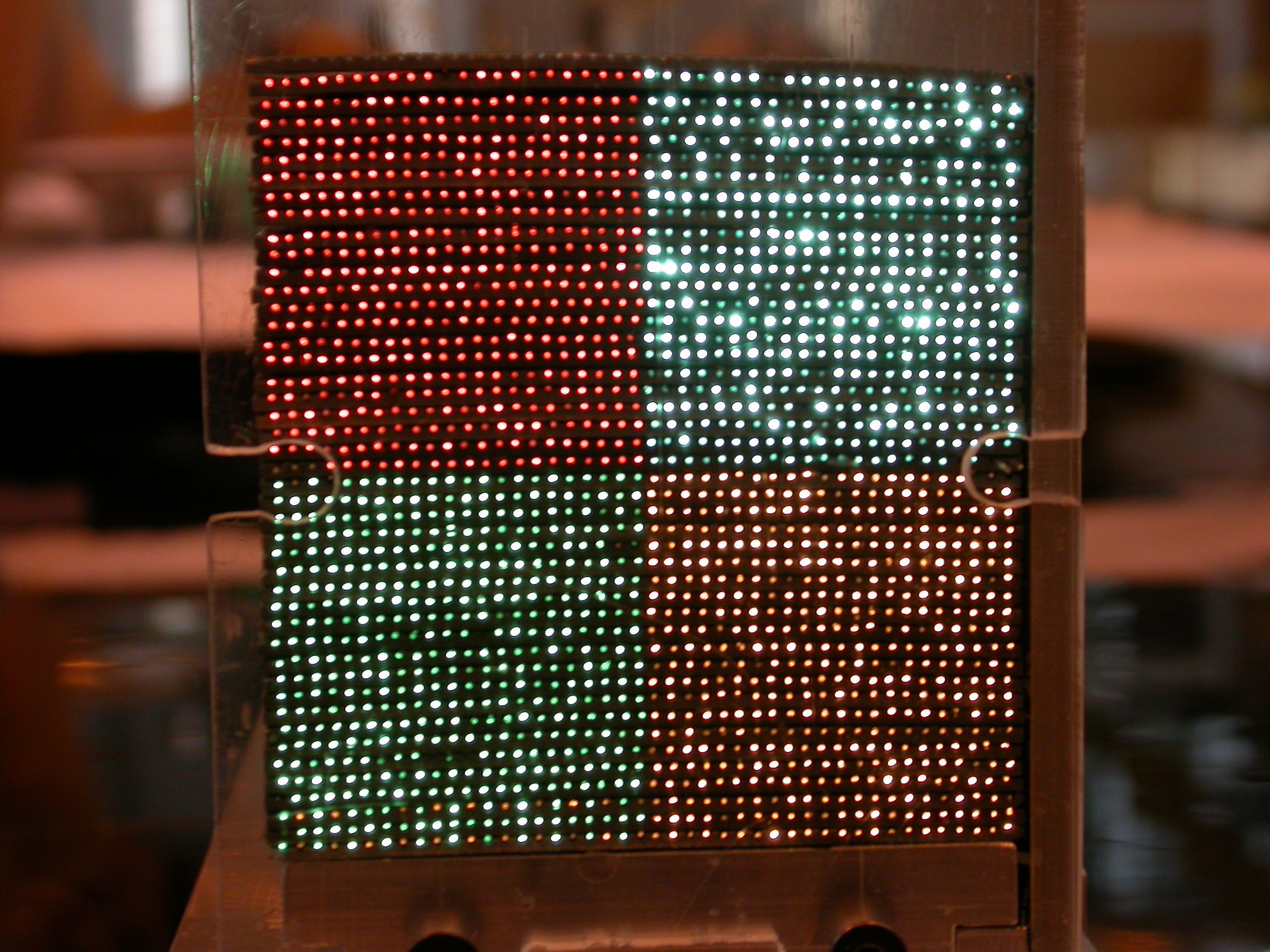}
        \caption{The matrix of quartz fibers.}
        \label{fig:zdc_3}
    \end{subfigure}
    \hfill
    \begin{subfigure}[]{0.34\textwidth}
        \includegraphics[width=\linewidth]{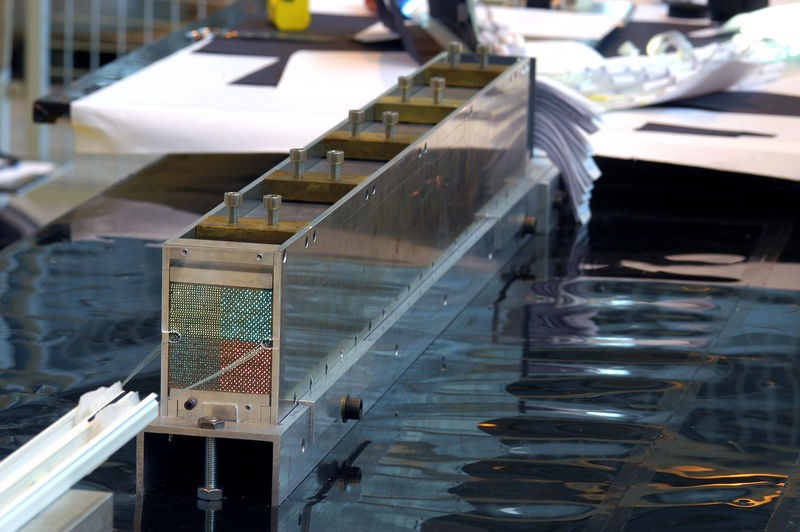}
        \caption{The ZN detector.\\ \mbox{}}
        \label{fig:zdc_2}
    \end{subfigure}
    \caption{The \ac{ZDC} neutron detector construction~\cite{zdc_image_1, zdc_image_2}.}
    \label{fig:zdc}
\end{figure}

The \ac{ZDC}~\cite{zdc_technical_report} in the \ac{ALICE} experiment measures particle showers to determine the centrality of collisions, helping to understand their dynamics~\cite{zdc_centrality}. Comprising four calorimeters, the system includes two dedicated for proton detection (ZP) and two for neutron detection (ZN). An image of ZN is provided in Figure~\ref{fig:zdc}.

\begin{figure}[h!]
    \centering
    \includegraphics[width=0.5\textwidth]{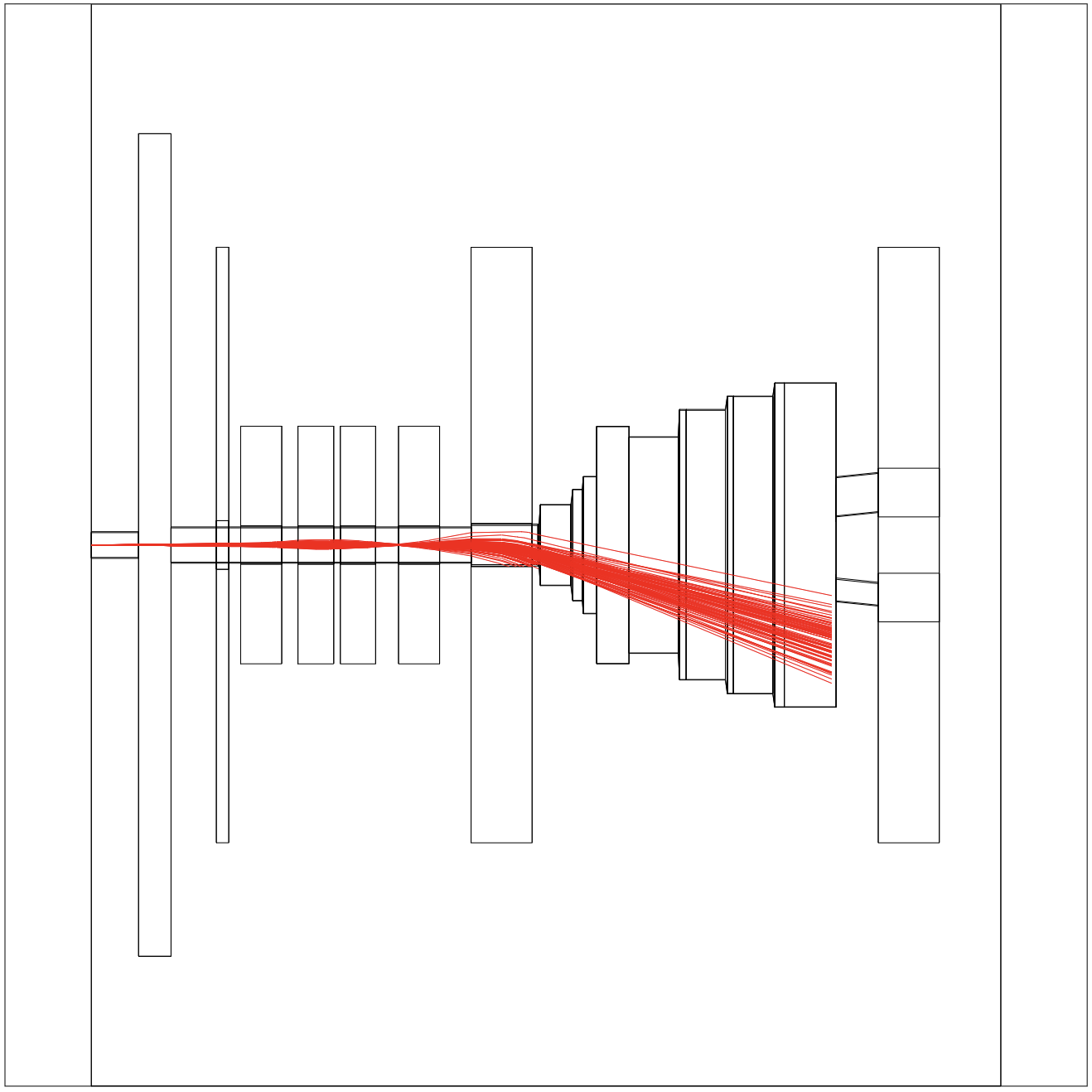}
    \caption{Proton tracks passing through the \ac{ALICE} detector simulated using GEANT~\cite{zdc_technical_report}.}
    \label{fig:geant}
\end{figure}

\begin{figure}[!h]
    \centering
    \includegraphics[width=\textwidth]{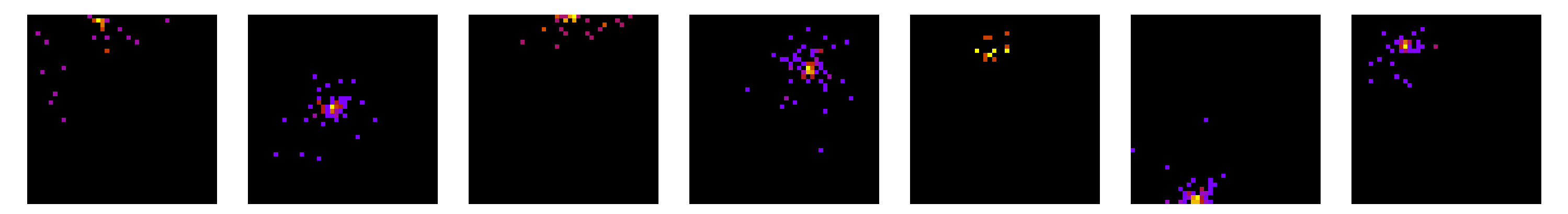}
    \caption{Example \ac{ZDC} neutron detector responses simulated using Monte Carlo techniques.}
    \label{fig:zdc_sim}
\end{figure}

The conventional method for simulating \ac{ZDC} responses relies on GEANT, a Monte Carlo toolkit designed to simulate the trajectory of particles through matter~\cite{geant4, geant}. The procedure involves two primary steps using GEANT. First, it tracks the generated particles and nucleons as they pass through the beam elements up to the calorimeters (Figure~\ref{fig:geant}). Then, it models the particle shower inside the calorimeters. The Cherenkov light production and the transport of light to the photomultipliers are simulated through a separate program~\cite{zdc_technical_report}. Figure~\ref{fig:zdc_sim} shows examples of simulated detector responses. The existing techniques involve intricate processing to calculate all the potential interactions between particles and matter. Although this approach gives precise results, it comes with a significant computational cost.

\newpage
Recent research has shown that generative neural networks are a promising alternative to traditional methods for fast simulation, particularly in the field of high-energy physics~\cite{deOliveira2017, vae_jet, calo_flow}. These networks have been applied to fast \ac{ZDC} simulation~\cite{zdc_sinkhorn, zdc_diffusion, zdc_gan_diversity}, but their use has been limited and often does not leverage the latest advancements in computer vision.

The purpose of this thesis is to investigate the application of state-of-the-art generative neural networks to the simulation of the \ac{ZDC} neutron detector. This study explores novel architectures and frameworks, such as \ac{ViT} and MLP-Mixer, which have not been applied in this domain. In addition, modern generative frameworks like \ac{VQ} and diffusion models are examined. The aim is to evaluate the performance of these models compared to established methods, highlight their potential advantages, and identify any challenges they may present.

\section{Contributions}

The contribution of this thesis can be summarized as follows:

\begin{itemize}
    \item Implementation and evaluation of various state-of-the-art neural network architectures and generative frameworks with comparison to existing methods.
    \item Achieving a competitive to state-of-the-art Wasserstein metric score of 3.15 in the \ac{ZDC} neutron detector simulation task using a diffusion model, with a low generation time of 5 milliseconds per sample.
    \item Formulating a set of recommendations for developing models that enable fast \ac{ZDC} simulation.
    \item Providing open-source code and detailed hyperparameter settings for reproducibility and further research\footnote{\url{https://github.com/m-wojnar/zdc}}.
\end{itemize}

\section{Thesis overview}

The following chapter provides an overview of the literature on generative neural networks and fast simulations in high-energy physics. Chapter~\ref{sec:theoretical} describes the generative neural networks discussed in this thesis. Subsequently, in Chapter~\ref{sec:experiments} the methodology and results of the experiments are presented, and finally the thesis is concluded in Chapter~\ref{sec:summary}\footnote{During the preparation of this work the OpenAI’s ChatGPT and Writefull were used to correct grammar or spelling mistakes and to improve readability.}.

\chapter{Related work}
\label{sec:related_work}

The aim of this chapter is to provide an overview of the current state of the art, review the key literature, and identify unsolved problems regarding the scope of this work. The main focus is on generative neural networks in computer vision 
and fast simulations, including an application of \ac{ML} in physical simulations. Throughout this work an ``architecture'' refers to the specific type and organization of layers that constitute a neural network. In contrast, the term ``generative framework'' is used to denote a system of networks that together form a generative model.

\section{Generative neural networks}
\label{sec:generative_neural_networks}

The scope of this work is closely related to generative neural networks, therefore the following is a detailed overview of the relevant literature. This section focuses on a review of the existing work, while the theoretical description of the presented models is given in Chapter~\ref{sec:theoretical}.

\subsection{Autoencoders}

The latent space of classical autoencoders lacks a structured and continuous nature, preventing the effective generation of new data points. To solve this problem and enable sampling from the latent space, \ac{VAE}~\cite{vae} was proposed in 2014. 
Classical \acp{VAE} often encounter issues with blurred images and low output resolution, leading to the development of modern architectures enabling the generation of large, high-quality samples. This is achieved through the application of an extremely deep hierarchical structure, as in \ac{VDVAE}~\cite{vdvae} introduced in 2020 and further improved in Efficient-VDVAE~\cite{efficient_vdvae} from 2022. The authors of~\cite{vdvae} argue that a sufficiently deep \ac{VAE} can match the performance of deep autoregressive models in terms of maximum likelihood estimation.

Note that there exists another family of autoencoders that does not rely on \ac{KL} loss. For instance, the Wasserstein autoencoder~\cite{wasserstein_ae}, introduced in 2017, uses the Wasserstein distance instead. 
Another method is to employ the Sinkhorn algorithm, as in the \ac{SAE}~\cite{sinkhorn_ae} from 2018. 
The subsequent family of autoencoders are the discrete autoencoders, described in more detail in Section~\ref{sec:quantization_related_work}.

\subsection{Generative adversarial networks}

\Ac{GAN}~\cite{gan}, presented in 2014, was a significant milestone in the field of generative neural networks. 
For a long time, \acp{GAN} have held the leading position in the area of image generation, with the 2018 BigGAN~\cite{big_gan} demonstrating unprecedented performance.

\acp{GAN} constitute a versatile framework suitable for a range of applications, therefore it is often combined with other models. For example, the adversarial autoencoder~\cite{adversarial_ae} replaces \ac{KL} loss in favor of the adversarial training procedure. On the other hand, BigBiGAN~\cite{gan_representation_learning} enables \ac{GAN} to simultaneously learn latent representations, which have been the domain of autoencoders. The discriminators are often used to implement an adversarial loss function, as in~\cite{vq_gan}.

\acp{GAN} frequently suffer from training instability and mode collapse. Various solutions to these problems are documented in the literature, such as the introduction of the Wasserstein distance into the optimization goal in \ac{WGAN}~\cite{wgan}, the use of normalization layers~\cite{dcgan}, the application of gradient penalties~\cite{gan_gp}, and the application of a label smoothing. The authors of~\cite{gan_convergence} analyze common techniques and argue that zero-centered gradient penalties lead to stable training, while many popular \ac{GAN} variants still struggle with convergence issues. Another problem encountered by \acp{GAN} is the high fidelity of local patterns while facing difficulties in accurately reproducing the overall structure of the image, which is crucial for image-to-image tasks. The authors of~\cite{gan_loss} propose to combine loss functions that capture global (e.g., $l1$ or $l2$ loss) and local structure (adversarial loss). They also use a high dropout during both training and testing as a form of regularization.

\subsection{Autoregressive models}

Autoregressive models~\cite{nade}
excel in time series analysis and are also effective in image generation. One possible approach is the pixel by pixel generation, like in Pixel Recurrent Neural Network~\cite{pixel_rnn} introduced in 2016. This concept was further extended by PixelCNN~\cite{pixel_cnn} by leveraging gated convolutional layers. 
Similarly, iGPT~\cite{igpt} from 2020 treats images as sequences of raw pixels. Despite its strengths, iGPT faces several challenges, particularly the high computational cost due to the quadratic complexity of the attention mechanism, which limits its scalability and efficiency. 
\ac{MaskGIT}~\cite{mask_git}, presented in 2022, is a modern generative framework based on a bidirectional transformer decoder. Unlike iGPT, \ac{MaskGIT} employs a two-stage design that includes an image tokenizer and functions within the latent space. 
The 2024 Visual AutoRegressive modeling~\cite{var} significantly differs from conventional autoregressive methods by implementing a multi-scale, coarse-to-fine ``next-scale prediction''. This technique allows to achieve performance comparable to that of diffusion models with a substantial improvement in inference speed. Furthermore, the model demonstrates strong scalability and zero-shot generalization abilities.

\subsection{Normalizing flows}

\Acp{NF}~\cite{normalizing_flows}, introduced in 2015, often employ an autoregressive structure, for example, Masked Autoregressive Flow~\cite{maf} or Inverse Autoregressive Flow~\cite{iaf}. Despite their strong mathematical foundation, they are computationally inefficient during forward or backward pass (depending on the model), making large-scale deployment difficult. An advanced theory of continuous normalizing flows~\cite{neural_ode} also exists, which employs differential equation solvers. Dedicated generative frameworks for images have also been developed, such as RealNVP~\cite{real_nvp} or Glow~\cite{glow}. However, they have not been established as a state-of-the-art solution in the field of computer vision.

\subsection{Vector quantization}
\label{sec:quantization_related_work}

Introduced in 2017, \ac{VQVAE}~\cite{vq_vae} provided a novel approach to discrete representation learning.
In 2021, the model was expanded to VQ-GAN~\cite{vq_gan}, where \ac{VQVAE} serves as a generator and the model leverages a transformer as a learnable prior, instead of PixelCNN suggested by the authors of \ac{VQVAE}. VQ-GAN model incorporates a patch-based adversarial loss and perceptual loss~\cite{perceptual}, and adopts sampling techniques from natural language processing such as \textit{top-k} sampling, \textit{top-p} (nucleus) sampling~\cite{nucleus}, and rejection sampling.
The development of \ac{ViTVQGAN}~\cite{vit_vq_gan} further enhanced the framework by leveraging a \ac{ViT}-based generator. The authors modify the loss function and improve codebook utilization by matching vectors projected to a lower dimension and applying $l2$ normalization to both the encoder output and the codebook vectors.
DALL-E~\cite{dalle}, also launched in 2021, shares similar concepts and enables text-to-image capabilities while scaling up the model significantly. 

\subsection{Diffusion models}

Introduced as a concept in non-equilibrium thermodynamics~\cite{nonequilibrium}, diffusion models have gained significant attention in generative modeling through the foundational work on \acp{DDPM}~\cite{ddpm} from 2020. 
Diffusion models quickly achieved superior performance and became the new state-of-the-art in image generation~\cite{diffusion_beats_gan}. Furthermore, conditional diffusion models without the need for a classifier guidance~\cite{classifier_free} have simplified their implementation while maintaining performance.
\ac{DDIM}~\cite{ddim} extended the framework by providing a more efficient sampling. %
Significant developments in this domain include the introduction of the latent diffusion model~\cite{stable_diffusion} in 2022 and the upscaling of models such as \ac{SDXL}~\cite{sdxl} from 2023. Techniques to accelerate the diffusion process, such as adversarial diffusion distillation in \ac{SDXL} Turbo~\cite{sdxl_turbo}, have reduced computational overhead by decreasing the number of required denoising steps. Moreover, the diffusion framework has adapted \ac{VQ}~\cite{vq_diffusion} along with the \ac{ViT} architecture~\cite{dit}. The current state-of-the-art diffusion model is Stable Diffusion 3.0~\cite{sd3} presented in 2024, which is based on the concept of rectified flows~\cite{flow_matching}.

\section{Fast simulations}
\label{sec:fast_sim}

Numerical simulations using \ac{HPC} have become integral to scientific research, forming the third pillar of science alongside theory and experiments. The \Ac{HPC} enables a detailed exploration of complex systems that are otherwise infeasible to study using analytical methods. These simulations are essential across various fields, allowing researchers to model phenomena ranging from subatomic particle interactions to large-scale climate systems.

Numerical simulations can be categorized into several main types according to their methodologies and applications. Continuous simulations involve solving differential equations that describe the continuous behavior of systems over time and space, such as fluid dynamics, electromagnetic fields, and heat transfer problems. Techniques such as finite element analysis~\cite{fem} and finite difference method~\cite{fdm} are frequently used in these contexts. In contrast, discrete simulations represent systems in which changes occur at specific intervals, employing techniques such as cellular automata and agent-based models~\cite{discrete}. These methods are widely used in fields such as epidemiology, ecology, and social sciences. Another essential category is the discrete-event simulation~\cite{discrete_event}, which is heavily used in logistics and telecommunications. Mixed simulations combine aspects of both continuous and discrete simulations to model systems that include both continuous processes and discrete events. Additionally, multi-scale simulations address systems operating across multiple spatial or temporal scales.

Despite their power and versatility, modeling complex systems using the \ac{HPC} presents significant challenges, mainly related to computational time and memory usage. To mitigate these issues, researchers employ surrogate models -- simplified versions developed from complex simulations that approximate the system's behavior~\cite{surrogate}. Using surrogate models has become a standard research practice to balance accuracy and computational efficiency. By reducing computational complexity, these models enable more extensive exploration of parameter spaces, uncertainty quantification, and real-time decision making. Surrogate models are also frequently used in high-energy physics, particularly at CERN~\cite{fast_sim_cern}.

\subsection{Machine learning for fast physical simulations}

The emergence of \ac{ML} and neural networks has transformed the creation of surrogate models through significant success in approximating complex systems combined with computational versatility.
The growing role of \ac{ML} in the physical sciences is highlighted in~\cite{ml_and_physical}, where the authors discuss the range of applications, emphasizing the importance of interpretability. Another work~\cite{theory_inspired} investigates the fusion of theoretical physics and \ac{ML}. The synergy between physics-driven and data-driven models leads to outcomes that are often more precise, simpler, and capable of better extrapolation.

The application of \ac{ML} in physical simulations can take various forms, such as supervised learning, incorporating a physical loss term, or leveraging differentiable numerical simulations~\cite{pbdl}. Supervised learning, as a conventional technique, involves training algorithms to replicate real-world behaviors using experimental or simulated data. Although straightforward, this approach has limitations, particularly in modeling physical constraints and generalizing beyond the training data. However, its simplicity makes it a fundamental starting point in many \ac{ML} projects. Another approach is the introduction of the physical loss terms, where existing knowledge of physical processes is used to guide the model towards producing realistic effects. This method often yields better generalizing models, but it can be slower and does not guarantee the reliability of the model. The next step is to combine differentiable numerical simulations with gradient-based \ac{ML} techniques, which further improves learning feedback and generalization. However, it is more complex to implement and requires external dependencies.

The wide application of \ac{ML} in physics makes a comprehensive coverage of all aspects challenging, so several key developments have been selected to illustrate the extensive role of neural networks in the physical sciences.
One of the most significant and recognizable neural networks in the field of physics are \acp{PINN} presented in a two-part article~\cite{pinn_1, pinn_2}. \Ac{PINN}, designed to solve differential equations, integrates the equations directly into the network training process. As~\cite{pinn_where_we_are} details, \ac{PINN} offers several benefits over conventional methods, such as being mesh-free, enabling on-demand computation after the training, and handling complex geometries and high-dimensional domains. However, they suffer from generalizability issues, discontinuities, and complex boundaries. Despite this, they are successfully used in numerous fields, including fluid dynamics, optics, electromagnetics, molecular dynamics, and industrial applications.

Physics simulations can also be efficiently performed by graph neural networks. The introduction of Graph Network-based Simulators~\cite{complex_physics_gnn} allowed for the simulation of various physical domains, including fluids, rigid solids, and deformable materials. This model has shown remarkable generalization capabilities, scaling from single-timestep predictions to complex, multi-timestep scenarios with a significantly higher number of particles.

\subsection{Generative neural networks for fast simulations in CERN}

CERN predicts a significant increase in its computational power needs unless it incorporates Research \& Development improvements, especially in the area of fast simulations~\cite{calochallenge}. Consequently, various efforts have been made to accelerate the simulation of experiments at CERN~\cite{fast_mc}. The emergence of generative neural networks has opened up new possibilities for the development of fast simulations, independent of Monte Carlo methods. However, most of these studies use Monte Carlo simulation results as a training dataset, due to their widespread availability, with many relying on data generated by GEANT.

The first articles presenting the use of generative neural networks for fast simulations at CERN date back to 2017. The authors of~\cite{deOliveira2017} introduce \ac{LAGAN}, a set of guidelines for learning \acp{GAN} in sparse regimes where location within the image is crucial. The work presents a simulation of jet images, which are observable outcomes of quarks and gluons scattering at high energy.

The same team presented CaloGAN~\cite{calo_gan} in 2018, which employs a combination of \ac{LAGAN} networks to model electromagnetic showers in longitudinally segmented calorimeters, achieving a speedup of approximately $10^5$ times on \acp{GPU} compared to traditional simulation methods. That year also saw the introduction of 3D GAN~\cite{3d_gan}, focusing on simulations for the Compact Linear Collider detector. This model demonstrates the potential of \acp{GAN} to efficiently simulate complex three-dimensional structures. Additionally,~\cite{Musella2018} presented a \ac{GAN} model trained on CMS collaboration data to simulate hadronic jets in proton-proton collisions, highlighting the versatility of neural networks in modeling complex particle interactions.

In 2019, further progress has been observed. The study of~\cite{tpc} explored the use of \acp{GAN} in simulating the operations of the Time Projection Chamber in the \ac{ALICE} experiment, achieving notable speed improvements while indicating a trade-off with simulation fidelity. Another study~\cite{mmd_gan} improved the accuracy of \acp{GAN} in simulating mass distributions by introducing the \ac{MMD}, a distance in the space of probability measures, in the generator loss.

The year 2020 saw the introduction of a modified \ac{VAE} framework~\cite{vae_jet} capable of generating realistic and diverse high-energy physics events trained using the dataset from~\cite{deOliveira2017}. This model allows for control over generated events with a continuous latent space.

CaloFlow~\cite{calo_flow}, a framework that uses \ac{NF} to simulate calorimeter responses, was presented in 2021. It also introduced a new metric for evaluating the quality of generative models: the performance of a classifier trained to distinguish between real and generated images. An improved version, CaloFlow II~\cite{calo_flow_ii}, increased the shower generation speed by a factor of 500, making it approximately $10^4$ times faster than GEANT4. Another development~\cite{getting_high} models electromagnetic showers in the International Large Detector, which is a proposed particle detector designed for the International Linear Collider, a planned high-energy electron-positron collider. The work is not directly related to the modeling of detectors at CERN but introduces interesting solutions. The approach involves integrating Bounded Information Bottleneck Autoencoder~\cite{bib_ae}, \ac{GAN}, a postprocessing network, and the \ac{MMD} as a loss function.

In 2022, the Calo Challenge~\cite{calochallenge2022} initiated by CERN emphasized the demand for more efficient simulation methods. This challenge led to the development of CaloMan~\cite{calo_man}, a method for modeling calorimeter showers by learning their manifold structure using a generalized autoencoder. It allowed faster training and generation due to reduced data dimensionality. In addition, the authors used a conditional \ac{NF} for the density estimation on the learned manifold.

In 2023, a technique combining an autoencoder with \ac{NF}~\cite{vae_nf} was introduced. This method allows for easy sampling from an analytical probability distribution and conversion to the latent space of the autoencoder using \ac{NF}.

\subsection{Generative neural networks for fast simulations of ZDC}

In the ongoing \ac{LHC} Run 3, \ac{ML} has been applied to a limited number of core tasks, including the calibration of particle-identification data from the \ac{ALICE} Time-Projection Chamber and the software trigger for heavy-flavour research~\cite{alpaca}. Despite the limited number of applications, Research \& Development efforts are continuously progressing. The \Ac{ZDC} appears to be particularly suitable for \ac{ML} applications due to its two-dimensional structure and computational complexity of simulation~\cite{lpcc}. The ZN simulation presents a significant challenge for classical generative frameworks: \acp{VAE} generate blurred results, whereas \acp{GAN} are often unable to learn the detector responses consistent with physics.

A 2021 study introduced a new approach, \ac{e2eSAE} with a noise generator~\cite{zdc_sinkhorn}. This model enhances the standard autoencoder design by incorporating a noise generator -- a neural network trained to transform samples from a specific distribution into the latent space of the autoencoder. This method is particularly effective for imbalanced conditional classes, enabling distinct classes to be encoded separately in the latent space. The presented \ac{e2eSAE} shows competitive performance in capturing physical relations, such as the collision center, which is crucial for realistic data simulation.

\ac{ZDC} has a distinctive feature -- some particles with identical properties (mass, charge) and dynamic parameters (such as energy, momenta, primary position) give consistent calorimeter responses, while others result in diverse responses. The limited diversity of samples generated by \acp{GAN} was addressed in 2023 by the authors of~\cite{zdc_gan_diversity}. They point out that conditional \acp{GAN} are especially prone to mode collapse and often disregard the input noise vector in favor of conditional information. The proposed method, SDI-GAN, selectively enhances the diversity of GAN-generated samples by incorporating an additional regularization factor into the training loss function. This approach encourages the generator to explore new data modes for particles with diverse responses while maintaining consistency for others.

Another 2023 study analyzes the ZN simulation with \ac{ML} models~\cite{zdc_ml}. The authors evaluated the basic \ac{VAE} and \ac{GAN} models and improved the \ac{GAN} framework with an additional regularization network (auxiliary regressor) and postprocessing step. The method additionally employs a neural network classifier to filter inputs that do not trigger the ZN response before passing the data to the generative network.

A different approach proposed in 2024 in~\cite{zdc_diffusion} leverages a diffusion model, which achieves high-fidelity results and provides control over the simulation quality by adjusting the number of denoising steps. The authors note a high generation time and propose a compromise by using a latent diffusion model to accelerate the simulation process. \cite{zdc_corr_vae} employs CorrVAE~\cite{corrvae} to encode different aspects of an object into two separate hidden variables. This is achieved by having separate encoders for properties and objects. Additionally, the correlation between properties is identified and processed by the mask pool layer, which consolidates the relevant information into a bridging latent variable. The authors of~\cite{zdc_gan_proton} integrate SDI-GAN with an auxiliary regressor and intensity regularizator to model the responses of the \ac{ZDC} proton detector. This work is a pioneering application of generative neural networks for fast ZP simulation. 

Table~\ref{tab:cern_fast_sim} summarizes the literature presented in this section in chronological order, indicating the used models.

\begin{table}[h!]
    \caption{Chronological overview of papers applying generative neural networks for fast simulations in high-energy physics.}
    \centering
    \begin{tabular}{|c|c|c|c|c|c|c|}
    \hline
    \textbf{Year} & \textbf{Reference} & \textbf{GAN} & \textbf{Autoencoder} & \textbf{NF} & \textbf{VQ} & \textbf{Diffusion} \\  \hline \hline
    2017 &~\cite{deOliveira2017}    & \checkmark &            &            &            &            \\ \hline
    2018 &~\cite{calo_gan}          & \checkmark &            &            &            &            \\ \hline
    2018 &~\cite{3d_gan}            & \checkmark &            &            &            &            \\ \hline
    2018 &~\cite{Musella2018}       & \checkmark &            &            &            &            \\ \hline
    2019 &~\cite{tpc}               & \checkmark &            &            &            &            \\ \hline
    2019 &~\cite{mmd_gan}           & \checkmark &            &            &            &            \\ \hline
    2020 &~\cite{vae_jet}           &            & \checkmark &            &            &            \\ \hline
    2021 &~\cite{calo_flow}         &            &            & \checkmark &            &            \\ \hline
    2021 &~\cite{calo_flow_ii}      &            &            & \checkmark &            &            \\ \hline
    2021 &~\cite{getting_high}      &            & \checkmark &            &            &            \\ \hline
    2021 &~\cite{zdc_sinkhorn}      &            & \checkmark &            &            &            \\ \hline
    2022 &~\cite{calo_man}          &            & \checkmark & \checkmark &            &            \\ \hline
    2023 &~\cite{vae_nf}            &            & \checkmark & \checkmark &            &            \\ \hline
    2023 &~\cite{zdc_gan_diversity} & \checkmark &            &            &            &            \\ \hline
    2023 &~\cite{zdc_ml}            & \checkmark & \checkmark &            &            &            \\ \hline
    2024 &~\cite{zdc_diffusion}     &            &            &            &            & \checkmark \\ \hline
    2024 &~\cite{zdc_corr_vae}      &            & \checkmark &            &            &            \\ \hline
    2024 &~\cite{zdc_gan_proton}    & \checkmark &            &            &            &            \\ \hline
    \end{tabular}
    \label{tab:cern_fast_sim}
\end{table}

\section{Conclusions}

Significant advances in \ac{ML}, especially in the area of deep learning, have enabled the simulation of physical phenomena. CERN is a prime example of an organization benefiting from the development of generative neural networks for fast simulations, as they facilitate ongoing and enable expansion of research in high-energy physics. 

In particular, exploring the \ac{ZDC} simulation is emerging as an essential area with practical applications in upcoming \ac{LHC} runs. Despite the existing literature on this subject, it does not comprehensively address the application of the most recent advances in the field of generative neural networks. None of the works uses models from the MetaFormer family and, with a single exception, the existing approaches are not based on modern \ac{VQ} or diffusion frameworks (Table~\ref{tab:cern_fast_sim}). Furthermore, the proposed solutions rarely include any physical loss terms. Therefore, the following objectives were established within this work in the task of ZN simulation:

\begin{itemize}
    \item evaluation of the surrogates based on state-of-the-art neural network architectures, specifically \ac{ViT} and MLP-Mixer,
    \item application of novel generative frameworks, in particular \ac{VQ} and diffusion,
    \item comparison with existing models.
\end{itemize}

\chapter{Generative neural networks}
\label{sec:theoretical}

This chapter presents the most relevant aspects of the generative neural networks discussed in this thesis. It covers a theoretical introduction, a discussion of potential challenges, and technical remarks.

\section{Variational autoencoders}
\label{sec:autoencoders_background}

The classic autoencoder framework consists of two neural networks: an encoder and a decoder. The former encodes the input data into a latent space, while the latter maps the points from the latent space back to the original data space. Autoencoders can be used for representation learning, but need additional extensions to become generative models, as it is not possible to directly sample from the non-regularized latent space.

To construct a generative model, \ac{VAE} incorporates variational inference of a posterior distribution of a latent variable~\cite{vae}. The main objective of \ac{VAE} is to represent the probability distribution $p(x)$ of the data $x$ by estimating a latent variable model $p(x|z)$, where $z$ is the latent variable. Since the true posterior $p(z|x)$ is usually intractable, \ac{VAE} approximates it with a tractable distribution $q(z|x)$ determined by the output of the encoder. Typically, a Gaussian distribution is selected for $q$, although the \ac{VAE} framework itself is more general and allows for any distribution. The training goal of a VAE, also referred to as the \ac{ELBO}, can be formulated as:

\begin{equation}
    \mathcal{L}(\theta, \phi; x) = \mathbb{E}_{q_{\theta}(z|x)}[\log p_{\phi}(x|z)] - D_{KL}(q_{\theta}(z|x) || p(z)),
    \label{eq:elbo}
\end{equation}

\noindent where $\theta$ and $\phi$ represent the parameters of the encoder $E$ and decoder $D$, respectively, $D_{KL}$ denotes the \ac{KL} divergence, and $p(z)$ stands for the prior distribution over the latent variables. The first term of \ac{ELBO} encourages the decoded samples to resemble the original data, acting as a reconstruction loss. The second component serves as a regularization term, ensuring that the encoded latent variables remain close to the prior distribution. Assuming that the data follows a Gaussian distribution with some standard deviation $\sigma_d$, the reconstruction term can be represented in the form of the $l2$ loss (also known as the \ac{MSE} loss):

\begin{equation}
\begin{aligned}
    \mathbb{E}_{q_{\theta}(z|x)}[\log p_{\phi}(x|z)] 
    &= \mathbb{E}_{q_{\theta}(z|x)}[\log \mathcal{N}(x | D_\phi(z), \sigma^2_{d})] \\
    &= \mathbb{E}_{q_{\theta}(z|x)} \left[ - \frac{\| x - D_{\phi}(z) \|^2}{2 \sigma^2_{d}} - \frac{d}{2} \log (2\pi \sigma^2_{d}) \right] \\
    &= -\frac{1}{2 \sigma^2_{d}} \mathbb{E}_{q_{\theta}(z|x)} \left[ \| x - D_{\phi}(z) \|^2 \right] - \frac{d}{2} \log (2\pi \sigma^2_{d}),
\end{aligned}
\end{equation}

\noindent where $\mathcal{N}$ denotes a Gaussian distribution of $x$ with a mean of $D_\phi(z)$ and a variance of $\sigma^2_{d}$, $D_\phi$ is the decoder parameterized by $\phi$, and $||x||^2$ is the squared $l2$ norm of the vector $x$. Omitting the constants, the reconstruction term is formulated as:

\begin{equation}
    \mathcal{L}_{rec}(\theta, \phi; x) = ||D_\phi(E_\theta(x)) - x||^2
\end{equation}

\begin{figure}[h!]
    \centering
    \includegraphics[width=0.7\linewidth]{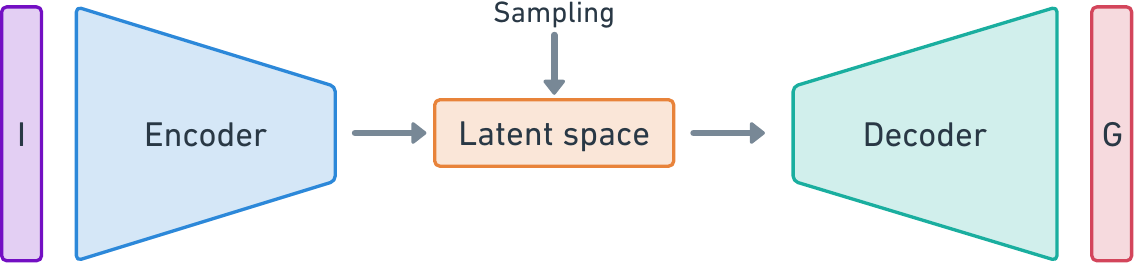}
    \caption{The variational autoencoder design.}
    \label{fig:vae}
\end{figure}

This approach enables \acp{VAE} to simultaneously learn a generative model to produce new data samples and an inference model to derive latent representations of the data, as illustrated in Figure~\ref{fig:vae} (where $I$ represents input data and $G$ represents generated data). Conditioning in \acp{VAE} can be achieved by incorporating conditional variables in the decoder input.

\subsection{Reparametrization trick}

The reparameterization trick~\cite{vae} is a key element in the \ac{VAE} training process, as it enables the propagation of gradients through a stochastic sampling operation. It involves separating randomness from the parameters of the neural network -- instead of directly sampling a latent variable $z$ from the posterior approximation $q_{\theta}(z|x)$, \acp{VAE} represent $z$ as a deterministic function of the input data $x$, parameters $\theta$, and an independently sampled random variable $\epsilon$. In the case of a Gaussian distribution, the reparameterization of $z$ can be expressed as: 

\begin{equation} 
    z(\theta; x) = \mu_{\theta}(x) + \sigma_{\theta}(x) \odot \epsilon, \quad \epsilon \sim \mathcal{N}(0, I)
\end{equation} 

\noindent where $\mu_{\theta}$ and $\sigma_{\theta}$ represent the outputs of the encoder.

\subsection{Posterior collapse}
\label{sec:posterior_collapse}

\begin{figure}[h!]
    \centering
    \begin{subfigure}[]{0.49\textwidth}
        \includegraphics[width=\linewidth]{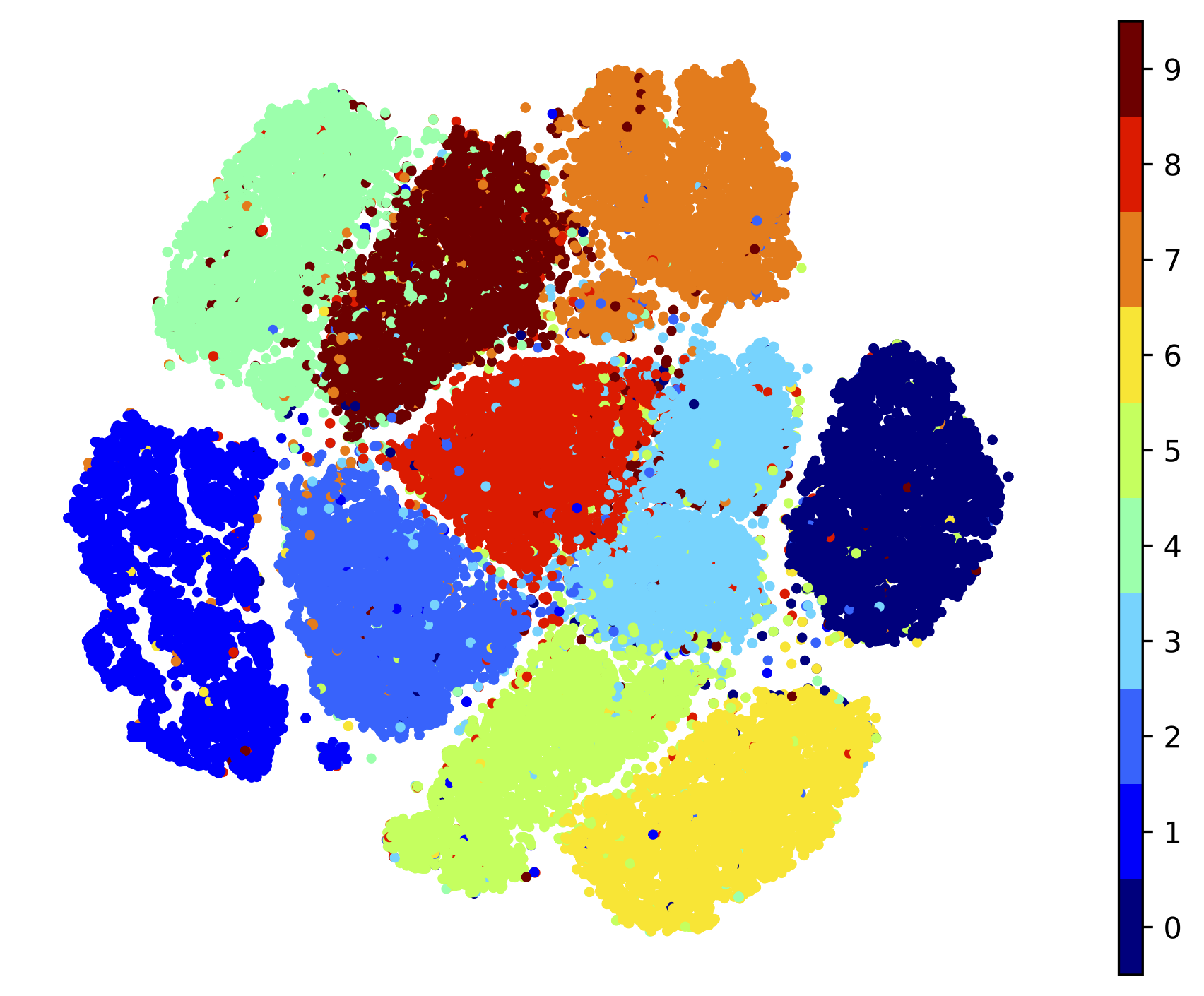}
    \end{subfigure}
    \hfill
    \begin{subfigure}[]{0.49\textwidth}
        \includegraphics[width=\linewidth]{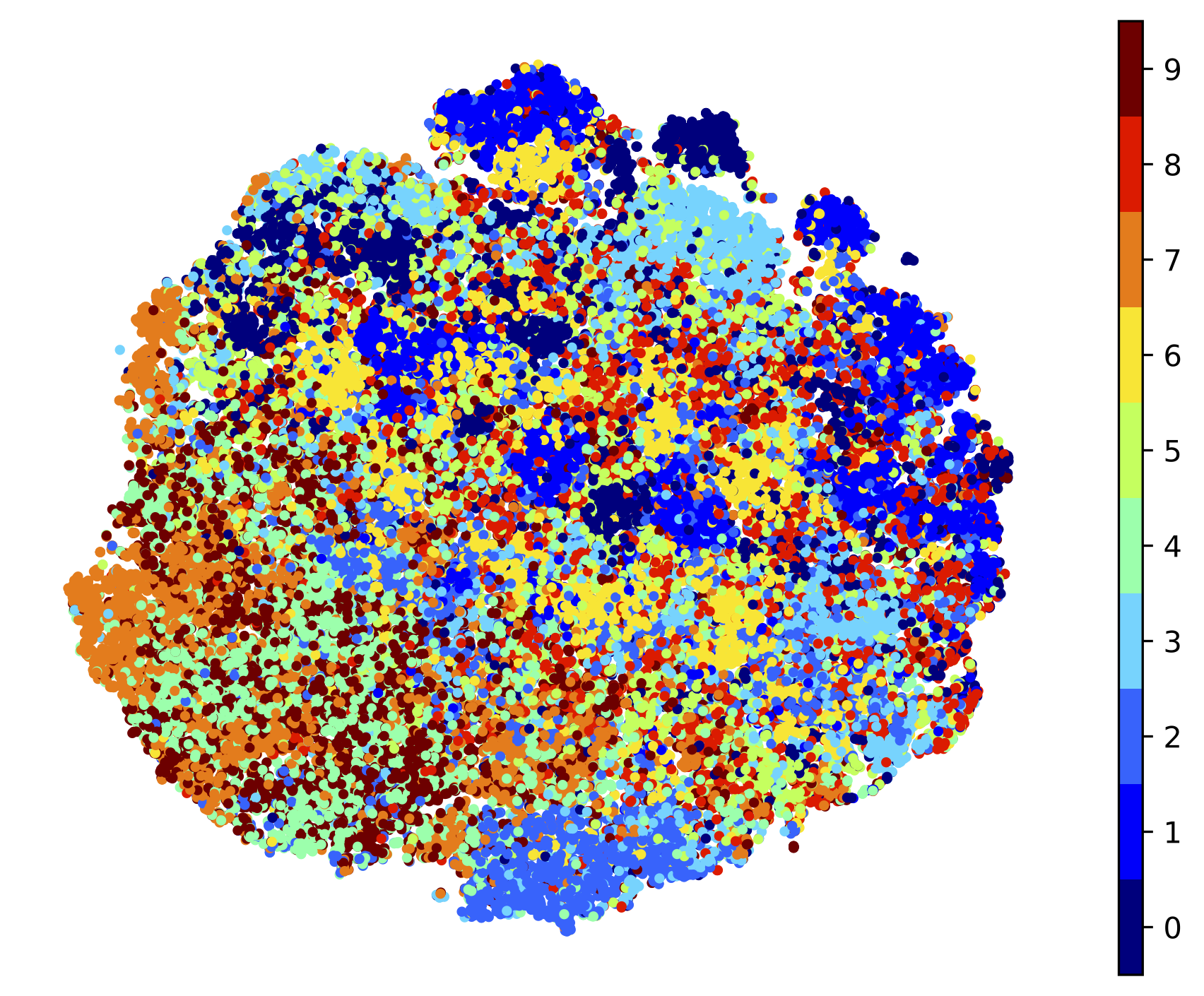}
    \end{subfigure}
    \caption{Visualization of a properly structured latent space (left) and a collapsed latent space of a \ac{VAE} (right).}
    \label{fig:posterior_collapse}
\end{figure}

Posterior collapse in a \ac{VAE} occurs when the variational posterior distribution of a latent variable becomes indistinguishable from the prior distribution. This means that the model starts ignoring latent variables, the visible sign of which is the \ac{KL} loss term dropping to zero. Figure~\ref{fig:posterior_collapse} shows the \ac{tSNE} projection of the \ac{VAE} latent space that was trained to reconstruct the MNIST dataset. The left side illustrates a latent space where the classes are correctly encoded, preserving the separation and placing similar classes in close proximity. In contrast, the right side shows a latent space of a model that has experienced posterior collapse.

Several methods have been proposed to address posterior collapse in \acp{VAE}. These include adjusting the weight of the \ac{KL} term to ensure a trade-off between reconstruction and regularization, implementing \ac{KL} annealing where the significance of the \ac{KL} term is changing during training~\cite{kl_annealing}, and incorporating skip connections to maintain information flow throughout the network.

\section{Supervised autoencoders}

\begin{figure}[h!]
    \centering
    \includegraphics[width=0.7\linewidth]{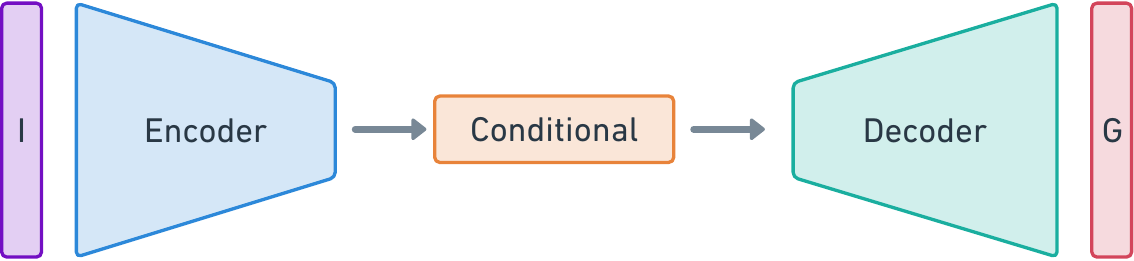}
    \caption{The supervised autoencoder design.}
    \label{fig:supervised_autoencoder}
\end{figure}

As stated in~\cite{supervised_example}, the supervised autoencoder can be viewed as a way to incorporate the physical loss term into the model~\cite{pbdl}. The objective of the supervised autoencoder is to improve the generalization of the neural network by simultaneously predicting targets and inputs~\cite{supervised_autoencoder}. The encoder, illustrated in Figure~\ref{fig:supervised_autoencoder}, transforms the input into conditional variables, while the decoder reconstructs the original data. Note the absence of the \ac{KL} loss term, which distinguishes it from the \ac{VAE}. The loss function of the supervised autoencoder is:

\begin{equation}
    \mathcal{L}(\theta, \phi; x, c) = ||D_\phi(E_\theta(x)) - x||^2 +  \beta ||E_\theta(x) - c||^2,
    \label{eq:supervised_ae}
\end{equation}

\noindent $E_\theta$ represents the encoder with parameters $\theta$, $D_\phi$ represents the decoder with parameters $\phi$, $c$ denotes the conditional variables, and $\beta$ is a weight of the regularization term. Even without the \ac{KL} term in its loss function, the supervised autoencoder can function as a generative model by regularizing the latent space to match the conditional variables space. Note that while conditional variables can be standardized, the latent space may not necessarily be continuous or strictly Gaussian because of the non-normally distributed conditional variables.

\section{Autoencoders with noise generator}

\begin{figure}[h!]
    \centering
    \includegraphics[width=0.7\linewidth]{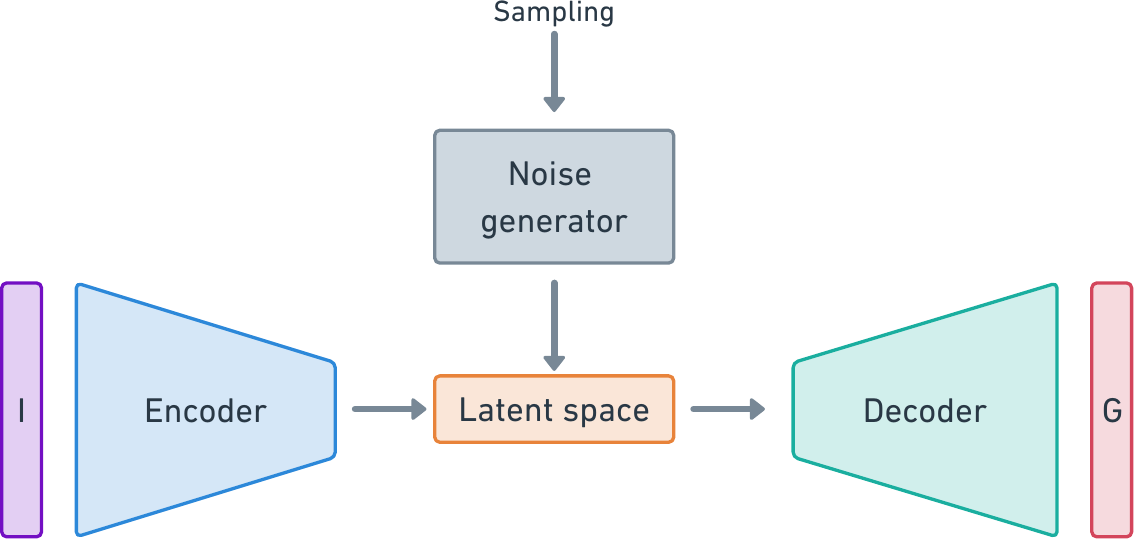}
    \caption{The autoencoder with noise generator design.}
    \label{fig:ae_ng}
\end{figure}

A \ac{VAE}, as a generative model, constrains the latent space to match a particular distribution, potentially limiting the expressiveness of the model and placing conditional variables on a single manifold in the latent space~\cite{zdc_sinkhorn}. Additionally, the widely used Gaussian distribution worsens the linear separability of complex data distributions and does not support the sparse latent representation achieved through the \ac{ReLU} activation~\cite{sparse_vae}. In response to this, autoencoders with a noise generator were proposed (Figure~\ref{fig:ae_ng}) where the model does not regularize the hidden space, allowing more freedom while preserving generative capabilities through an additional neural network known as the noise generator. This auxiliary network is responsible for mapping conditional variables into the latent space.

The authors of \ac{e2eSAE}~\cite{zdc_sinkhorn} propose a loss function that combines reconstruction loss $\mathcal{L}_{rec}$, Sinkhorn loss $\mathcal{L}_S$ aligning the noise generator and encoder outputs, and a regularization term $\mathcal{L}_{reg}$:

\begin{equation}
    \mathcal{L}(\theta, \phi, \psi; x, c) = \mathcal{L}_{rec}(\theta, \phi; x) + \beta \mathcal{L}_{S}(\theta, \psi; x, c) + \gamma \mathcal{L}_{reg}(\theta, \phi, \psi),
    \label{eq:e2e_sae}
\end{equation}

\noindent where $\psi$ denotes the noise generator parameters, $\beta$ is the Sinkhorn loss weight, and $\gamma$ is the regularization weight. The Sinkhorn algorithm is a fast approximation of the Wasserstein distance used in optimal transport theory~\cite{sinkhorn}. The original \ac{e2eSAE} incorporates a complex Laplacian pyramid loss~\cite{laplacian_loss} combined with $l2$ loss as a reconstruction loss term. However, to simulate \ac{ZDC} responses the authors use only $l2$ loss and omit the regularization term\footnote{\url{https://gitlab.cern.ch/swenzel/zdcfastsim/}}\textsuperscript{,}\footnote{\url{https://github.com/KamilDeja/e2e_sinkhorn_autoencoder}}, therefore this thesis follows this approach. 

A more straightforward formulation of the autoencoder with a noise generator, similar to~(\ref{eq:supervised_ae}), incorporates the $l2$ loss instead of the Sinkhorn loss:

\begin{equation}
    \mathcal{L}(\theta, \phi, \psi; x, c) = ||D_\phi(E_\theta(x)) - x||^2 +  \beta ||E_\theta(x) - NG_\psi(c)||^2,
\end{equation}

\noindent where $NG$ denotes the noise generator.

\section{Generative adversarial networks}
\label{sec:gan_background}

\begin{figure}[h!]
    \centering
    \includegraphics[width=0.8\linewidth]{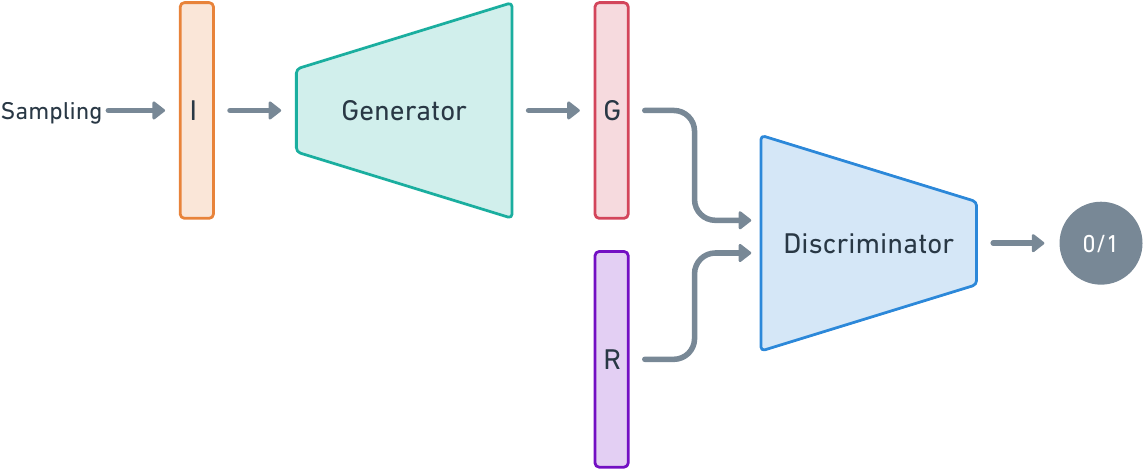}
    \caption{The generative adversarial networks design.}
    \label{fig:gan}
\end{figure}

\Acp{GAN}~\cite{gan} consist of two neural networks competing in a minimax game (Figure~\ref{fig:gan}, where $R$ is a sample from the real data distribution). The generators objective is to perfect data generation, while the role of the discriminator is to develop the ability to distinguish between real and fake samples. The training objective can be written as:

\begin{equation}
    \min_{G_\phi} \max_{D_\theta} \mathbb{E}_{x \sim p(x)}[\log D_\theta(x)] + \mathbb{E}_{z \sim p(z)}[\log(1 - D_\theta(G_\phi(z)))],
\end{equation}

\noindent where $G_\phi$ denotes the generator parameterized by $\phi$ and $D_\theta$ refers to the discriminator parameterized by $\theta$, which produces outputs in the $[0, 1]$ interval. Conditioning in \acp{GAN} is implemented by passing conditional variables to the generator and possibly the discriminator. Moreover, both the generated and real samples can be conditioned on the same features.

\subsection{Training stability and mode collapse}
\label{sec:mode_collapse}

One of the significant challenges in training \acp{GAN} is maintaining stability and preventing mode collapse, where the generator produces limited varieties of samples. Among the many solutions proposed in the literature, the zero-centered gradient penalties~\cite{gan_gp} are known for their ability to promote stable convergence~\cite{gan_convergence}. The regularization term of the discriminator can be written as:

\begin{equation}
\begin{aligned}
    \Omega_{\text{GP}} (\theta, \phi; x) = &\mathbb{E}_{x \sim p(x)}\left[(1 - D_\theta(x))^2 ||\nabla l_\theta(x)||^2 \right] \\ + &\mathbb{E}_{z \sim p(z)}\left[D_\theta(G_\phi(z))^2 ||\nabla l_\theta(\Tilde{x}) |_{\Tilde{x}=G_\phi(z)} ||^2 \right],
\end{aligned}
\end{equation}

\noindent where $l_\theta(x) = \sigma^{-1}(D_\theta(x))$ is the logit of the discriminator. Then the \ac{GAN} training can be formulated as a stochastic gradient descent with respect to:

\begin{equation}
\begin{aligned}
    \theta^{t+1} &= \theta^t - \alpha_D \nabla_\theta \left[ \log D_\theta(x) + \log(1 - D_\theta(G_\phi(z))) + \beta(i) \Omega_{\text{GP}} (\theta, \phi; x) \right] \\
    \phi^{t+1} &= \phi^t - \alpha_G \nabla_\phi \left[ \log(1 - D_\theta(G_\phi(z))) \right]
\end{aligned}
\end{equation}

\noindent where $\alpha_D$ and $\alpha_G$ are the learning rates of the discriminator and generator, and $\beta(i)$ denotes the weight of the regularization term that varies over time. The weight may be fixed to constant or gradually annealed during the training according to the formula $\beta(i) = \beta_0 \cdot \gamma^{i/T}$, where $\beta_0$ is the base weight, $\gamma \in (0, 1)$ is the decay rate, $i$ denotes the current epoch number, and $T$ represents the total number of epochs.

Another widely recognized technique is \ac{WGAN}. The training objective is derived using the Wasserstein distance, leading to the following optimization problem with a critic $D$ parameterized by $\theta$ (unlike the discriminator, the critic generates a continuous output):

\begin{equation}
    \max_{D_\theta} \mathbb{E}_{x \sim p(x)} [D_\theta (x)] - \mathbb{E}_{z \sim p(z)} [D_\theta (G_\phi (z))]
    \label{eq:wgan}
\end{equation}

\noindent The optimization criterion assumes that $D_\theta$ is 1-Lipschitz. A real-valued function $f$ is 1-Lipschitz continuous if $\forall x, y \in \mathbb{R}, \, |f(x) - f(y)| \leq |x - y|$. To enforce this, the critic is constrained using weight clipping or gradient penalty methods. A variant of \ac{WGAN} that utilizes a gradient penalty introduces the following penalty term:

\begin{equation}
    \Omega_{\text{WGAN}} (\theta; x) = \lambda \mathbb{E}_{\hat{x} \sim p(\hat{x})} \left[ \left( \|\nabla D_\theta (\hat{x})\| - 1 \right)^2 \right]
\end{equation}

\noindent where $\lambda$ is a regularization weight, and $\hat{x}$ is uniformly drawn from paths connecting points taken from the real and generated data distributions.

\subsection{GAN extensions}
\label{sec:gan_extensions}

SDI-GAN~\cite{zdc_gan_diversity} used in the ZCD simulation adjusts the diversity of the generated samples depending on the degree of randomness of the particle responses. The model incorporates a regularization term that is dynamically adjusted based on the observed diversity of the training data. Specifically, the diversity measure $f_{div}(c)$ is calculated as:

\begin{equation}
    f_{div}(c) = \sum_{i,j} \sqrt{\frac{\sum_{t} (x^t_{tij} - \mu_{ij})^2}{|X_c|}},
    \label{eq:diversity}
\end{equation}

\noindent where $x_{tij}$ represents the pixel value at coordinates $i, j$ for sample $t$, and $\mu_{ij}$ is the mean pixel value across all samples conditioned on $c$. The regularization term scales the diversity proportionally to the data variance associated with specific conditional variables and is defined as:

\begin{equation}
    \Omega_{\text{SDI}} (\theta, \phi; x, c) = f_{div}(c) \cdot  \left( \frac{d_I(G_\phi(c, z_1), G_\phi(c, z_2))}{d_z(z_1, z_2)} \right)^{-1},
\end{equation}

\noindent where $z_1$ and $z_2$ are two different latent codes, $d_z$ is a distance metric in the latent space, and $d_I$, calculated using pixel-wise or perceptual similarity metrics, measures the dissimilarity between the images generated with the same conditional variable $c$.

The authors of~\cite{zdc_ml} propose to extend the \ac{GAN} framework with an auxiliary regressor and a postprocessing step to improve the quality of the generated \ac{ZDC} responses. The auxiliary regressor is trained to predict the coordinates of the peak photon counts in the input data. The network is then used to provide an additional loss term during training, acting as a physical loss term. The accuracy of the simulation is further improved through a postprocessing step involving scalar multiplication, which adjusts the output images to improve validation metrics.

One of the methods to ensure the global coherence of \ac{GAN}-generated outputs involves incorporating the $l1$ or $l2$ loss into the generator loss function~\cite{gan_loss}. This approach aims at capturing both the global structure of the image and the local details with adversarial loss.

\section{Vector quantization}
\label{sec:quantization}

\ac{VQVAE}~\cite{vq_vae} introduce a discrete latent representation through vector quantization. This design involves the use of quantized vectors $z_q$ that are selected as the nearest neighbors from the codebook $\mathcal{Z}$:

\begin{equation}
    z_q(x) = \underset{z \in \mathcal{Z}}{\operatorname{argmin}} ||E_\theta(x) - z||_2.
    \label{eq:codebook}
\end{equation}

\begin{figure}[h!]
    \centering
    \includegraphics[width=0.7\linewidth]{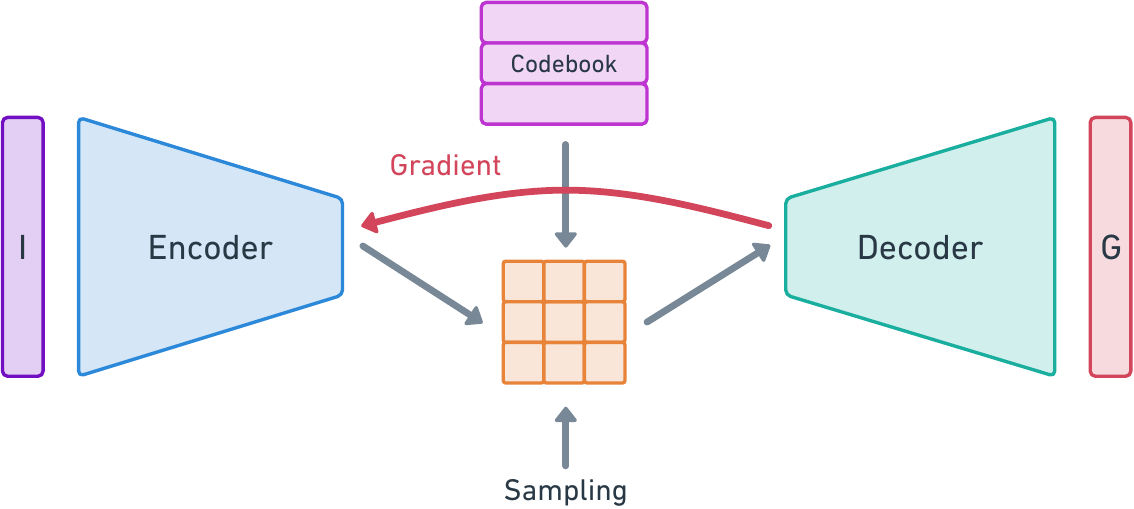}
    \caption{The \ac{VQVAE} design.}
    \label{fig:vq_vae}
\end{figure}

\noindent The decoder reconstructs the input based on the quantized vectors. As there is no defined gradient for a quantization operation, a straight-through gradient estimator is used during the training. It means that the gradient value is copied from the decoder input to the encoder output, as depicted in Figure~\ref{fig:vq_vae}.

The first training stage focuses on the joint training of the discrete autocoder and codebook. The loss function~(\ref{eq:vqvae_loss}) consists of three terms: the first is the reconstruction loss, the second term is dedicated to optimizing the codebook, and the third term, known as the commitment loss, ensures that the encoder commits to an embedding $z$ without excessive growth of the output. The \textit{sg} operator, referred to as ``stop gradient'', stop the optimizer updates for selected parts of the neural network. Note that while \ac{VQVAE} is part of the \ac{VAE} family, it does not include the \ac{KL} term in the loss function. This is due to the assumption of a uniform prior, resulting in a constant \ac{KL} divergence value equal to $\log K$.

\begin{equation}
    \mathcal{L}(\theta, \phi, \mathcal{Z}; x) = \mathbb{E}[\log p_{\theta,\phi}(x|z_q(x))] + ||\operatorname{sg}[E_\theta(x)] - z_q(x) ||^2 + \beta || E_\theta(x) - \operatorname{sg}[z_q(x)] ||^2
    \label{eq:vqvae_loss}
\end{equation}

\begin{figure}[h!]
    \centering
    \includegraphics[width=0.5\linewidth]{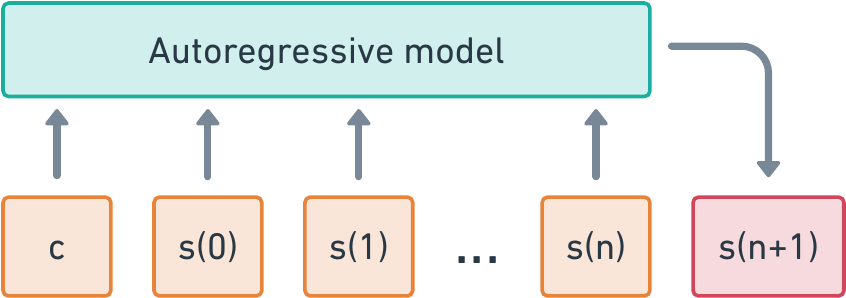}
    \caption{The autoregressive models design.}
    \label{fig:autoregressive}
\end{figure}

The goal of the second stage is to develop a generative model by integrating a learnable prior. Since the \ac{VQVAE} produces discrete outputs (i.e., the indices of the codebook entries), often arranged in a grid-like fashion for images, the learnable prior aims to sample from a discrete latent space. While the authors of \ac{VQVAE} propose the use of the PixelCNN model~\cite{pixel_cnn}, current best practices involve utilizing transformers. The autoregressive design of the transformers, illustrated in Figure~\ref{fig:autoregressive}, implies the generation of a single element of the sequence $s \in \{0, 1, ..., |\mathcal{Z}| - 1\}^m$ per forward pass. The process is repeated $m$ times, where $m$ is the length of the sequence. Autoregressive modeling can be formally defined as the probability distribution of the current element given all previous elements as in~(\ref{eq:autoregressive}). A similar formula can be used to describe conditional generation~(\ref{eq:cond_autoregressive}). The corresponding loss functions are~(\ref{eq:autoregressive_loss}) and~(\ref{eq:cond_autoregressive_loss}). In practice, conditioning in transformers is achieved by adding conditional variables as a prefix to the input, as shown in Figure~\ref{fig:autoregressive}. 

\begin{multicols}{2}
    \begin{equation}
        \begin{split}
        p(s) = \prod_i p(s_i|s_{<i})
        \end{split}
        \label{eq:autoregressive}
    \end{equation}\break
    \begin{equation}
        p(s|c) = \prod_i p(s_i|s_{<i}, c)
        \label{eq:cond_autoregressive}
    \end{equation}
\end{multicols}
\vspace{-14mm}
\begin{multicols}{2}
    \begin{equation}
        \mathcal{L}(\theta; s) = \mathbb{E}[\log p_{\theta}(s)]
        \label{eq:autoregressive_loss}
    \end{equation}\break
    \begin{equation}
        \mathcal{L}(\theta; s, c) = \mathbb{E}[\log p_{\theta}(s|c)]
        \label{eq:cond_autoregressive_loss}
    \end{equation}
\end{multicols}

\subsection{VQ-VAE extensions}
\label{sec:vq_extensions}

\begin{figure}[h!]
    \centering
    \includegraphics[width=0.7\linewidth]{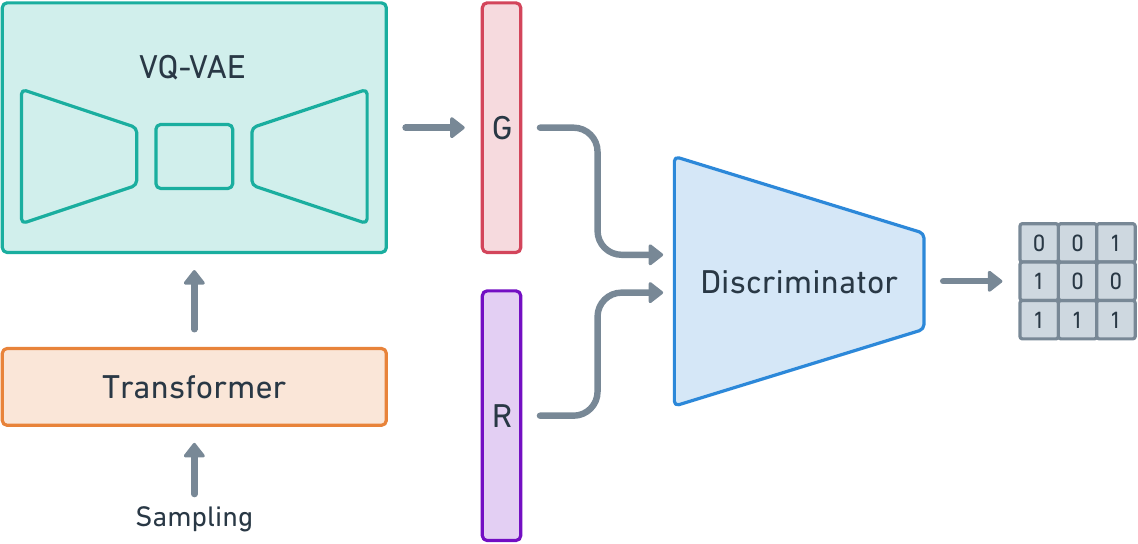}
    \caption{The VQ-GAN design.}
    \label{fig:vq_gan}
\end{figure}

The \ac{VQVAE} was further extended to the VQ-GAN~\cite{vq_gan}. Apart from \ac{VQVAE} functioning as a generator, the authors leverage a transformer as a learnable prior, and to address the challenge of processing very long sequences, they introduce a sliding attention window. The VQ-GAN loss function includes a patch-based adversarial loss. The discriminator attempts to distinguish fragments of the generated image from the corresponding fragments of the real image, instead of evaluating the entire output. Consequently, a grid of values ranging from $0$ to $1$ is produced, as depicted in Figure~\ref{fig:vq_gan}.

\begin{figure}[h!]
    \centering
    \includegraphics[width=0.8\linewidth]{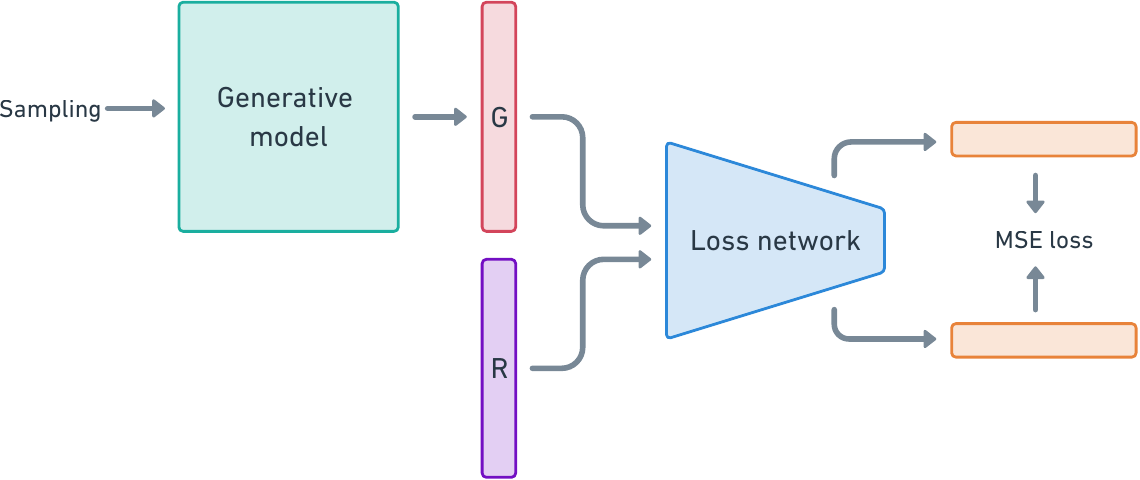}
    \caption{Schematic illustration of the perceptual loss.}
    \label{fig:perceptual}
\end{figure}

Instead of $l2$ loss as a reconstruction loss, VQ-GAN employs a perceptual loss~\cite{perceptual}. This method evaluates the similarity between two images, leveraging feature representations extracted from a pre-trained \ac{CNN} to better capture textural and structural differences. The concept is depicted in Figure~\ref{fig:perceptual}, where \ac{CNN} embeddings of real and generated images are compared using the $l2$ loss.

Moreover, the VQ-GAN authors propose to incorporate transformer sampling methods, including \textit{top-k} sampling, \textit{top-p} (nucleus) sampling~\cite{nucleus}, and rejection sampling. \textit{Top-k} sampling involves selecting the next token from $k$ most probable tokens, while \textit{top-p} sampling assumes selecting from the smallest set whose cumulative probability exceeds a threshold $p$. Finally, rejection sampling involves choosing the highest quality sample from a pool of generated samples.

Vision transformer-based VQ-GAN (ViT-VQ-GAN)~\cite{vit_vq_gan} extends the VQ-GAN design by incorporating \ac{ViT} into the generator and enhancing the utilization of the codebook (as detailed in Section~\ref{sec:codebook_collapse}). Regarding the loss function, ViT-VQ-GAN integrates logit-laplace loss~\cite{dalle}, $l2$ loss, perceptual loss, and adversarial loss in the reconstruction loss term.

\subsection{Codebook collapse}
\label{sec:codebook_collapse}

Vector quantization helps to overcome the issue of posterior collapse that is often encountered in \acp{VAE}. However, a new challenge arises, known as codebook collapse, which refers to insufficient utilization of the codebook embeddings. The metric that allows us to measure codebook utilization is the perplexity: 

\begin{equation}
    \exp{-\sum_{i} \left( \frac{n_i}{N} \log \frac{n_i}{N} \right)}
    \label{eq:perplexity}
\end{equation}

\noindent where $n_i$ is the number of times the $i$-th code appears in the latent representations, and $N$ is the total number of latent codes used in the dataset. There are several strategies to mitigate this issue, one of which is the codebook update technique utilizing \ac{EMA}~\cite{vq_vae}:

\begin{equation}
    \begin{split}
        N_i^{(t)} &= \gamma N_i^{(t - 1)} + (1 - \gamma) n_i^{(t)}, \\
        m_i^{(t)} &= \gamma m_i^{(t - 1)} + (1 - \gamma)\sum_j z_{i, j}^{(t)}, \\
        z_i^{(t)} &= \frac {m_i^{(t)}}{N_i^{(t)}},
    \end{split}
    \label{eq:vqvae_ema}
\end{equation}

\noindent where $\{z_{i,1}, z_{i,2}, ... , z_{i,n_i}\}$ represents the set of $n_i$ outputs of the encoder that are most similar to codebook element $z_i$, and $\gamma$ is a smoothing parameter. The approach is not performing gradient codebook, thereby simplifying the loss function to:

\begin{equation}
    \mathcal{L}(\theta, \phi; x) = \mathbb{E}[\log p_{\theta,\phi}(x|z_q(x))] + \beta || E_\theta(x) - \operatorname{sg}[z_q(x)] ||^2.
    \label{eq:vqvae_ema_loss}
\end{equation}

An alternative method proposed in~\cite{vit_vq_gan} involves a linear projection of the codebook elements and the encoder outputs. The separation of the lookup and embedding spaces divides the process into two steps: first, finding the nearest neighbors in a lower-dimensional space and then matching the corresponding latent codes to a higher-dimensional embedding space. Additionally, the authors implement $l2$ normalization during both steps, which results in performing lookup with cosine similarity, thus improving the training stability. Finally, the quantization is carried out according to:

\begin{equation}
    z_q(x) = \underset{z \in \mathcal{Z}}{\operatorname{argmin}} \left\lVert \frac{WE(x)}{||WE(x)||_2} - \frac{Wz}{||Wz||_2} \right\rVert_2,
\end{equation}

\noindent where $W$ is the projection matrix.

\section{Diffusion}
\label{sec:diffusion_background}

\begin{figure}[h!]
    \centering
    \includegraphics[width=0.7\linewidth]{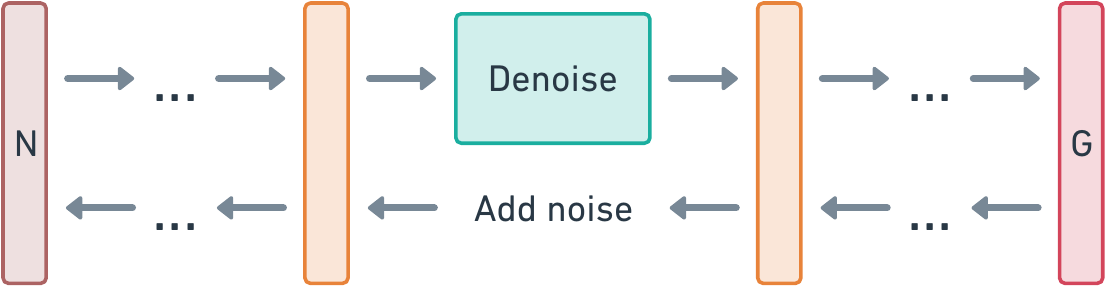}
    \caption{The diffusion models design.}
    \label{fig:diffusion}
\end{figure}

\Ac{DDPM}~\cite{ddpm} generates new samples by progressively refining noisy data. The model reverses a diffusion process, turning Gaussian noise back into the original data distribution (Figure~\ref{fig:diffusion}). The forward process generates a sequence of data points $x_1, x_2, \ldots, x_T$ with increasing noise, starting from the original data $x_0$. At each step, Gaussian noise is added as follows:

\begin{equation}
    q(x_t|x_{t-1}) = \mathcal{N}(x_t | \sqrt{\alpha_t} x_{t-1}, (1 - \alpha_t)I),
\end{equation}

\noindent where $\alpha_t = 1 - \beta_t$ and $\beta_t$ is a variance schedule. The variance schedule controls the amount of noise added at each step, ensuring a gradual and stable diffusion, which is essential for the model's performance and the stability of the reverse process. The data point any time step $t$ can be derived analytically. Given $\bar{\alpha}_t = \prod_{s=1}^t \alpha_s$, the expression is:

\begin{equation}
    q(x_t | x_0) = \mathcal{N}(x_t; \sqrt{\bar{\alpha}_t} x_0, (1 - \bar{\alpha}_t) I).
\end{equation}

The forward and reverse transitions form the core of the diffusion model, creating a multistep encoder-decoder structure. Hence, \ac{ELBO} for the \ac{DDPM} can be defined as:

\begin{equation}
\begin{aligned}
    \mathcal{L}(\theta; x) &= \mathbb{E}_{q(x_1|x_0)} \left[ \log p_\theta (x_0|x_1) \right] \\
    &- \mathbb{E}_{q(x_{T-1}|x_0)} \left[ D_{KL}\left(q(x_T|x_{T-1}) \| p(x_T)\right) \right] \\ 
    &- \sum_{t=1}^{T-1} \mathbb{E}_{q(x_{t-1}, x_{t+1}|x_0)} \left[ D_{KL}\left(q(x_t|x_{t-1}) \| p_\theta(x_t|x_{t+1})\right) \right],
\end{aligned}
\end{equation}

\noindent where $\theta$ are the parameters of the denoising model. This \ac{ELBO} has three terms: reconstruction, prior matching, and consistency. They evaluate how well the model reconstructs data, matches a Gaussian prior, and performs denoising transitions, respectively. Rewriting this equation and leveraging the known forward process allows certain terms to be omitted, leading to the following optimization goal:

\begin{equation}
    \theta^* = \arg\min_\theta \sum_{t=1}^{T} \frac{1}{2\sigma^2_q(t)} \frac{(1 - \alpha_t)^2\bar{\alpha}_{t-1}}{(1 - \bar{\alpha}_t)^2} \mathbb{E}_{q(x_t|x_0)} \left[ \|\hat{x}_\theta(x_t) - x_0\|^2 \right].
\end{equation}

\noindent where $\hat{x}_\theta(x_t)$ is the model's prediction at step $t$. In practice, the weighting term is often negligible, simplifying the optimization to:

\begin{equation}
    \arg\min_\theta \|\hat{x}_\theta(x_t) - x_0\|.
\end{equation}

Diffusion models generally predict the noise instead of the actual data, allowing the data to be retrieved by subtracting the predicted noise. Then the loss function can be written as:

\begin{equation}
    \mathcal{L}(\theta; x_t, \epsilon_t) =  \|\hat{\epsilon}_\theta(x_t) - \epsilon_t\|^2 ,
\end{equation}

\noindent where $\hat{\epsilon}_\theta(x_t)$ is the noise predicted by the model at step $t$ and $\epsilon_t$ is drawn from a Gaussian distribution. Additionally, instead of directly predicting $x_{t-1}$, the network first predicts $\hat{x}_0$ and then the noise is added to get $x_{t-1}$:

\begin{multicols}{2}
\noindent
  \begin{equation}
    \hat{x}_0 = \frac{x_t - \sqrt{1 - \bar{\alpha}_t} \hat{\epsilon}_\theta(x_t)}{\sqrt{\bar{\alpha}_t}},
  \end{equation}
  \begin{equation}
    x_{t-1} = \sqrt{\bar{\alpha}_{t-1}} \hat{x}_0 + \sqrt{1 - \bar{\alpha}_{t-1}} \epsilon_{t-1}.
  \end{equation}
\end{multicols}

\subsection{Improvements to the sampling process}
\label{sec:ddim}

The \ac{DDIM}~\cite{ddim} extends the \ac{DDPM} by improving the sampling efficiency. The authors introduce a non-Markovian process that allows for fewer denoising steps without losing sample quality. The reverse process of the \ac{DDIM} is formulated as:

\begin{equation}
    x_{t-1} = \sqrt{\bar{\alpha}_{t-1}} \hat{x}_0 + \sqrt{1 - \bar{\alpha}_{t-1} - \eta_t^2} \hat{\epsilon}_\theta(x_t) + \eta_t \epsilon_{t-1},
\end{equation}

\noindent where $\eta_t$ controls the noise added during the reverse process, balancing the quality and diversity of samples. Setting $\eta_t = 0$ makes the process deterministic, leading to the \ac{DDIM}. If $\eta_t = \sqrt{(1 - \bar{\alpha}_{t-1}) / (1 - \bar{\alpha}_t)} \sqrt{1 - \bar{\alpha}_t / \bar{\alpha}_{t-1}}$, the process becomes Markovian, resulting in the \ac{DDPM}.

\chapter{Experiments, results, and discussion}
\label{sec:experiments}

This chapter presents the findings of this study, highlighting the performance of various models for fast \ac{ZDC} simulation. The dataset (Section~\ref{sec:dataset}), metrics (Section~\ref{sec:metrics}), and training setup (Section~\ref{sec:setup}) are described first, followed by the main results (Section~\ref{sec:architecture_experiments} to~\ref{sec:speed}). The implementation details of the models, together with the parameter studies, can be found in Sections~\ref{sec:autoencoders} to~\ref{sec:diffusion}.

\section{Dataset analysis}
\label{sec:dataset}

\begin{figure}[h!]
    \centering
    \begin{subfigure}[]{0.19\textwidth}
        \includegraphics[width=\linewidth]{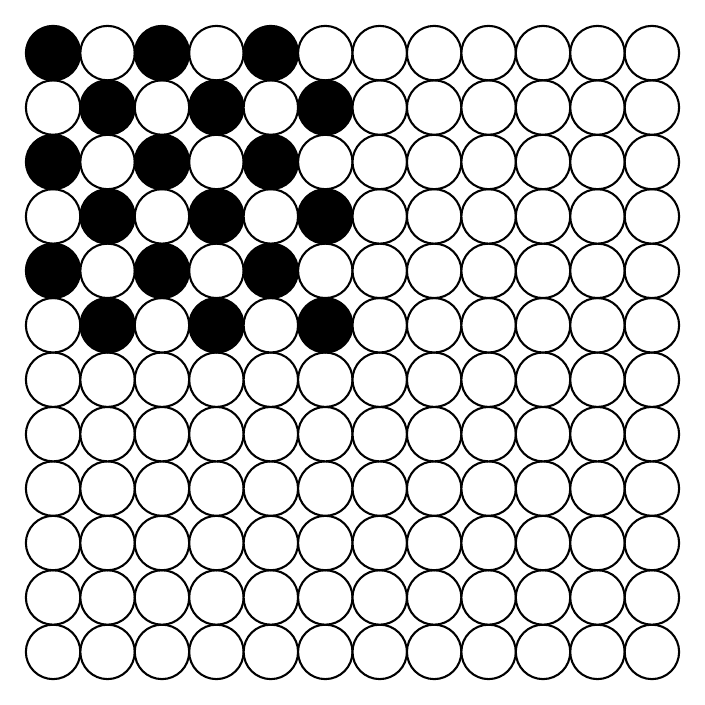}
        \caption{Channel 1.}
    \end{subfigure}
    \hfill
    \begin{subfigure}[]{0.19\textwidth}
        \includegraphics[width=\linewidth]{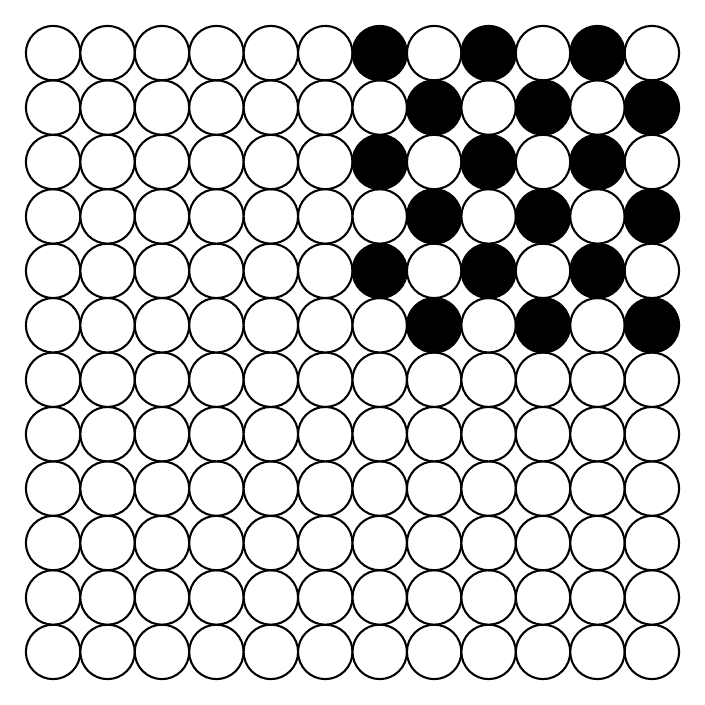}
        \caption{Channel 2.}
    \end{subfigure}
    \hfill
    \begin{subfigure}[]{0.19\textwidth}
        \includegraphics[width=\linewidth]{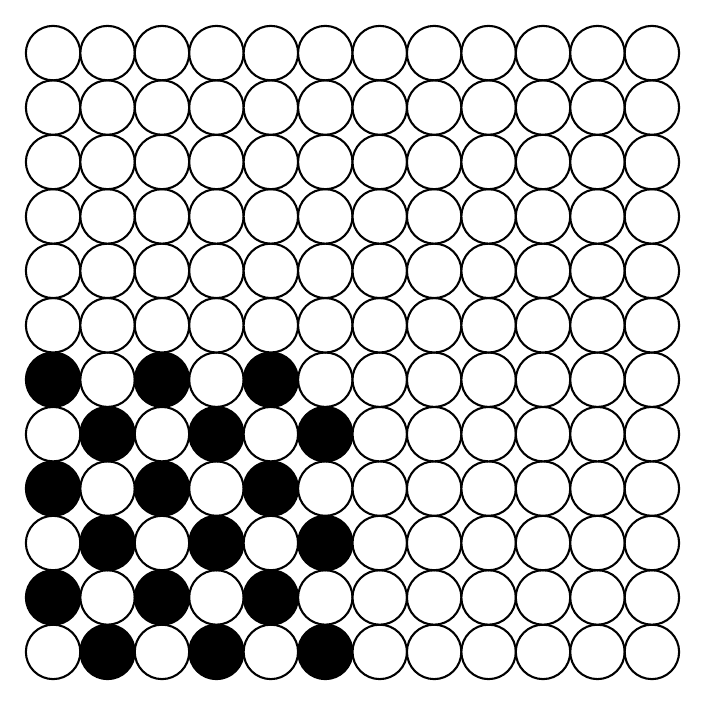}
        \caption{Channel 3.}
    \end{subfigure}
    \hfill
    \begin{subfigure}[]{0.19\textwidth}
        \centering
        \includegraphics[width=\linewidth]{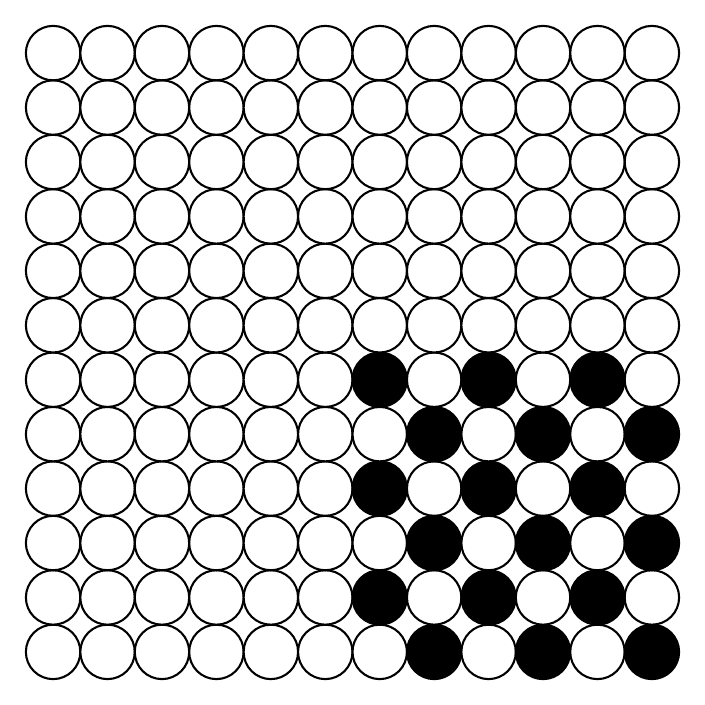}
        \caption{Channel 4.}
    \end{subfigure}
    \hfill
    \begin{subfigure}[]{0.19\textwidth}
        \centering
        \includegraphics[width=\linewidth]{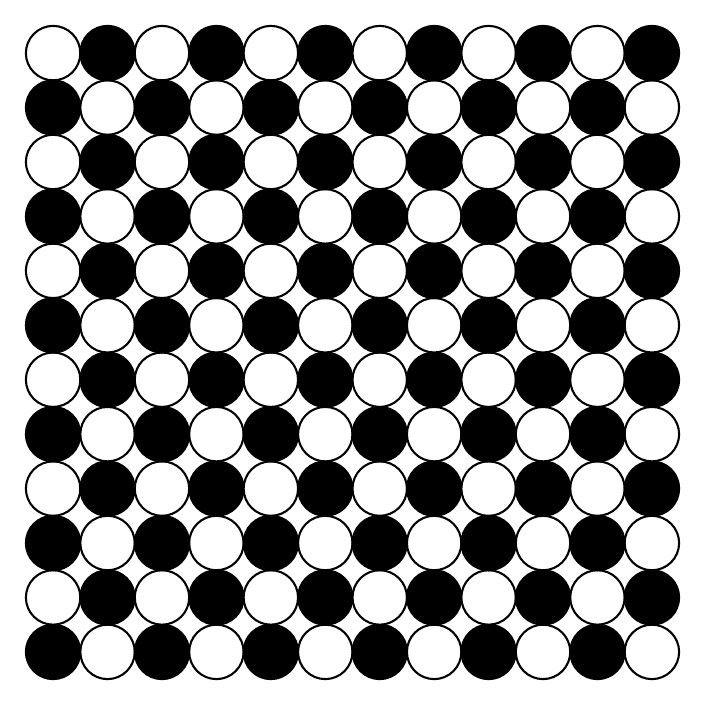}
        \caption{Channel 5.}
    \end{subfigure}
    \hfill
    \caption{Visualization of the division of quartz fibers into channels in the ZN detector.}
    \label{fig:zdc_channels}
\end{figure}

The dataset used in this work is a set of Monte Carlo simulations of the \ac{ZDC} neutron detector responses. The \ac{ZDC} itself employs a design known as ``spaghetti calorimeters''. These calorimeters consist of stacked heavy metal plates with a matrix of quartz fibers inside. The basic principle of operation is based on the detection of Cherenkov light emitted by charged particles of the shower as they pass through the fibers. To facilitate photon counting, the fibers are organized into five distinct channels. Each channel is connected to a dedicated photomultiplier that accurately estimates the number of photons. Figure~\ref{fig:zdc_channels} shows the schematic division into channels, where black circles indicate fibers that belong to the selected channel. Note that the figure shows coarser grids for clarity, whereas the real ZN detector has a 44 by 44 grid.

\subsection{Statistical analysis}

The dataset consists of two parts: particle properties, which serve as conditional variables, and ZN responses, which have dimensions $44 \times 44 \times 1$, making them suitable to be treated as images by generative models. The particle properties are represented as 9-dimensional vectors that contain information on energy, primary vertex positions in the x, y, and z dimensions, momenta in the x, y, and z dimensions, mass, and charge. Furthermore, PDG identifiers can be included to specify the type of particle.

The dataset contains 306780 samples, with all detector responses having at least 10 photons (filtering out zero responses is a separate problem and is not covered in this work, but can be found in the literature~\cite{zdc_ml}). Among all examples, 99695 have a different sum of photons, and there are only 1805 unique feature vectors, indicating that the dataset might not cover the entire space of possible particle features. Additionally, the dataset is highly imbalanced -- 60.83\% are photons ($\gamma$), 23.34\% are neutrons ($n$), 5.30\% are protons ($p$), 3.08\% are lambda baryons ($\Lambda$), 2.09\% are short-lived kaons ($K_S^0$), 1.79\% are long-lived kaons ($K_L^0$), 1.30\% are pions ($\pi+$), and the others represent less than 1\% of the dataset size. In general, the dataset includes 21 different types of particles.

\begin{figure}[h!]
    \centering
    \includegraphics[width=0.9\linewidth]{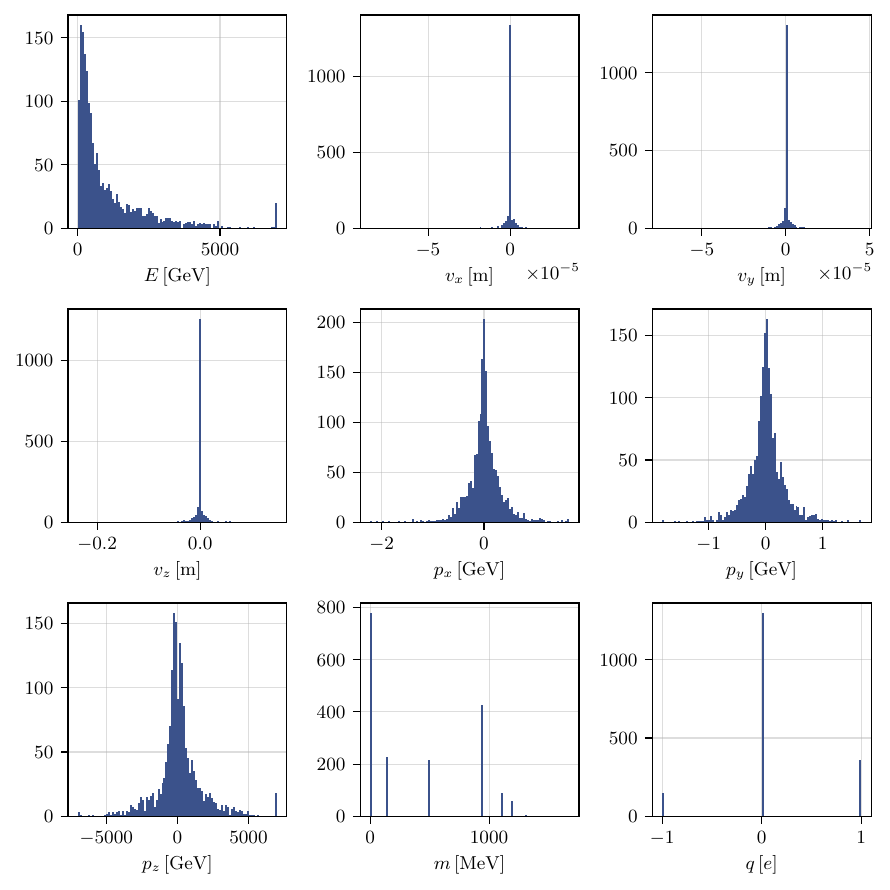}
    \caption{Histograms of particle features. $E$ stands for energy, $v$ for primary vertex positions, $p$ for momenta, $m$ for mass, and $q$ for charge.}
    \label{fig:histograms}
\end{figure}

Figure~\ref{fig:histograms} presents histograms of particle features. The energy distribution features a sharp peak at lower values, gradually decreasing toward higher energies. Note the small peak around 7000 [GeV], representing protons with exceptionally high momenta along the $z$ axis, which can also be seen in the $p_z$ histogram. The primary vertex positions are tightly centered around zero, suggesting minimal spatial displacement from the origin. The momenta exhibit bell-shaped distributions with peaks at zero, indicating balanced particle movements in various directions, with $p_z$ showing a broader range. The mass histogram shows distinct peaks that correspond to quantized particle masses. Finally, the charge histogram shows that the particles have charges of -1, 0, or 1, representing typical states. Note that momenta and mass formally have units of $\frac{\text{GeV}}{c}$ and $\frac{\text{GeV}}{c^2}$, respectively, where $c$ is the speed of light in a vacuum. However, in scientific software, the values are often normalized to GeV for simplicity\footnote{Source code of the Monte Carlo simulation framework, where mass is represented in GeV units: \url{https://github.com/vmc-project/vmc/blob/master/source/include/TVirtualMC.h}}. Under this assumption, the energy can be expressed as:

\begin{equation}
\begin{aligned}
    E^2 &= \left( \frac{p}{c}c \right)^2 + \left( \frac{m}{c^2}c^2 \right)^2 \\
    &= p^2 + m^2 \\
    &= p_x^2 + p_y^2 + p_z^2 + m^2,
    \label{eq:energy}
\end{aligned}
\end{equation}

\noindent where $E$ denotes the energy, $p_x$, $p_y$, and $p_z$ are momenta in x, y, and z directions, respectively, and $m$ stands for the mass.

\subsection{Visualizations}

\begin{figure}[h!]
    \centering
    \begin{subfigure}{0.45\textwidth}
        \includegraphics[width=\textwidth]{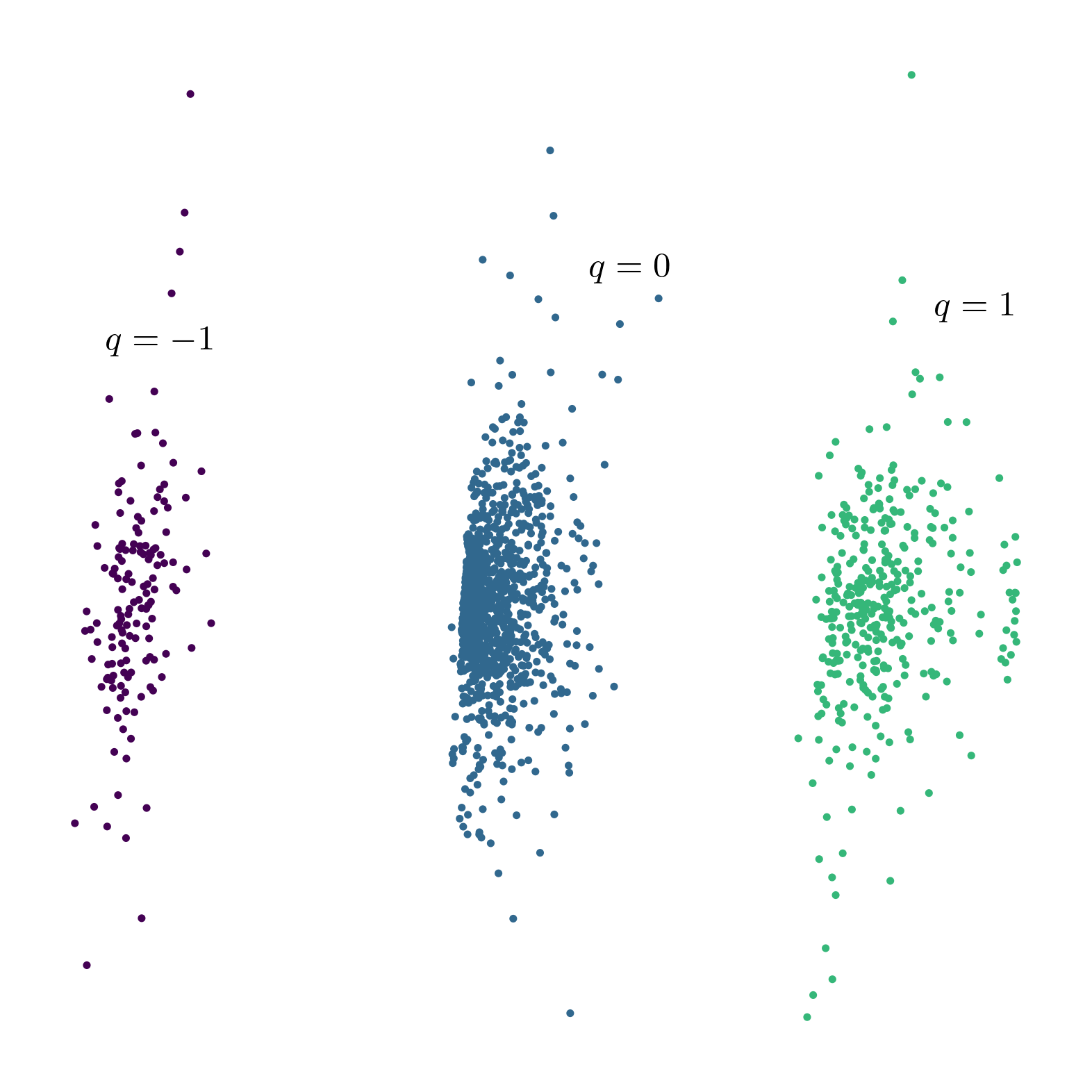}
        \caption{Particle features visualization using linear \ac{PCA}. The colors indicate the charge.}
        \label{fig:pca}
    \end{subfigure}
    \hspace{5mm}
    \begin{subfigure}{0.45\textwidth}
        \includegraphics[width=\textwidth]{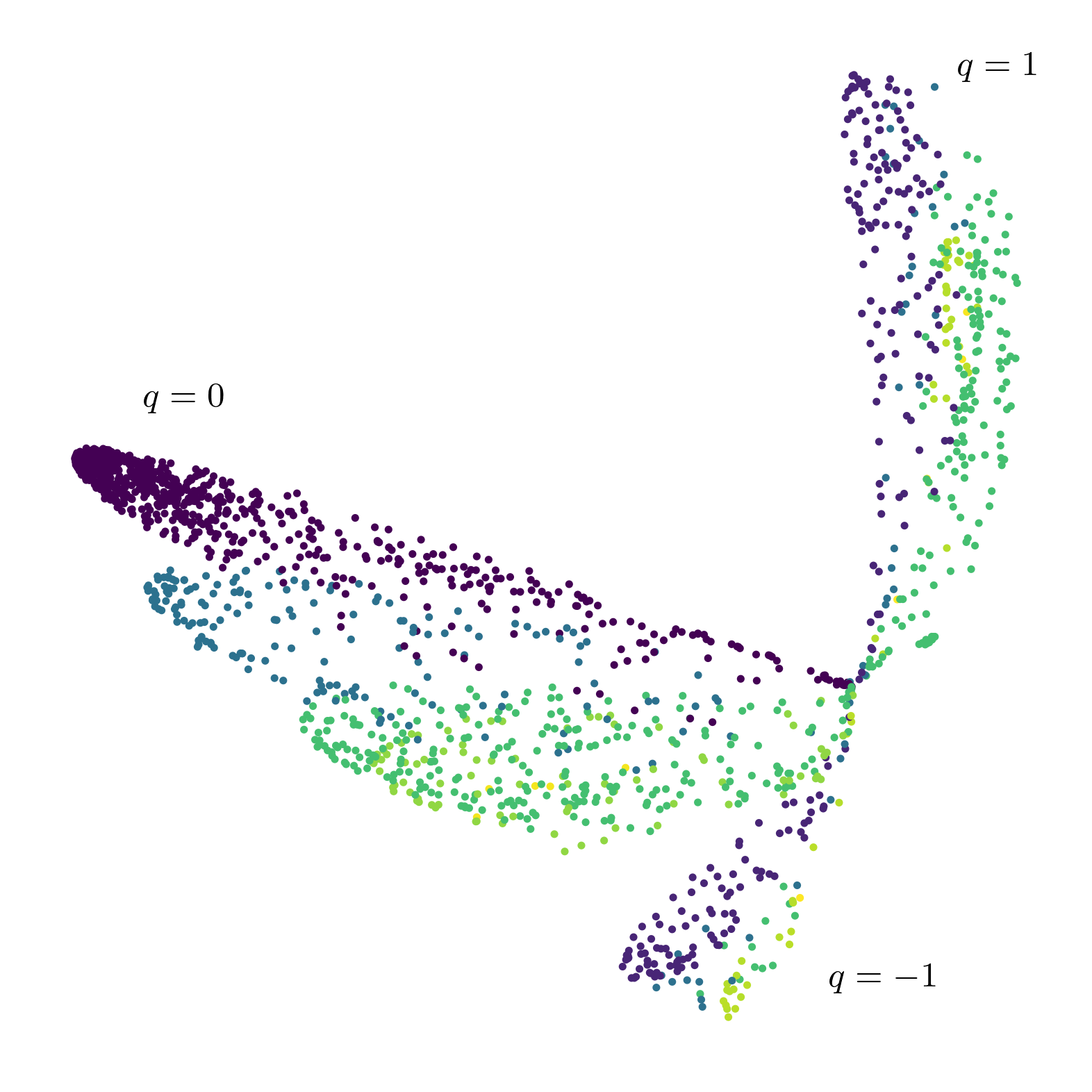}
        \caption{Particle features visualization using kernel \ac{PCA}. The colors indicate the mass, and the labels identify clusters with different charges.}
        \label{fig:kernel_pca}
    \end{subfigure}
    \\
    \begin{subfigure}{0.45\textwidth}
        \includegraphics[width=\textwidth]{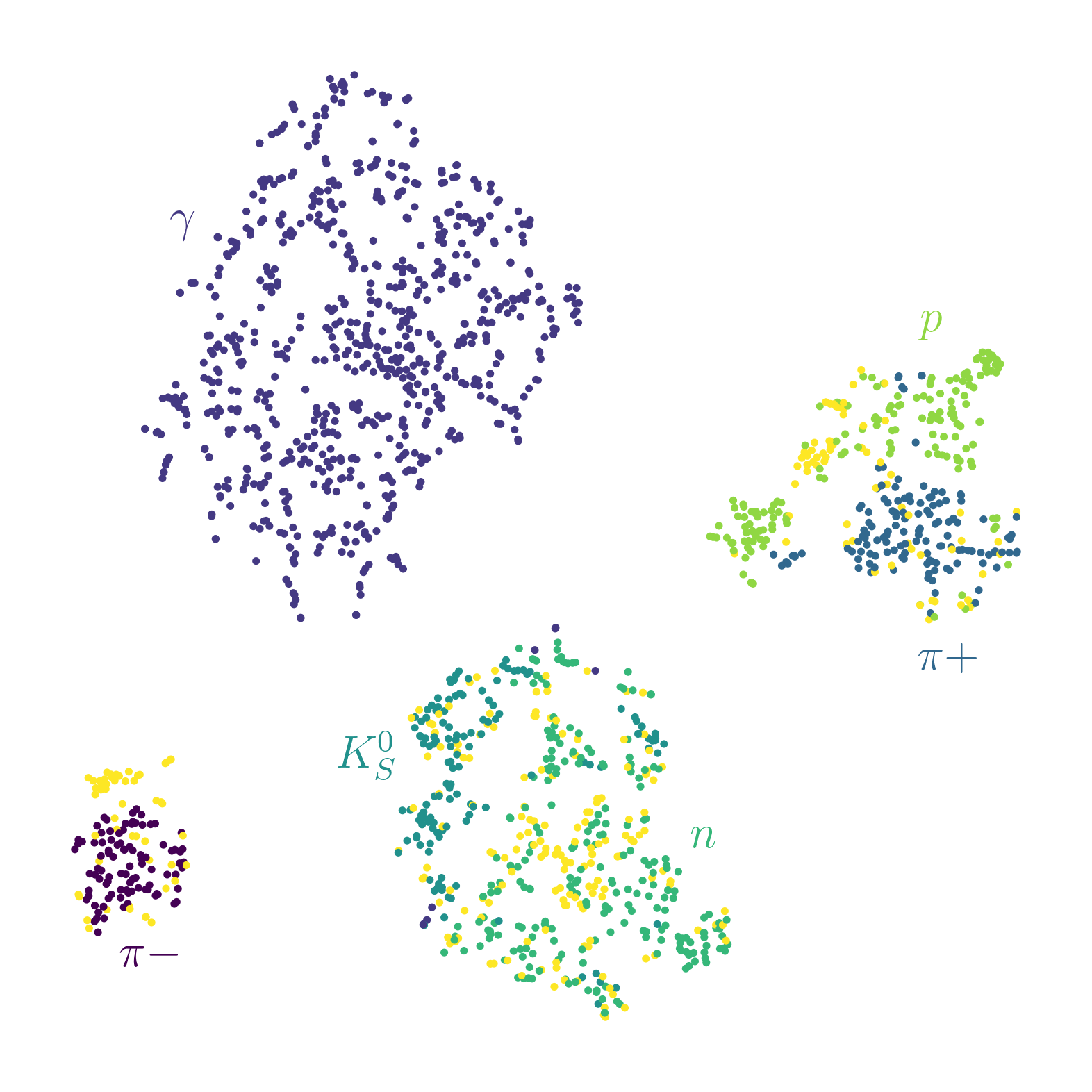}
        \caption{Particle features visualization using \ac{tSNE}. The colors denote the types of the particles.}
        \label{fig:tsne}
    \end{subfigure}
    \hspace{5mm}
    \begin{subfigure}{0.45\textwidth}
        \includegraphics[width=\textwidth]{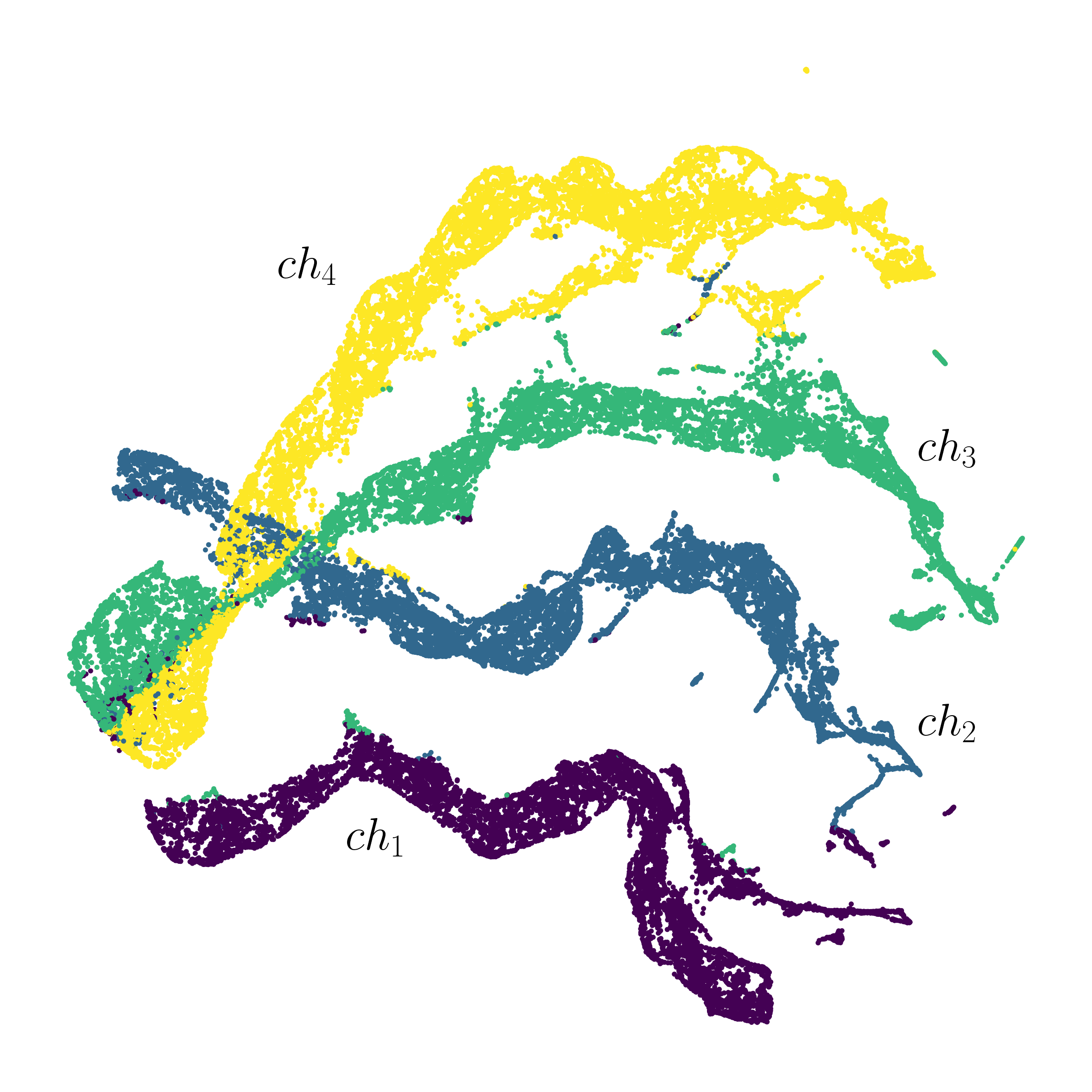}
        \caption{Response channels visualization using \ac{UMAP}. The colors indicate the channel with the highest value.}
        \label{fig:umap}
    \end{subfigure}
\caption{Dataset visualizations with various methods.}
\label{fig:visualization}
\end{figure}

Figure~\ref{fig:visualization} shows dataset visualizations made using \ac{PCA}, kernel \ac{PCA} with \ac{RBF} kernel, \ac{tSNE}, and \ac{UMAP}. Particle features were normalized prior to generating the embeddings and only unique particles were used for visualization, thus there are 1805 points on the first three visualizations. 

The \ac{PCA} visualization shown in Figure~\ref{fig:pca} clearly separates the particles into three groups that reflect the charge of the particles, indicating its importance in the variance of the dataset. The first two components account for 47.9\% of the explained variance. 

The visualization made using kernel \ac{PCA} with \ac{RBF} kernel also divides the points into three large clusters according to charge (negative at the bottom, neutral on the left, positive at the top). Interestingly, within these clusters, groups with different masses can be easily distinguished, as illustrated in Figure~\ref{fig:kernel_pca}.

The \ac{tSNE} embedding, shown in Figure~\ref{fig:tsne}, separates the particles into four clusters. Each cluster contains different particles, which are color-coded. The figure uses colors to represent the six most common particle types, while all other types are depicted in a uniform color (yellow). Note that the most frequently represented $\gamma$ particles have their own cluster.

Figure~\ref{fig:umap} presents a \ac{UMAP} visualization of the channel values of a random 20\% subset of detector responses (61356 samples). Different colors indicate which of the first four channels has the highest value (i.e., which quadrant is the most active). Note that most of the points are clearly separated, implying that detection often occurs in only one of the quadrants. There is also an area of intersection of points from the last three channels. Interestingly, the points associated with the first channel are completely separated from the others. Efforts were made to directly visualize the \ac{ZDC} responses, but the results did not reveal any observable structure.

\subsection{Responses diversity}

\begin{figure}[h!]
    \centering
    \includegraphics[width=0.8\linewidth]{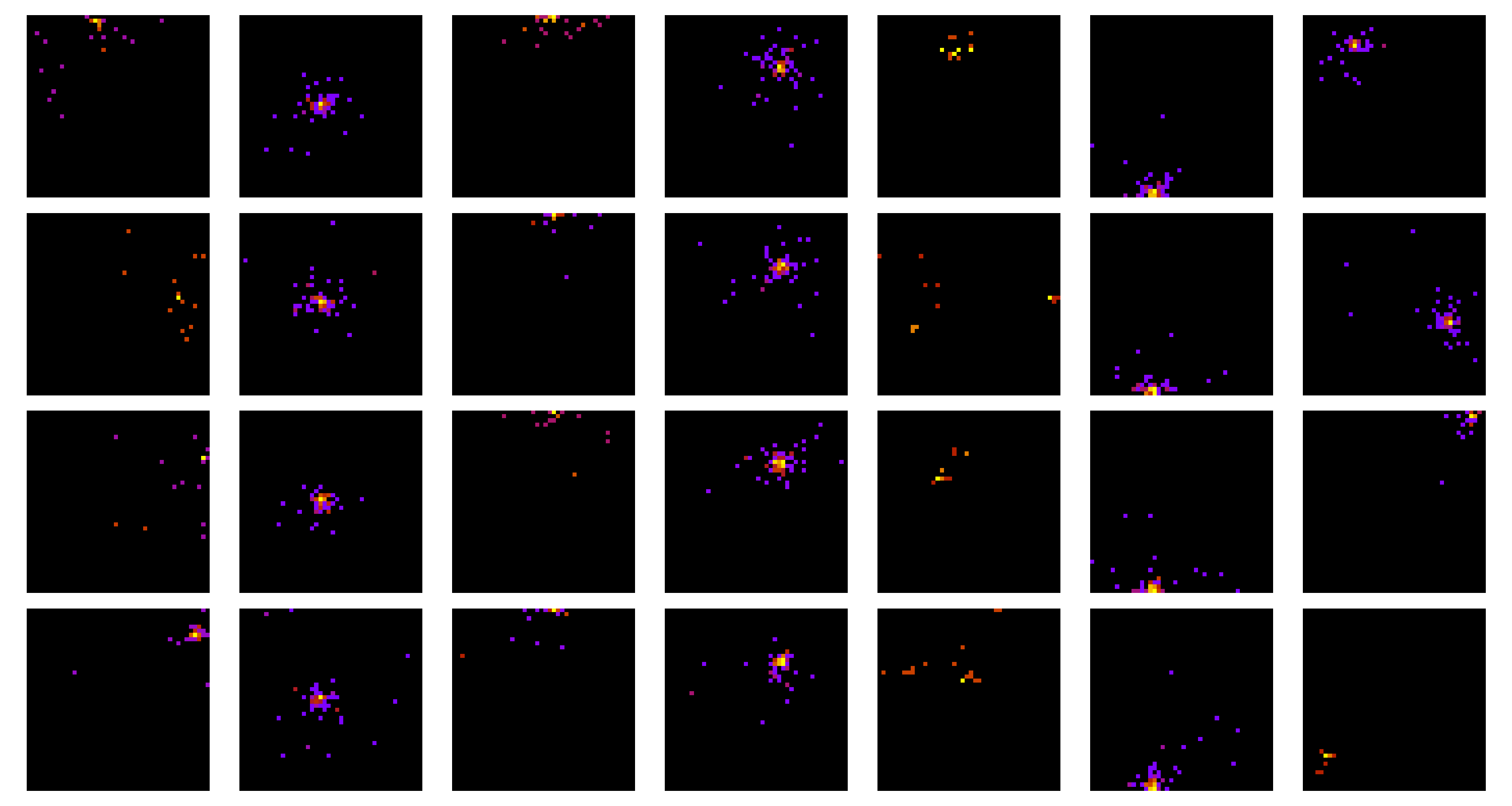}
    \caption{Example \ac{ZDC} neutron detector simulations generated with GEANT software. The following columns are the responses to different particles, the first from the left is $\pi+$, followed by $\gamma$ particles (the most common in this dataset), and the last is $K_S^0$. The rows show independent runs for the same particles.}
    \label{fig:samples}
\end{figure}

In the \ac{ZDC} simulation, the varying diversity of the detector responses plays an important role, i.e., for some particles the detector gives consistent responses in independent runs, while for others they are diversed. As illustrated in the Monte Carlo simulations shown in Figure~\ref{fig:samples}, the second, fourth, and sixth particles from the left have consistent responses, while the first, fifth, and seventh particles produce diverse responses. %
Table~\ref{tab:diversity} presents the feature importances determined using a random forest algorithm. The energy and mass of the particle have the most significant impact on the diversity (and indirectly the momenta, as the energy depends on it according to~(\ref{eq:energy})).

\begin{table}[h!]
    \caption{The features importance for particle diversity.}
    \label{tab:diversity}
    \centering
    \begin{tabular}{|c||c|}
        \hline
        \textbf{Feature} & \textbf{Importance} \\
        \hline \hline
        $E$              & 0.324               \\ \hline
        $v_x$            & 0.001               \\ \hline
        $v_y$            & 0.005               \\ \hline
        $v_z$            & 0.001               \\ \hline
        $p_x$            & 0.050               \\ \hline
        $p_y$            & 0.054               \\ \hline
        $p_z$            & 0.013               \\ \hline
        $m$              & 0.506               \\ \hline
        $q$              & 0.023               \\ \hline
        PDGID            & 0.021               \\ \hline
    \end{tabular}
\end{table}

\section{Evaluation metrics}
\label{sec:metrics}

\begin{figure}[h!]
    \centering
    \includegraphics[width=\linewidth]{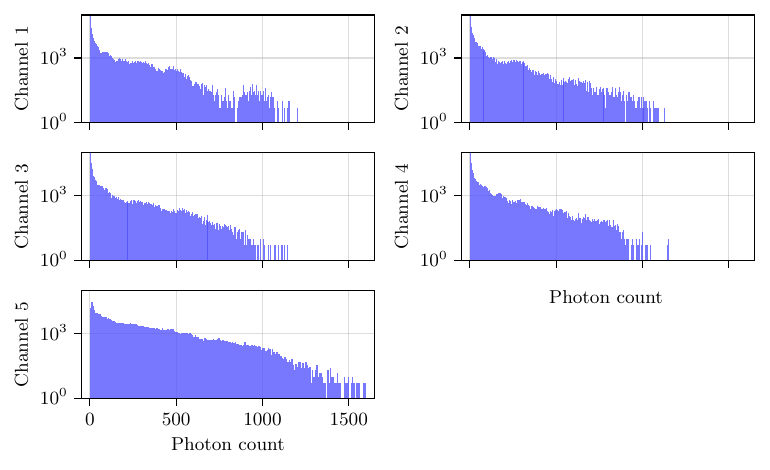}
    \caption{Histogram of the sum of photons in the ZN channels. Note the logarithmic scale on the y-axis.}
    \label{fig:channel_histograms}
\end{figure}

In the literature on fast \ac{ZDC} simulation, the Wasserstein metric is the most commonly used measure of model performance. This metric is calculated as the mean Wasserstein-1 distance between the original and generated histograms of the photon sums. The histograms of a test subset of the GEANT-generated dataset are shown in Figure~\ref{fig:channel_histograms}. Note the higher values in channel 5, which are due to its coverage of the whole detector area, in contrast to the other channels that only cover their respective quadrants. Employing a metric that assesses the global characteristics of the samples is justified, as the detector responses often vary between runs, and variations at the level of individual pixels should not influence the metric's value. A second metric that also appears in the literature is \ac{MAE}, which compares the $l1$ distance of the channel values. Unlike the Wasserstein metric, this is a local metric that directly compares the channel values of the original and generated samples. The last metric is the pixel-wise \ac{RMSE}, which directly compares pixels. This metric indicates the extent to which the model attempts to match the original data. The metrics used can be defined as follows:

\begin{equation}
    \text{Wasserstein-1}(w, \hat{w}) = \frac{1}{5} \sum_{i=1}^5  \int_0^1 \left| F^{-1}_{w_i}(z) - F^{-1}_{\hat{w}_i}(z) \right| dz,
\end{equation}
\begin{equation}
    \text{MAE}(w, \hat{w}) = \frac{1}{n} \sum_{k=0}^n \frac{1}{5} \sum_{i=1}^5 |w_i^k - \hat{w}_i^k|,
\end{equation}
\begin{equation}
    \text{RMSE}(x, \hat{x}) = \sqrt{ \frac{1}{n} \sum_{k=0}^n \frac{1}{44 \cdot 44} \sum_{i=1}^{44} \sum_{j=1}^{44} (x_{ij}^k - \hat{x}_{ij}^k)^2 },
\end{equation}

\noindent where $F^{-1}_q$ is the inverse cumulative distribution function of the distribution $q$, $w_i$ denotes the distribution of the $i$-th channel, $n$ refers to the number of evaluated examples, $w_i^k$ represents the value of the $i$-th channel of the $k$-th response, $x_{ij}^k$ is the value of the pixel with $i$ and $j$ coordinates of the $k$-th response, and $\hat{w}$ and $\hat{x}$ are the corresponding predicted values.

Note that while lower \ac{MAE} and \ac{RMSE} metrics are generally considered better, it does not necessarily imply that the model is superior in the \ac{ZDC} simulation task. Achieving the lowest possible value is not always desired, as it may suggest that the model frequently produces consistent responses, even when the actual physical process yields diverse results. The optimal values of these metrics should be close to those derived from the original dataset. 

\begin{figure}[h!]
    \centering
    \includegraphics[width=\linewidth]{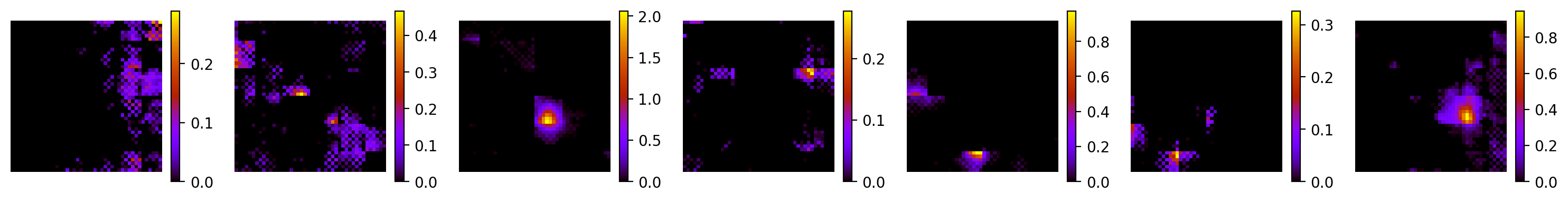}
    \caption{Example \ac{ZDC} responses generated by an autoencoder directly optimizing Wasserstein distance. Corresponding Monte Carlo simulations can be found in the first row of Figure~\ref{fig:samples}.}
    \label{fig:optimize_channels}
\end{figure}

Although it may be appealing to directly optimize the primary metric (i.e., the Wasserstein distance), the practical implementation of this task is challenging. The initial difficulty lies in the computational complexity of the Wasserstein loss function, which requires the use of the Sinkhorn algorithm as an approximation~\cite{sinkhorn}. Even with the use of the fast loss function executed on \ac{GPU}, the direct optimization does not produce satisfactory results, as illustrated in Figure~\ref{fig:optimize_channels}. The images contain unusual and blurry clouds that do not reflect the typical \ac{ZDC} response. In addition, multiple activation centers are often present and the intensity of the responses is disrupted compared to the original simulations shown in Figure~\ref{fig:samples}. Therefore, modeling the detector relies on the use of traditional generative methods and potentially incorporating physical loss terms to guide the model toward output that aligns with the laws of physics.

\section{Training setup}
\label{sec:setup}

Table~\ref{tab:setup} lists the settings used in model training and evaluation. The dataset was divided into training, validation, and testing subsets in proportions of 70\%, 10\%, and 20\%, respectively. Training was carried out on the training set, the parameters were tuned on the validation set, and the final results were calculated on the test set. Conditional variables (i.e., particle parameters) were standardized to a zero mean and a standard deviation of 1, and the \ac{ZDC} responses were logarithmized. Note that the metrics were calculated after inverse transforming the data to their original scale. Due to the high stochasticity of the simulated process, the reported metric values are the average of five runs.

The AdamW optimizer~\cite{adamw} was used, with hyperparameters fine-tuned through more than 100 trials per model with Optuna software~\cite{optuna} optimizing the Wasserstein metric. The optimization procedure leveraged the \ac{TPE} sampler without pruning and considered the following parameters: learning rate, $\beta_1$, $\beta_2$, $\epsilon$, and the use of the cosine learning rate schedule, weight decay, and Nesterov momentum. The optimized hyperparameters can be found in Appendix~\ref{sec:optimizer_settings}. The repository linked to this thesis contains both the hyperparameters and the source code of the models.

The training was performed on a single NVIDIA A100 \ac{GPU} with 40 GB of memory on the Athena supercomputer\footnote{\url{https://www.cyfronet.pl/en/19073,artykul,athena.html}}. All models were trained for 100 epochs in batches of size 256. At the end of each epoch, the weights were stored to allow for reconstruction at any time during the training. The metrics for the training and validation subsets were calculated after each epoch throughout the training, whereas the metrics for the test subset were computed after the completion of the training (except for models explicitly mentioned, particularly in cases of divergent training near the end of the training).

\begin{table}[h!]
    \caption{Training and evaluation settings.}
    \label{tab:setup}
    \centering
    \begin{tabular}{|c|c||c|}
        \hline
        \textbf{Category}          & \textbf{Parameter}    & \textbf{Value}                      \\
        \hline \hline
        Dataset                    & Split                 & 70\% train -- 10\% val -- 20\% test \\ \hline
        \multirow{7}{*}{Optimizer} & Type                  & AdamW                               \\ \cline{2-3}
                                   & Optimization software & Optuna with \ac{TPE} sampler        \\ \cline{2-3}
                                   & Optimized parameters  & \makecell{learning rate, $\beta_1$, $\beta_2$, $\epsilon$, \\ cosine schedule, weight decay, \\ Nesterov momentum} \\ \cline{2-3}
                                   & Number of optimization trials & $\geq 100$                  \\ \cline{2-3}
                                   & Optimization results  & See Appendix~\ref{sec:optimizer_settings} \\ \hline
        \multirow{2}{*}{Training}  & Number of epochs      & 100                                 \\ \cline{2-3}
                                   & Batch size            & 256                                 \\ \hline
        \multirow{2}{*}{Evaluation} & Metrics              & Wasserstein, MAE, RMSE              \\ \cline{2-3}
                                   & Number of evaluation iterations & 5                         \\ \hline
    \end{tabular}
\end{table}

\section{Model architecture}
\label{sec:architecture_experiments}

As part of the experiment, encoder-like and decoder-like networks were implemented in the \ac{CNN}, \ac{ViT}, and MLP-Mixer architectures. The networks have been used as building blocks in autoencoders, \acp{GAN}, and \ac{VQ} models, with additional layers specific to each framework. All models discussed in this thesis operate in the raw data space. For each model based on a given architecture, an optimizer was tuned, whose parameters are listed in Appedix~\ref{sec:optimizer_settings}.

The \ac{CNN} encoder-like network employs convolutional blocks inspired by \ac{VDVAE}, utilizing \ac{GELU} and a bottleneck design composed of four convolutional layers with a residual connection. The image is processed in three stages, where each stage involves a convolution increasing the number of channels, two convolution blocks, \ac{BN}, and average pooling. The network is designed to transform the inputs of $44 \times 44 \times 1$ into a volume of $6 \times 6 \times c$, where $c$ represents the number of output channels. Conditional variables are concatenated to each channel. The decoder-like network architecture is symmetric: initially, the conditional variables are appended, and then the volume is processed in three stages. A convolution that decreases the number of channels, upsampling using the nearest-neighbor method, two convolution blocks, and \ac{BN} are applied in each stage. Finally, a convolution layer reduces the number of channels to one, followed by \ac{ReLU} to facilitate the network to output exactly zero.

The encoder-like network in \ac{ViT} architecture initially segments the image into $8 \times 8$ patches, followed by linear projection and the application of learned positional encoding. The network operates on a volume of size $6 \times 6 \times c$. Conditional variables are projected linearly and are prepended as the first element of the sequence. The volume is then processed through multiple stacked transformer blocks. The decoder-like network, given the input volume, applies linear projection and adds learned positional embeddings. Similarly to the encoder, conditional variables are prepended, and stacked transformer blocks are applied. Finally, a transposed convolution followed by \ac{ReLU} is used to obtain an output with the desired size and number of channels. The networks in the MLP-Mixer architecture are identical, with two distinctions: positional embedding is omitted, and MLP-Mixer mixing blocks are applied instead of transformer blocks.

In this experiment, in addition to comparing neural network architectures, different models within the autoencoder framework were assessed. The first model is \ac{VAE}, followed by \ac{VAE} with conditional variables containing learned embeddings depending on the particle type. Another model is a supervised autoencoder, and finally, autoencoders with a noise generator trained using the Sinkhorn or $l2$ loss. Each model had approximately one million parameters and a latent space of size 10 or 20 for the noise generator autoencoders. The choice of these values is based on the results of the experiments detailed in Sections~\ref{sec:size} and~\ref{sec:latent_space_size}, respectively.

The results are presented in Tables~\ref{tab:architectures_I} and~\ref{tab:architectures_II}. \ac{VAE} consistently produces similar results across various architectures. Using the Sinkhorn loss improves the results for \ac{CNN} models but does not provide gains over the $l2$ loss in \ac{ViT} models. Interestingly, the incorporation of the physical loss term by the supervised autoencoder did not improve performance. The autoencoder with a noise generator and the $l2$ loss achieved the best score of 11.19 in the Wasserstein metric, slightly surpassing \ac{CNN}-based \ac{VAE}. As \ac{ViT} architecture delivers the best and most consistent performance, it was selected for all further experiments.

\begin{table}[h!]
    \caption{Performance of autoencoders across various architectures (part I).\protect\footnotemark}
    \label{tab:architectures_I}
    \centering
    \begin{tabular}{|c||c|c|c||c|c|c|}
        \hline
        \textbf{Architecture} & \multicolumn{3}{c||}{\textbf{CNN}}                  & \multicolumn{3}{c|}{\textbf{ViT}}                    \\ \hline
        \textbf{Metric}       & \textbf{Wasserstein} & \textbf{MAE} & \textbf{RMSE} & \textbf{Wasserstein} & \textbf{MAE} & \textbf{RMSE}  \\
        \hline \hline
        \ac{VAE}              & 11.52                & 17.76        & 50.38         & 11.90                & 18.05        & 49.48          \\ \hline
        \ac{VAE} + Embedding  & $15.93^*$            & $20.16^*$    & $49.59^*$     & 11.61                & 17.98        & 49.47          \\ \hline
        Supervised AE         & 23.71                & 31.90        & 72.32         & 20.43                & 30.60        & 74.64          \\ \hline
        AE + Sinkhorn NG      & 26.53                & 29.07        & 66.16         & 11.34                & 15.88        & 44.17          \\ \hline
        AE + \ac{MSE} NG      & 37.56                & 39.32        & 92.28         & \textbf{11.19}       & 15.47        & 43.49          \\ \hline
    \end{tabular}
\end{table}

\begin{table}[h!]
    \caption{Performance of autoencoders across various architectures (part II).\protect\footnotemark}
    \label{tab:architectures_II}
    \centering
    \begin{tabular}{|c||c|c|c|}
        \hline
        \textbf{Architecture} & \multicolumn{3}{c|}{\textbf{MLP-Mixer}}              \\ \hline
        \textbf{Metric}       & \textbf{Wasserstein} & \textbf{MAE} & \textbf{RMSE}  \\
        \hline \hline
        \ac{VAE}              & 12.22                & 18.00        & 49.51          \\ \hline
        \ac{VAE} + Embedding  & 12.12                & 18.20        & 49.73          \\ \hline
        Supervised AE         & 17.08                & 26.90        & 104.83         \\ \hline
        AE + Sinkhorn NG      & $\times$             & $\times$     & $\times$       \\ \hline
        AE + \ac{MSE} NG      & $\times$             & $\times$     & $\times$       \\ \hline
    \end{tabular}
\end{table}

\footnotetext{The $^*$ symbol denotes that the results are based on earlier checkpoints due to a collapse occurring at the end of the training.}
\footnotetext{The $\times$ symbol indicates that the model did not converge during the training.}

\section{Model size}
\label{sec:size}

The aim of the following experiment was to investigate the effect of model size on simulation quality. To achieve this, the reconstruction performance of the \ac{VQVAE} autoencoder was evaluated. This model was chosen because of the non-regularized latent space. The \ac{ViT} architecture was used, with a codebook size of 256 as specified in Section~\ref{sec:codebook_size}. A gradient update technique was applied, and to improve codebook utilization, projection and $l2$ normalization were used. Five different model sizes, ranging from 0.25 million to 52 million parameters, were tested as shown in Table~\ref{tab:vq_specs}. Optimizer tuning (Appendix~\ref{sec:optimizer_settings}) was performed for each model.

\begin{figure}[h!]
    \centering
    \includegraphics[width=0.8\linewidth]{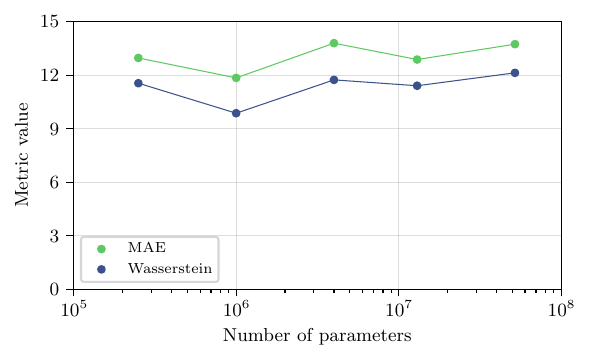}
    \caption{Performance of \ac{VQVAE} depending on model size.}
    \label{fig:size}
\end{figure}

In this scenario, since the objective is reconstruction rather than image generation, the objective is the lowest value across all three metrics. The results presented in Table~\ref{tab:size} and Figure~\ref{fig:size} show that the medium-sized model (one million parameters) achieved the best performance with a Wasserstein score of 9.86. The larger models did not show significant improvement, suggesting that one million parameters are sufficient for the task. Note that the medium-sized model may have achieved the best results due to the small codebook size (larger models might perform better with a larger codebook). Another possible reason is the dataset size. According to transformer scaling laws, achieving better performance requires both a larger model and a larger dataset~\cite{scaling_vit}.

\vspace{-3mm}
\begin{table}[h!]
    \caption{\ac{VQVAE} autoencoder specification depending on model size.}
    \label{tab:vq_specs}
    \centering
    \begin{tabular}{|c||c|c|c|c|}
        \hline
        \textbf{Model size}   & \textbf{\#blocks} & \textbf{\#heads} & \textbf{Hidden dim} & \textbf{Embedding dim} \\
        \hline \hline
        0.25M                 & 3                 & 3                & 48                  & 128                    \\ \hline
        1M                    & 4                 & 4                & 96                  & 256                    \\ \hline
        4M                    & 4                 & 4                & 192                 & 256                    \\ \hline
        13M                   & 6                 & 6                & 288                 & 512                    \\ \hline
        52M                   & 8                 & 8                & 512                 & 512                    \\ \hline
    \end{tabular}
\end{table}

\vspace{-6mm}
\begin{table}[h!]
    \caption{\ac{VQVAE} reconstruction performance depending on model size.}
    \label{tab:size}
    \centering
    \begin{tabular}{|c||c|c|c|}
        \hline
        \textbf{Model size}   & \textbf{Wasserstein} & \textbf{MAE} & \textbf{RMSE} \\
        \hline \hline
        0.25M                 & 11.54                & 12.96        & 38.46         \\ \hline
        1M                    & \textbf{9.86}        & \textbf{11.84} & \textbf{37.22} \\ \hline
        4M                    & 11.73                & 13.78        & 43.54         \\ \hline
        13M                   & 11.40                & 12.87        & 37.90         \\ \hline
        52M                   & 12.12                & 13.73        & 39.74         \\ \hline
    \end{tabular}
\end{table}

\vspace{-6mm}

\section{Generative frameworks}
\label{sec:frameworks}

For comparison of generative frameworks, the best representatives from each family were selected. All models were based on the \ac{ViT} architecture and had a size of about one million, except for the diffusion model which had four million parameters and used a combination of \ac{CNN} and \ac{ViT}. The autoencoders were represented by the autoencoder with a noise generator trained with the $l2$ loss (Section~\ref{sec:architecture_experiments}), \ac{GAN} was a classical model with postprocessing (Section~\ref{sec:gan_experiments}), \ac{VQVAE} trained with the $l2$ loss, codebook of size 256, gradient update, projection and $l2$ normalization (Section~\ref{sec:codebook_size}) was paired with a transformer with adjusted sampling temperature of 1.4 (Section~\ref{sec:vq_transformer}), VQ-GAN employed the same autoencoder and transformer, but added adversatial loss and \ac{EMA} codebook update (Section~\ref{sec:vq_loss_functions}), while diffusion models were represented by DDIM with 50 denoising steps and the $\eta$ parameter set to 0.7 (Seciton~\ref{sec:diffusion}).

To calculate the metrics of the original data, the dataset was randomly split. One subset was considered as generated samples, while the other was treated as actual responses. To calculate \ac{MAE} and \ac{RMSE}, it was essential to partition the dataset to match the particle features in both splits. The evaluation was repeated five times, following the procedure used for generative models.

\begin{figure}[h!]
    \begin{subfigure}{\textwidth}
        \includegraphics[width=\textwidth]{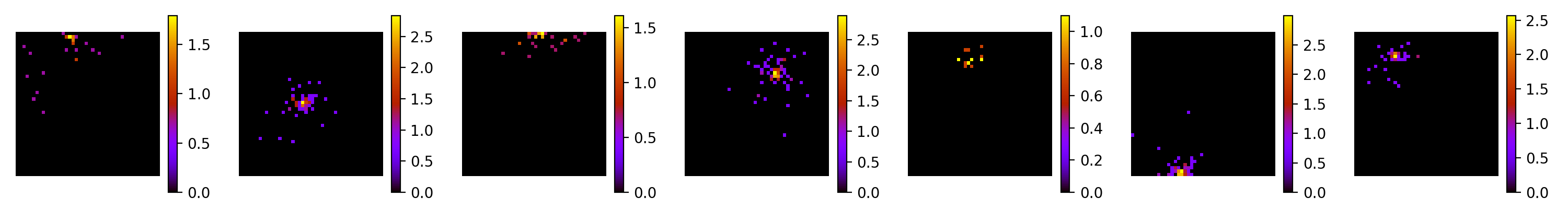}
        \caption{GEANT (same as the first row in Figure~\ref{fig:samples}).}
    \end{subfigure}
    \\
    \begin{subfigure}{\textwidth}
        \includegraphics[width=\textwidth]{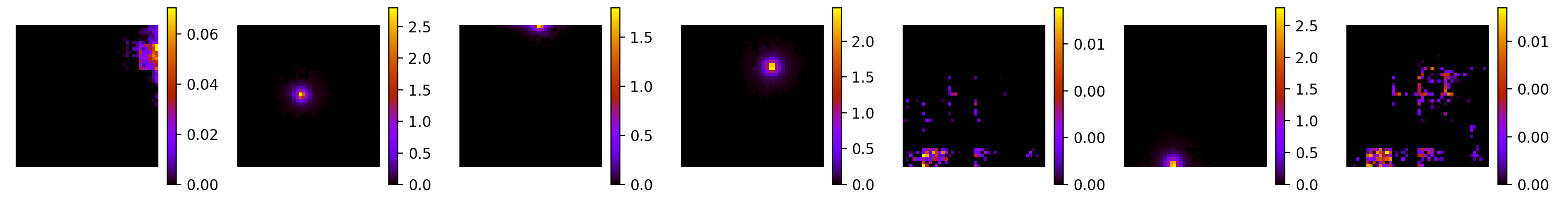}
        \caption{Autoencoder with a noise generator and MSE loss.}
    \end{subfigure}
    \\
    \begin{subfigure}{\textwidth}
        \includegraphics[width=\textwidth]{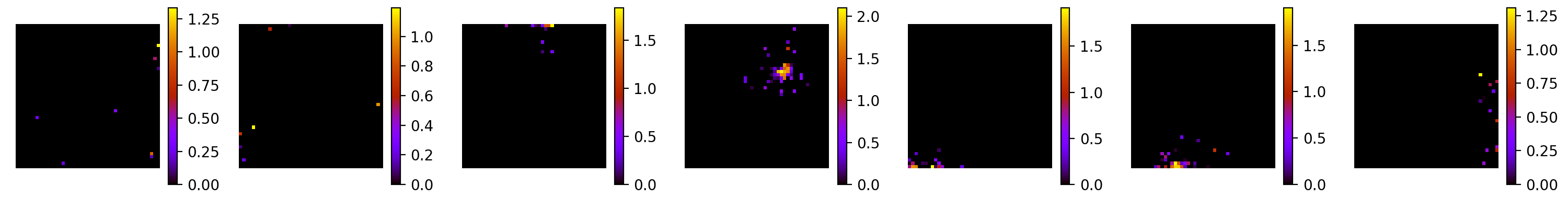}
        \caption{GAN with a postprocessing step.}
    \end{subfigure}
    \\
    \begin{subfigure}{\textwidth}
        \includegraphics[width=\textwidth]{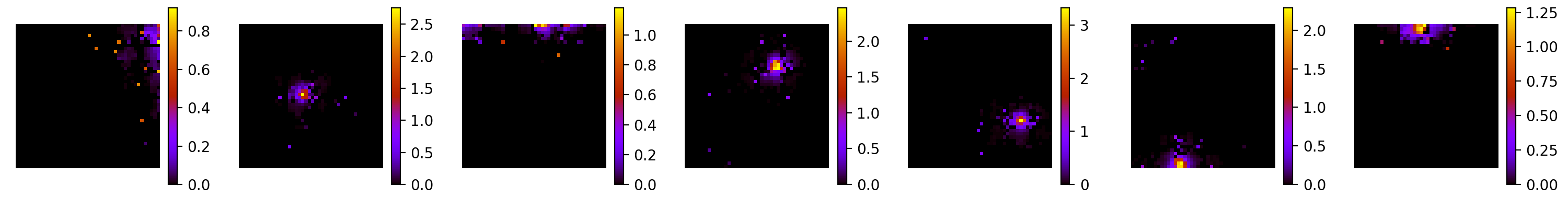}
        \caption{\ac{VQVAE} with a transformer as a learnable prior and adjusted sampling temperature.}
    \end{subfigure}
    \\
    \begin{subfigure}{\textwidth}
        \includegraphics[width=\textwidth]{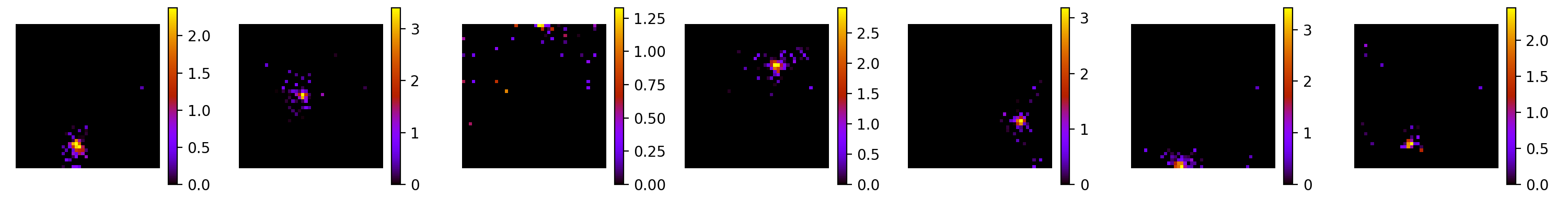}
        \caption{VQ-GAN with a transformer as a learnable prior.}
    \end{subfigure}
    \\
    \begin{subfigure}{\textwidth}
        \includegraphics[width=\textwidth]{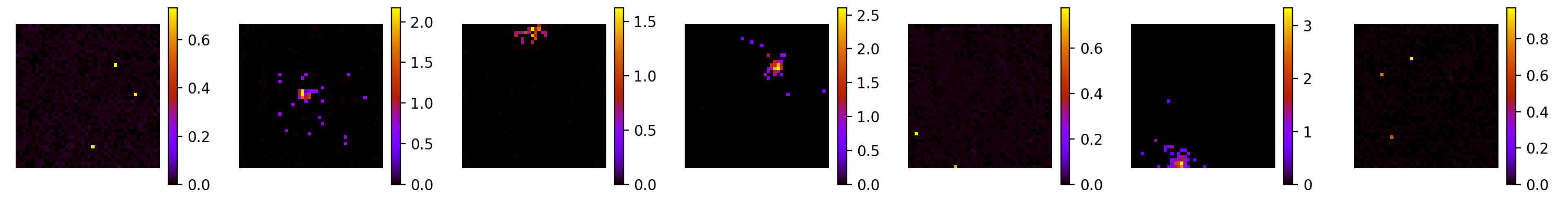}
        \caption{DDIM after 50 denoising steps and adjusted $\eta$ parameter.}
    \end{subfigure}
    \\
    \caption{Example simulations generated by the models discussed in this thesis.}
    \label{fig:example_sim}
\end{figure}

\begin{figure}[h!]
    \centering
    \includegraphics[width=0.7\linewidth]{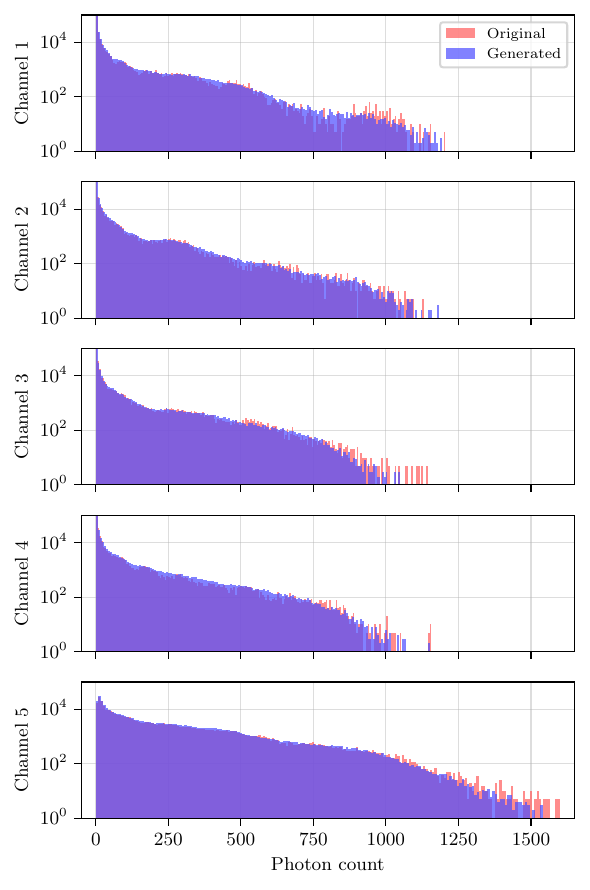}
    \caption{Histogram of the sum of photons in the ZN channels for responses generated by GEANT (Original) and the diffusion model (Generated). Note the logarithmic scale on the y-axis.}
    \label{fig:diffusion_histogram}
\end{figure}

Table \ref{tab:frameworks} provides a comparison of the generative frameworks discussed. The diffusion model excels with the lowest Wasserstein distance of 3.15. The histograms of the values generated by this model closely match those of the Monte Carlo simulations in all channels (Figure~\ref{fig:diffusion_histogram}), further confirming the effectiveness of the model for the ZN simulation. VQ-GAN also performs well with the Wasserstein metric of 4.58, whereas \ac{GAN} shows average performance. \ac{VQVAE} and autoencoder perform the worst, as indicated by the highest Wasserstein distance. Note that \ac{GAN} has a significantly higher \ac{RMSE} value compared to the original data, suggesting that while accurately captures the photon count distribution, it may produce detector responses that are too varied from the original. In contrast, the autoencoder shows a lower \ac{RMSE} compared to GEANT, probably due to smoothed images and issues with particles having diverse responses, resulting in the model generating output close to zero. Example model outputs are shown in Figure~\ref{fig:example_sim}. A comparison that additionally includes \acp{NF} is available in~\cite{zdc_fast_sim}.

\begin{table}[h!]
    \caption{Performance comparison of generative frameworks.}
    \label{tab:frameworks}
    \centering
    \begin{tabular}{|c||c|c|c|}
        \hline
        \textbf{Model}        & \textbf{Wasserstein} & \textbf{MAE} & \textbf{RMSE} \\
        \hline \hline
        GEANT (original data) & 0.53                 & 16.41        & 59.87         \\ 
        \hline \hline
        Autoencoder           & 11.19                & 15.47        & 43.49         \\ \hline
        \ac{GAN}              & 5.70                 & 24.71        & 100.98        \\ \hline
        \ac{VQVAE}            & 9.61                 & 21.95        & 65.82         \\ \hline
        VQ-GAN                & 4.58                 & 22.90        & 85.45         \\ \hline
        Diffusion             & \textbf{3.15}        & 20.10        & 73.58         \\ \hline
    \end{tabular}
\end{table}

\section{Speed of event generation}
\label{sec:speed}

The generation time was measured on a single NVIDIA A100 \ac{GPU}, in batches of 256 images, excluding the compilation time. Table~\ref{tab:framework_speed} shows that autoencoder and \ac{GAN} achieved exceptionally low generation time of around 0.02 milliseconds per sample. \ac{VQ} models had a longer time of 0.1 milliseconds, while diffusion, despite its excellent performance, was significantly slower with more than 5 milliseconds.

The results from Tables~\ref{tab:frameworks} and~\ref{tab:framework_speed} indicate that \ac{VQ} offers the best compromise between performance and sample generation time. Although the autoencoder achieved the lowest generation time, it is disregarded because of the low quality of its generated samples. The diffusion model has a much slower generation time, however, remains sufficiently quick for practical use in the fast \ac{ZDC} simulation. Note that this model did not operate in latent space and requires multiple forward passes, which may be the reason for the extended generation time.

\begin{table}[h!]
    \caption{Generation time of various generative models.}
    \label{tab:framework_speed}
    \centering
    \begin{tabular}{|c||c|}
        \hline
        \textbf{Model}        & \textbf{Time [ms]} \\
        \hline \hline
        Autoencoder           & \textbf{0.015}     \\ \hline
        \ac{GAN}              & \textbf{0.023}     \\ \hline
        VQ                    & 0.091              \\ \hline
        Diffusion             & 5.360              \\ \hline
    \end{tabular}
\end{table}

\section{Autoencoder experiments}
\label{sec:autoencoders}

In the next sections, additional experiments are described that focus on specific generative frameworks and improve the understanding of \ac{ZDC} neutron detector simulation with these models. In the following results, the \ac{RMSE} metric has been omitted for the sake of clarity, as the emphasis is on the Wasserstein metric.

\subsection{Latent space size and posterior collapse}
\label{sec:latent_space_size}

\begin{figure}[h!]
    \begin{subfigure}{0.5\textwidth}
        \includegraphics[width=\linewidth]{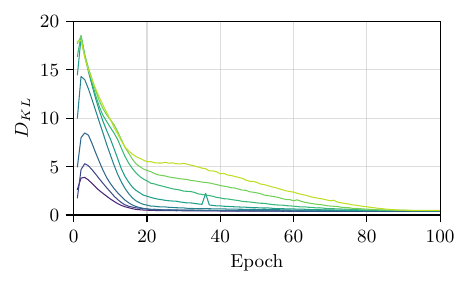}
        \caption{KL loss term during training across various latent space sizes.}
        \label{fig:kl_loss}
    \end{subfigure}
    \hfill
    \begin{subfigure}{0.5\textwidth}
        \includegraphics[width=\textwidth]{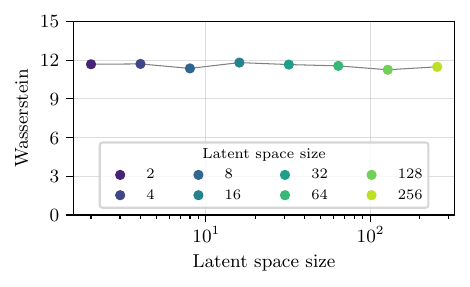}
        \caption{Performance of \ac{VAE} depending on the latent space size.}
        \label{fig:latent_size}
    \end{subfigure}
    \caption{Training of \ac{VAE} with different latent space sizes.}
\end{figure}

An important parameter of the classical autoencoders is the latent space size, as it determines the amount of information that can be encoded. Moreover, it is crucial to avoid posterior collapse when employing autoencoders (further described in Section~\ref{sec:posterior_collapse}). In the experiments, \ac{VAE} based on \ac{ViT} with one million parameters and a latent space of size ranging from 2 to 256 were tested (since the latent space was implemented as a flattened vector). Initially, a posterior collapse was encountered, which was resolved by adjusting the model architecture. The signal strength from the encoder to the decoder had to be increased, requiring an enlargement of the layers responsible for compressing and decompressing the information. Evidence that this issue was addressed is the non-zero \ac{KL} loss term observed in all cases, as shown in Figure~\ref{fig:kl_loss}.

The experimental results indicate that the size of the latent space does not significantly affect the performance (Figure~\ref{fig:latent_size}). Given that the purpose of the model is to generate inputs based on particle parameters, this result is expected and indicates that the model relies mainly on conditional variables. Interestingly, even though the latent space information seems unused during the generation process, using the full autoencoder is advantageous during training, possibly because the decoder receives ``dark knowledge'' from the encoder. Training the decoder alone (without the encoder) resulted in a noticeable drop in performance. Consequently, a latent space size of 10 was chosen as in~\cite{zdc_ml}, and for the autoencoders with a noise generator a size of 20 was selected according to~\cite{sinkhorn_ae}.

\section{Generative adversarial network experiments}
\label{sec:gan_experiments}

\Acp{GAN}, previously considered as state-of-the-art models, have been extensively used for \ac{ZDC} simulation. Various modifications of this model have been proposed, tailored specifically for ZDC~\cite{zdc_gan_diversity, zdc_diffusion, zdc_ml}. The experiments evaluated whether these extensions improve simulation fidelity, and proposed new methods. \ac{ViT}-based models of size around one million parameters were used. To generate diverse outputs, the models utilized samples drawn from a Gaussian distribution along with the conditional variables. Before starting the experiments, the optimizers for the discriminator and the generator were tuned jointly (Appendix~\ref{sec:optimizer_settings}).

\subsection{Improving training stability}

Due to unstable \ac{GAN} training and the common issue of mode collapse (Section~\ref{sec:mode_collapse}), different methods were investigated that stabilize training and lead to better equilibrium. Comparisons were made between standard \ac{GAN}~\cite{gan}, a model with 50\% dropout applied during the training and testing phases~\cite{gan_loss}, \ac{GAN} with a zero-centered gradient penalty using constant and annealed schedules~\cite{gan_gp}, and \ac{WGAN}~\cite{wgan}.

The results indicate that the classical formulation performs best in the \ac{ZDC} simulation task (Table~\ref{tab:gan_stability}). Other methods did not converge to an equilibrium point or did not improve the fidelity of the simulation. The reason for this could be that the optimizer tuning was performed for the classical \ac{GAN} and, in order to obtain better results for the other models, the tuning would have to be repeated (however, this is a computationally expensive task as it requires up to 1000 GPU hours per run). Another explanation could be that the \acp{GAN} are already good fitted for this task and there is no need to apply additional methods that improve stability.

\begin{table}[h!]
    \caption{Performance of various \ac{GAN} models improving stability.}
    \label{tab:gan_stability}
    \centering
    \begin{tabular}{|c||c|c|}
        \hline
        \textbf{Model}             & \textbf{Wasserstein} & \textbf{MAE} \\
        \hline \hline
        \ac{GAN}                   & \textbf{7.09}        & 25.65        \\ \hline
        \ac{GAN} + dropout         & 9.05                 & 37.74        \\ \hline
        \ac{GAN} + GP [fixed]      & $\times$             & $\times$     \\ \hline
        \ac{GAN} + GP [annealed]   & 18.83                & 39.35        \\ \hline
        WGAN                       & $\times$             & $\times$     \\ \hline
    \end{tabular}
\end{table}

\subsection{Improving fidelity of simulations}

In a follow-up experiment, the effectiveness of different \ac{GAN} variants tailored for the \ac{ZDC} simulation task was assessed. In addition to the standard formulation, \acp{GAN} with extra loss terms~\cite{gan_loss}, a GAN with an auxiliary regressor~\cite{zdc_ml}, and SDI-GAN~\cite{zdc_gan_diversity} (described in Section~\ref{sec:gan_extensions}) were tested. The \ac{MLP}-based auxiliary regressor was trained to predict the maximum value coordinates of the detector responses. Moreover, each model was tested with an additional postprocessing step which involves finding a suitable scalar to adjust the model response.

Table~\ref{tab:gan_variants} indicates that the classic \ac{GAN} with postprocessing step produced the best results. \ac{GAN} trained with $l2$ loss and SDI-GAN also performed well. The poor performance of the model with an auxiliary regressor might be due to the need to choose a target different from the coordinates of the maximum value to better represent the activation centers.

\begin{table}[h!]
    \caption{Performance of different \ac{GAN} extensions for \ac{ZDC} simulation.}
    \label{tab:gan_variants}
    \centering
    \begin{tabular}{|c|c||c|c|}
        \hline
        \textbf{Model}             & \textbf{Postprocessing} & \textbf{Wasserstein} & \textbf{MAE} \\
        \hline \hline
        \ac{GAN}                   &                         & 7.09                 & 25.65        \\ \hline
        \ac{GAN}                   & \checkmark              & \textbf{5.70}        & 24.71        \\ \hline
        \ac{GAN} + $l1$ loss       &                         & 53.34                & 63.33        \\ \hline
        \ac{GAN} + $l1$ loss       & \checkmark              & 30.56                & 74.90        \\ \hline
        \ac{GAN} + $l2$ loss       &                         & 6.44                 & 27.37        \\ \hline
        \ac{GAN} + $l2$ loss       & \checkmark              & 6.07                 & 26.78        \\ \hline
        \ac{GAN} + aux regressor   &                         & 20.74                & 43.40        \\ \hline
        \ac{GAN} + aux regressor   & \checkmark              & 19.51                & 44.53        \\ \hline
        SDI-GAN                    &                         & 6.57                 & 27.01        \\ \hline
        SDI-GAN                    & \checkmark              & 6.36                 & 26.58        \\ \hline
    \end{tabular}
\end{table}

\section{Vector quantization experiments}
\label{sec:vq}

The evaluation of the \ac{VQ} models was exceptionally extensive, as this family of models has not been previously documented in the ZN simulation literature. The experiments included analysis of codebook utilization and size, the application of various loss functions, and the evaluation of various transformer sizes and sampling techniques. All models were based on the \ac{ViT} architecture.

\subsection{Codebook utilization}
\label{sec:codebook_utilization}

To address the problem of codebook collapse (Section~\ref{sec:codebook_collapse}), a variety of techniques and strategies are used. The approaches outlined in~\cite{vq_vae, vq_gan, vit_vq_gan} were tested with a \ac{VQVAE} model containing approximately one million parameters and 256 codebook entries, trained using the $l2$ loss for image reconstruction. The codebook utilization was measured as (\ref{eq:perplexity}). An optimizer for the corresponding type of VQ model was used (Appendix~\ref{sec:optimizer_settings}).

\begin{figure}[h!]
    \centering
    \includegraphics[width=0.8\linewidth]{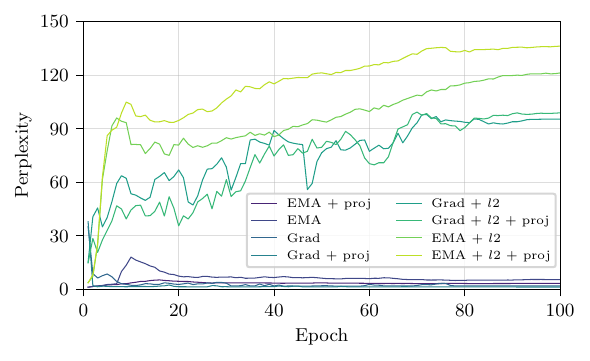}
    \caption{Codebook usage in \ac{VQVAE} with various codebook update rules.}
    \label{fig:codebook_perplexity}
\end{figure}

Figure~\ref{fig:codebook_perplexity} demonstrates that $l2$ normalization is crucial for effective codebook utilization. The \ac{EMA} update showed slightly better performance than the gradient update in terms of stability and codebook utilization. However, as shown in Table~\ref{tab:codebook}, the best results were obtained with the \ac{EMA} update combined with normalization and the gradient update with normalization and projection, thus these methods were used in other experiments. Note that the \ac{EMA} update with normalization and projection, although achieving the highest codebook utilization, produced slightly worse results.

\begin{table}[h!]
    \caption{\ac{VQVAE} reconstruction performance with various codebook update rules.}
    \label{tab:codebook}
    \centering
    \begin{tabular}{|c|c|c||c|c|}
        \hline
        \textbf{Update rule}    & \textbf{$l2$ norm}   & \textbf{Projection} & \textbf{Wasserstein} & \textbf{MAE} \\
        \hline \hline
        Gradient                &                      &                     & 11.74                & 12.99        \\ \hline
        Gradient                &                      & \checkmark          & 15.52                & 17.25        \\ \hline
        Gradient                & \checkmark           &                     & 10.55                & 12.43        \\ \hline
        Gradient                & \checkmark           & \checkmark          & 9.86                 & 11.84        \\ \hline
        EMA                     &                      &                     & 10.59                & 11.66        \\ \hline
        EMA                     &                      & \checkmark          & 10.72                & 11.78        \\ \hline
        EMA                     & \checkmark           &                     & \textbf{9.44}        & \textbf{11.26} \\ \hline
        EMA                     & \checkmark           & \checkmark          & 10.63                & 12.06        \\ \hline
    \end{tabular}
\end{table}

\subsection{Codebook size}
\label{sec:codebook_size}

Subsequently, the impact of codebook size on \ac{VQ} models performance was analyzed. A similar setup was used, with a gradient update incorporating normalization and projection, and the codebook size varied.

\begin{figure}[h!]
    \centering
    \includegraphics[width=0.8\linewidth]{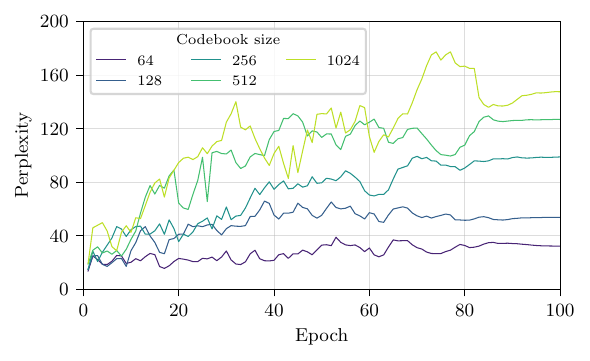}
    \caption{Codebook usage in \ac{VQVAE} depending on codebook size.}
    \label{fig:codebook_size_perplexity}
\end{figure}

The codebook utilization does not grow linearly with the number of entries (Figure~\ref{fig:codebook_size_perplexity}). As shown in Table~\ref{tab:codebook_size}, doubling the codebook size leads to a linear improvement in the Wasserstein metric for \ac{VQVAE} image reconstruction. With \ac{VQVAE} for particle features reconstruction, the \ac{MSE} metric decreases exponentially (Table~\ref{tab:codebook_size_cond}). However, this decrease eventually plateaus, resulting in reduced gains.

\begin{table}[h!]
    \begin{minipage}{0.5\linewidth}
        \caption{Performance of \ac{ZDC} responses reconstruction depending on codebook size.}
        \label{tab:codebook_size}
        \centering
        \begin{tabular}{|c||c|c|}
            \hline
            \textbf{Codebook size} & \textbf{Wasserstein} & \textbf{MAE} \\
            \hline \hline
            64                     & 11.09                & 12.67        \\ \hline
            128                    & 10.37                & 12.06        \\ \hline
            256                    & 9.86                 & 11.84        \\ \hline
            512                    & 9.09                 & 11.16        \\ \hline
            1024                   & \textbf{8.05}        & \textbf{10.49} \\ \hline
        \end{tabular}
    \end{minipage}
    \hfill
    \begin{minipage}{0.45\linewidth}
        \caption{Performance of particle features reconstruction depending on codebook size.}
        \label{tab:codebook_size_cond}
        \centering
        \begin{tabular}{|c||c|}
            \hline
            \textbf{Codebook size} & \textbf{MSE}  \\
            \hline \hline
            64                     & 0.44          \\ \hline
            128                    & 0.13          \\ \hline
            256                    & 0.052         \\ \hline
            512                    & 0.024         \\ \hline
            1024                   & \textbf{0.017} \\ \hline
        \end{tabular}
    \end{minipage}
\end{table}

\begin{figure}[h!]
    \centering
    \includegraphics[width=0.8\linewidth]{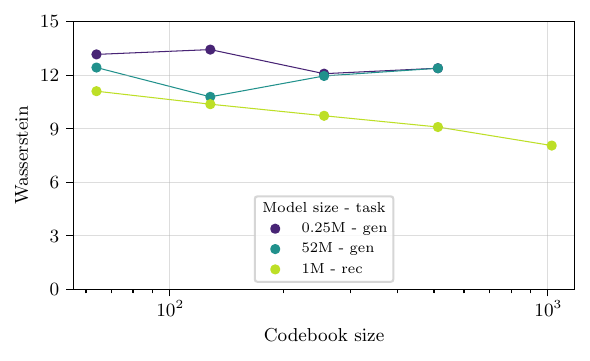}
    \vspace{-2mm}
    \caption{Performance of \ac{VQVAE} reconctruction and generation depending on model and codebook size.}
    \label{fig:codebook_size}
\end{figure}

The generation quality does not scale as well as the reconstruction quality with the codebook size. A large number of entries leads to poor performance of the learnable prior (a transformer with four million parameters was used). Figure~\ref{fig:codebook_size} shows the generation performance of the smallest and largest \ac{VQ} models compared to the reconstruction performance of the medium-sized model. The optimal codebook size is 128 for the large model and 256 for the small model; exceeding this value results in a drop in performance, even if the largest model is used.

\vspace{-2mm}

\vspace{-5mm}
\subsection{Loss functions}
\label{sec:vq_loss_functions}

In experiments with loss functions, the same \ac{VQVAE} model the with a codebook of size 256 was used. The weights for each loss term were either optimized with Optuna over at least 100 trials or empirically adjusted to ensure a similar influence on model training. Having a model to quantize particle features, \ac{VQ} models for reconstruction along with learnable priors were trained to obtain generative models.

The results of this experiment include both reconstruction (Table~\ref{tab:loss_res}) and generation (Table~\ref{tab:loss_gen}) tasks due to the difference in the best performing models. In both cases, incorporating perceptual loss generally led to worse results, while the impact of the $l1$ loss had various effects. However, $l2$ loss and adversarial loss (i.e., the use of VQ-GAN) consistently improved performance in almost all cases. Note that in the reconstruction task (where the aim is to minimize both metrics), different models achieved good results in the Wasserstein and MAE metrics. Interestingly, the generative model trained with $l2$ and adversarial losses outperformed the reconstruction-only model, indicating that the transformer further improved the sample quality. Finally, among models trained with and without adversarial loss (i.e., VQ-GAN and \ac{VQVAE}, respectively), the incorporation of the $l2$ loss allowed to achieve the best results in the generation task.

\begin{table}[h!]
    \caption{VQ models reconstruction performance with various loss functions.}
    \label{tab:loss_res}
    \centering
    \begin{tabular}{|c|c|c|c||c|c|}
        \hline
        \textbf{$l1$ loss}    & \textbf{$l2$ loss}   & \textbf{Perceptual loss} & \textbf{Adversarial loss} & \textbf{Wasserstein} & \textbf{MAE} \\
        \hline \hline
        \checkmark            &                      &                          &                           & $\times$             & $\times$     \\
        \hline
        \checkmark            &                      &                          & \checkmark                & 17.29                & 21.05        \\ 
        \hline
        \checkmark            &                      & \checkmark               &                           & 19.44                & 20.11        \\ 
        \hline
        \checkmark            &                      & \checkmark               & \checkmark                & 13.46                & 19.37        \\ 
        \hline
        \checkmark            & \checkmark           &                          &                           & $\times$             & $\times$     \\ 
        \hline
        \checkmark            & \checkmark           &                          & \checkmark                & \textbf{4.96}        & 16.30        \\ 
        \hline
        \checkmark            & \checkmark           & \checkmark               &                           & 16.88                & 17.50        \\ 
        \hline
        \checkmark            & \checkmark           & \checkmark               & \checkmark                & 8.59                 & 16.64        \\  
        \hline
                              &                      &                          &                           & --                   & --           \\  
        \hline
                              &                      &                          & \checkmark                & $\times$             & $\times$     \\ 
        \hline
                              &                      & \checkmark               &                           & 8.26                 & 17.11        \\
        \hline
                              &                      & \checkmark               & \checkmark                & $\times$             & $\times$     \\ 
        \hline 
                              & \checkmark           &                          &                           & 9.86                 & \textbf{11.84} \\
        \hline
                              & \checkmark           &                          & \checkmark                & 5.77                 & 15.29        \\ 
        \hline
                              & \checkmark           & \checkmark               &                           & 13.64                & 14.70        \\ 
        \hline
                              & \checkmark           & \checkmark               & \checkmark                & 10.48                & 15.36        \\ 
        \hline
    \end{tabular}
\end{table}

\begin{table}[h!]
    \caption{VQ models generation performance with various loss functions.}
    \label{tab:loss_gen}
    \centering
    \begin{tabular}{|c|c|c|c||c|c|}
        \hline
        \textbf{$l1$ loss}    & \textbf{$l2$ loss}   & \textbf{Perceptual loss} & \textbf{Adversarial loss} & \textbf{Wasserstein} & \textbf{MAE} \\
        \hline \hline
        \checkmark            &                      &                          &                           & $\times$             & $\times$     \\ 
        \hline
        \checkmark            &                      &                          & \checkmark                & 16.42                & 26.41        \\ 
        \hline
        \checkmark            &                      & \checkmark               &                           & 21.47                & 25.91        \\ 
        \hline
        \checkmark            &                      & \checkmark               & \checkmark                & 12.93                & 25.35        \\ 
        \hline
        \checkmark            & \checkmark           &                          &                           & $\times$             & $\times$     \\ 
        \hline
        \checkmark            & \checkmark           &                          & \checkmark                & 5.78                 & 24.58        \\ 
        \hline
        \checkmark            & \checkmark           & \checkmark               &                           & 18.04                & 24.02        \\ 
        \hline
        \checkmark            & \checkmark           & \checkmark               & \checkmark                & 8.35                 & 23.33        \\ 
        \hline
                              &                      &                          &                           & --                   & --           \\ 
        \hline
                              &                      &                          & \checkmark                & $\times$             & $\times$     \\ 
        \hline
                              &                      & \checkmark               &                           & 11.16                & 23.55        \\ 
        \hline
                              &                      & \checkmark               & \checkmark                & $\times$             & $\times$     \\ 
        \hline
                              & \checkmark           &                          &                           & 10.55                & 19.80        \\ 
        \hline
                              & \checkmark           &                          & \checkmark                & \textbf{4.58}        & 22.90        \\ 
        \hline
                              & \checkmark           & \checkmark               &                           & 15.94                & 22.72        \\ 
        \hline
                              & \checkmark           & \checkmark               & \checkmark                & 9.63                 & 22.36        \\ 
        \hline
    \end{tabular}
\end{table}

\subsection{Transformer settings}
\label{sec:vq_transformer}

The final experiment with \ac{VQ} models focused on the learnable prior of \ac{VQVAE} whose goal is to generate samples from the latent space of the discrete autoencoder. The transformer operated in the next-token prediction regime, which is considered the most efficient approach for small models~\cite{computational}. 

Initially, the effect of the model size on the transformer performance was examined. The particle features and \ac{ZDC} responses were discretized using the corresponding \ac{VQVAE} models. Optimizer tuning was performed for models of different sizes (Appendix~\ref{sec:optimizer_settings}) with their parameters specified in Table~\ref{tab:transformer_specs}. As shown in Table~\ref{tab:transformer_size}, model size does not significantly impact performance. The model with four million parameters has a slight advantage and was selected for further experiments.

\begin{table}[h!]
    \begin{minipage}{0.32\linewidth}
        \caption{Transformers test loss.}
        \label{tab:transformer_size}
        \centering
        \begin{tabular}{|c||c|c|c|}
            \hline
            \textbf{Model size}   & \textbf{Test loss}  \\
            \hline \hline
            0.25M                 & 3.03          \\ \hline
            1M                    & 2.98          \\ \hline
            4M                    & \textbf{2.97} \\ \hline
            11M                   & 2.98          \\ \hline
            50M                   & 3.01          \\ \hline
        \end{tabular}
    \end{minipage}
    \hfill
    \begin{minipage}{0.64\linewidth}
        \caption{Transformer specification depending on model size.}
        \label{tab:transformer_specs}
        \centering
        \begin{tabular}{|c||c|c|c|c|}
            \hline
            \textbf{Model size}   & \textbf{\#blocks} & \textbf{\#heads} & \textbf{Hidden dim} \\
            \hline \hline
            0.25M                 & 3                 & 3                & 72                  \\ \hline
            1M                    & 5                 & 4                & 128                 \\ \hline
            4M                    & 5                 & 4                & 256                 \\ \hline
            11M                   & 6                 & 6                & 384                 \\ \hline
            50M                   & 10                & 10               & 640                 \\ \hline
        \end{tabular}
    \end{minipage}
\end{table}

\begin{figure}[h!]
    \centering
    \begin{subfigure}{0.49\textwidth}
        \includegraphics[width=\textwidth]{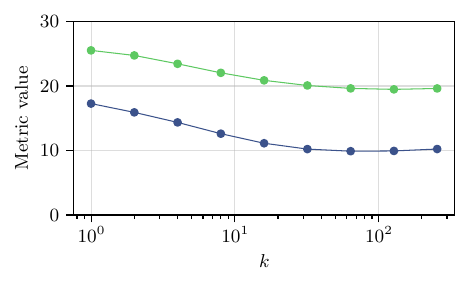}
        \caption{\textit{Top-k} sampling.}
    \end{subfigure}
    \hfill
    \begin{subfigure}{0.49\textwidth}
        \includegraphics[width=\textwidth]{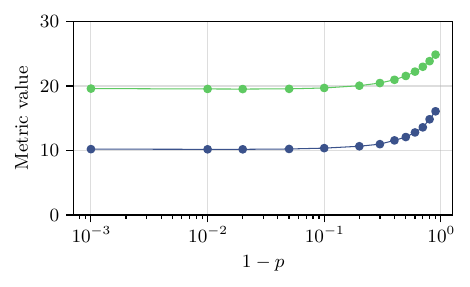}
        \caption{\textit{Top-p} sampling.}
    \end{subfigure}
    \\
    \begin{subfigure}{0.49\textwidth}
        \includegraphics[width=\textwidth]{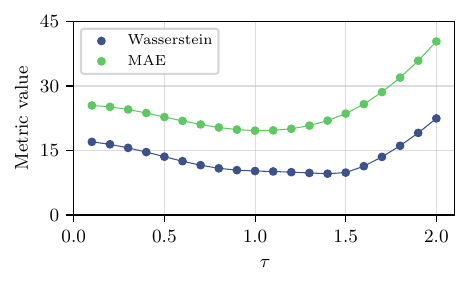}
        \caption{Sampling temperature.}
    \end{subfigure}
\caption{\ac{VQVAE} generation performance with various transformer sampling techniques.}
\label{fig:transformer_sampling}
\end{figure}

Methods for improving sampling from the transformer, such as \textit{top-k} sampling, \textit{top-p} sampling, and adjusting sampling temperature (described in Section~\ref{sec:vq_extensions}), were tested. The results in Figure~\ref{fig:transformer_sampling} show that \textit{top-k} and \textit{top-p} only slightly improve model performance. The best result was obtained by adjusting the temperature $\tau$ to 1.4, which reduced the Wasserstein metric score from 10.55 to 9.61.

\section{Diffusion experiments}
\label{sec:diffusion}

In the following experiments, a \ac{DDPM} model with four million parameters was used, incorporating convolution layers at the beginning and end of U-Net and attention layers in the middle. Similarly to earlier experiments, the optimizer was tuned, with the parameters outlined in Appendix~\ref{sec:optimizer_settings}. Notably, the optimization results indicated the use of a constant learning rate, which aligns with current best practices~\cite{dit}. However, the Wasserstein metric value on the validation set was unstable at the end of the training and to address this, cosine decay with an adjusted peak learning rate was applied, resulting in more stable training and improved performance\footnote{This observation aligns with the findings reported by the authors of~\cite{computational}, who observed that using a constant learning rate combined with \ac{EMA} parameter updates~\cite{dit} has a similar effect to employing a cosine learning rate schedule.}.

\subsection{Noise scheduler}

A comparison between two noise schedulers is presented in Table~\ref{tab:diffusion_schedule}. The training was performed with 1000 denoising steps, whereas the evaluation with 50 steps. The linear scheduler demonstrated superior performance and was therefore selected.

\begin{table}[h!]
    \caption{Performance of the diffusion model depending on the noise schedule.}
    \label{tab:diffusion_schedule}
    \centering
    \begin{tabular}{|c||c|c|}
        \hline
        \textbf{Noise schedule} & \textbf{Wasserstein} & \textbf{MAE} \\
        \hline \hline
        Linear                  & \textbf{3.69}        & 19.57        \\ \hline
        Squared cos             & 6.65                 & 21.57        \\ \hline
    \end{tabular}
\end{table}

\subsection{Number of denoising steps}

\begin{figure}[h!]
    \centering
    \includegraphics[width=0.8\linewidth]{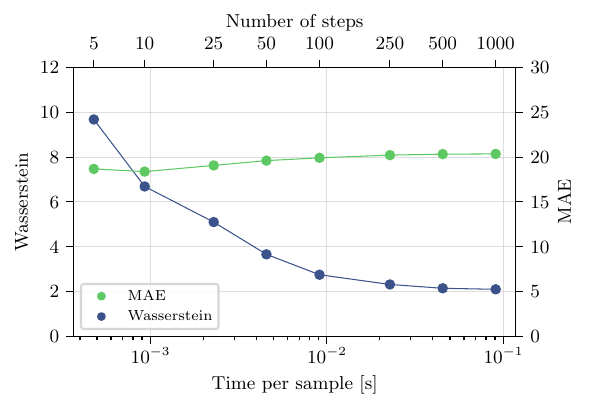}
    \caption{Diffusion generation performance depending on number of denoising steps.}
    \label{fig:diffusion_steps}
\end{figure}

The most important parameter of diffusion models is the number of denoising steps, which allows to balance between the speed and quality of the generated outputs (Section~\ref{sec:diffusion_background}). In this experiment, the \ac{DDIM} sampling was used and the value of $\eta$ was set to 1.0. The results are shown in Figure~\ref{fig:diffusion_steps}, where the generation time was obtained using a single NVIDIA A100 \ac{GPU} in batches of 2048 images, excluding the compilation time. 50 steps were selected as the optimal value. Note that while 1000 steps resulted in a Wasserstein score of 2.10, generating a single sample took around 100 milliseconds, which is considerable in the context of fast simulations.

\subsection{Denoising Diffusion Implicit Models}

The final experiment concentrated on determining the value of the $eta_t$ parameter for 50 denoising steps (Section~\ref{sec:ddim}). This parameter was fixed to a constant value, thus denoted as $eta$. By selecting the optimal value, the Wasserstein metric score improved from 3.69 to 3.15 for $\eta = 0.7$.

\begin{figure}[h!]
    \centering
    \includegraphics[width=0.8\linewidth]{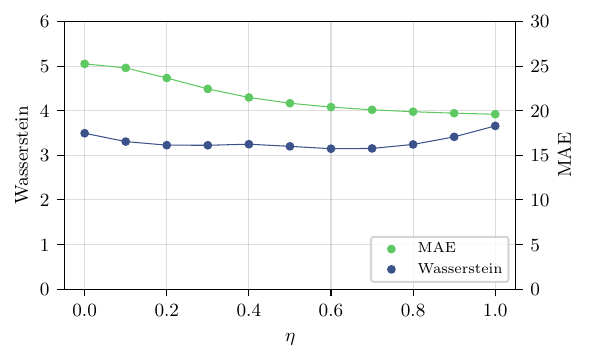}
    \caption{Diffusion generation performance depending on $\eta$ value.}
    \label{fig:diffusion_eta}
\end{figure}

Note that the optimal $\eta$ value might vary depending on the selected number of denoising steps. According to~\cite{ddim}, \ac{DDIM} is advantageous for a smaller number of steps, often with small $\eta$ values. When the number of sampling steps matches the number of denoising steps used during training, \ac{DDPM} is the recommended solution.

\chapter{Thesis summary}
\label{sec:summary}

This thesis presents a comprehensive analysis of the state-of-the-art neural network architectures and generative frameworks for fast simulation of the \ac{ZDC} neutron detector in the \ac{ALICE} experiment at CERN. This chapter summarizes the results of the study, provides recommendations for model design, and outlines future research directions.

\section{Key findings and conclusions}

The dataset analysis highlighted the difficulties posed by the imbalanced dataset and the variable diversity of the \ac{ZDC} responses, which are crucial elements that influence the performance of the models. Robust generative models are required to accurately capture these characteristics and produce high-fidelity simulations. The metrics used in this study include the Wasserstein distance, \ac{MAE}, and \ac{RMSE}. The Wasserstein distance, which is strongly related to the purpose of the \ac{ZDC} simulation, was particularly important.

The study of autoencoders was conducted to investigate the impact of latent space capacity on performance and apply methods to avoid posterior collapse. The \ac{GAN} research showed that a classically formulated \ac{GAN}, especially when combined with a postprocessing step, yields the best performance. Despite testing a variety of stabilization methods and \ac{ZDC}-tailored improvements, the classical approach surpassed the alternatives.

The thorough analysis of \ac{VQ} models highlighted the crucial role of the utilization and size of the codebook. Techniques like \ac{EMA} updates, $l2$ normalization, and projection significantly improved codebook utilization. Moreover, finding the optimal codebook size involves balancing the model's reconstruction performance with the expressiveness of the learnable prior. The combination of $l2$ and adversarial losses consistently improved the output quality. Furthermore, experiments with transformer settings demonstrated the benefits of incorporating sampling methods, particularly adjusting the sampling temperature.

Finally, the diffusion model experiments underscored their potential in the \ac{ZDC} simulation. The possibility of adjusting the number of denoising steps introduces a trade-off between quality and performance. Moreover, choosing the appropriate noise scheduler and fine-tuning \ac{DDIM} parameters further improved the results.

Experiments on model architectures have shown that \ac{ViT}-based autoencoders consistently achieve the highest performance. Furthermore, examining model sizes revealed that medium-sized models (around one million parameters) strike the best balance between performance and computational complexity. Various generative frameworks were compared, with diffusion models emerging as the top performer based on the Wasserstein distance. However, their slower generation time introduces a trade-off between quality and efficiency. VQ-GAN, while slightly less accurate, offers a good compromise with a faster generation time, making them suitable for scenarios that involve extremely fast simulations. \acp{GAN} show moderate performance, whereas \ac{VQVAE} and autoencoders exhibit the lowest performance.

Based on the results of this study, several recommendations can be made to improve the fidelity and efficiency of the fast \ac{ZDC} simulation employing generative neural networks. First, given an adequate computational budget, it is highly recommended to tune the optimizer hyperparameters, as this maximizes model performance and promotes stable training. Incorporating the attention mechanism to the models is advised along with the use of medium-sized models with several million parameters. Diffusion models outperform others, even with a limited number of denoising steps, and adjusting their parameters can further improve performance. Although diffusion models offer superior simulation fidelity, they have a longer generation time. To achieve a balance between speed and quality, VQ-GAN is recommended, with a particular focus on codebook size and techniques that improve utilization such as \ac{EMA} updates with normalization. Also, tuning the transformer sampling is advised, especially adjusting the sampling temperature.

\section{Future work}

The findings of this thesis enable the formulation of several goals for future research that will further improve the fast \ac{ZDC} simulation:

\begin{itemize}
    \item Future research could focus on further refining VQ-GAN due to its balance between performance and throughput. The goal is to improve simulation fidelity by integrating the latest \ac{ViT} developments~\cite{optimal_vit}, testing different sampling techniques, and evaluating new strategies for codebook updates~\cite{lfq}.
    \item Diffusion, as a state-of-the-art model, should be the subject of additional research. The key goal is to improve the generation speed by operating in the latent space~\cite{stable_diffusion} and reducing the number of denoising steps. Investigating knowledge distillation~\cite{sdxl_turbo} and rectified flows~\cite{flow_matching}, which have shown promise in recent models, might be beneficial.
    \item Exploring approaches to incorporate physical loss terms~\cite{pbdl} and regularize neural networks to produce physically consistent \ac{ZDC} responses would also be advantageous.
\end{itemize}

\printbibliography

\appendix

\chapter{Optimizer settings}
\label{sec:optimizer_settings}

This chapter details the settings of the AdamW optimizers used in each experiment. Furthermore, the parameter importances calculated using Optuna's default algorithm for all trials are shown in the following figures.

\newpage
\begin{table}[h!]
    \caption{The optimizer settings for autoencoders across various architectures.}
    \label{tab:optimizer_arch}
    \centering
    \begin{tabular}{|c||c|c|c|}
        \hline
        \textbf{Parameter}    & \textbf{CNN} & \textbf{ViT} & \textbf{MLP-Mixer}  \\
        \hline \hline
        Learning rate         & \num{1.4e-3} & \num{3.5e-3} & \num{5.7e-3}  \\ \hline
        $\beta_1$             & 0.76         & 0.64         & 0.89          \\ \hline
        $\beta_2$             & 0.88         & 0.73         & 0.88          \\ \hline
        $\epsilon$            & \num{6.1e-8} & \num{1.5e-7} & \num{1.2e-6}  \\ \hline
        Weight decay          & --           & 0.068        & 0.013         \\ \hline
        Nesterov momentum     & \checkmark   & --           & --            \\ \hline
        Cosine decay          & --           & \checkmark   & \checkmark    \\ \hline
        Initial div factor    & --           & 35           & 600           \\ \hline
        Final div factor      & --           & 770          & 1200          \\ \hline
        Warmup period [\%]    & --           & 32           & 34            \\ \hline
    \end{tabular}
\end{table}

\begin{figure}[h!]
    \centering
    \includegraphics[width=0.8\linewidth]{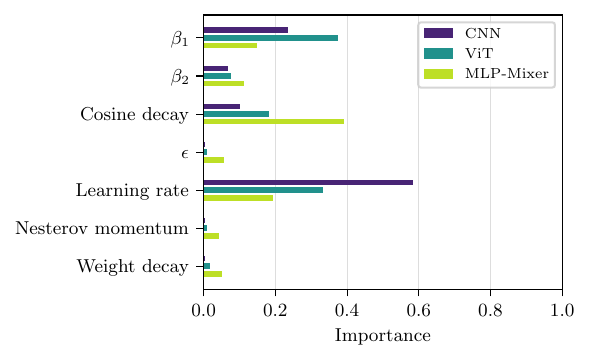}
    \caption{The importance of optimizer settings for autoencoders in various architectures.}
    \label{fig:importance_arch}
\end{figure}

\begin{table}[h!]
    \caption{The optimizer settings for \ac{VQVAE} depending on model size.}
    \label{tab:optimizer_size}
    \centering
    \begin{tabular}{|c||c|c|c|c|c|}
        \hline
        \textbf{Parameter}    & \textbf{0.25M} & \textbf{1M} & \textbf{4M} & \textbf{13M} & \textbf{52M}  \\
        \hline \hline
        Learning rate         & \num{1.4e-3} & \num{3.5e-3} & \num{5.7e-3} & \num{3.5e-3} & \num{5.7e-3}  \\ \hline
        $\beta_1$             & 0.79         & 0.71         & 0.54         & 0.57         & 0.84          \\ \hline
        $\beta_2$             & 0.76         & 0.88         & 0.80         & 0.85         & 0.82          \\ \hline
        $\epsilon$            & \num{2.0e-4} & \num{6.7e-9} & \num{1.4e-7} & \num{6.1e-5} & \num{2.5e-5}  \\ \hline
        Weight decay          & --           & 0.031        & 0.11         & --           & \num{1.0e-4}  \\ \hline
        Nesterov momentum     & --           & --           & --           & \checkmark   & \checkmark    \\ \hline
        Cosine decay          & \checkmark   & \checkmark   & \checkmark   & \checkmark   & \checkmark    \\ \hline
        Initial div factor    & 45           & 22           & 25           & 630          & 72            \\ \hline
        Final div factor      & 2100         & 44           & 4200         & 26           & 20            \\ \hline
        Warmup period [\%]    & 23           & 10           & 40           & 42           & 44            \\ \hline
    \end{tabular}
\end{table}

\begin{figure}[h!]
    \centering
    \includegraphics[width=0.8\linewidth]{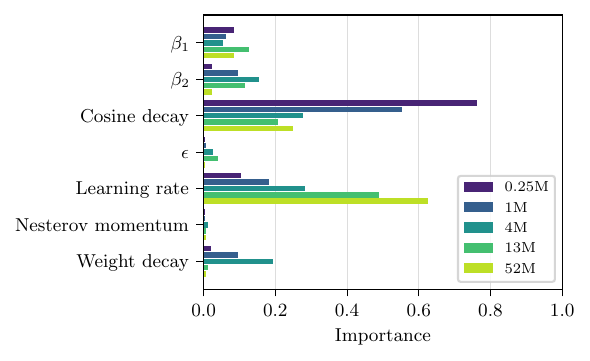}
    \caption{The importance of optimizer settings for \ac{VQVAE} depending on model size.}
    \label{fig:importance_size}
\end{figure}

\begin{table}[h!]
    \caption{The optimizer settings for transformers depending on model size.}
    \label{tab:optimizer_transformer}
    \centering
    \begin{tabular}{|c||c|c|c|c|c|}
        \hline
        \textbf{Parameter}    & \textbf{0.25M} & \textbf{1M} & \textbf{4M} & \textbf{11M} & \textbf{50M}  \\
        \hline \hline
        Learning rate         & \num{7.6e-3} & \num{5.2e-3} & \num{3.5e-3} & \num{5.3e-3} & \num{8.0e-3}  \\ \hline
        $\beta_1$             & 0.97         & 0.87         & 0.80         & 0.52         & 0.69          \\ \hline
        $\beta_2$             & 0.93         & 0.998        & 0.65         & 0.50         & 0.68          \\ \hline
        $\epsilon$            & \num{3.8e-7} & \num{1.0e-5} & \num{1.7e-9} & \num{7.0e-10} & \num{3.1e-8}  \\ \hline
        Weight decay          & --           & 0.14         & 0.33         & 0.40         & 0.23          \\ \hline
        Nesterov momentum     & \checkmark   & --           & \checkmark   & \checkmark   & \checkmark    \\ \hline
        Cosine decay          & \checkmark   & \checkmark   & \checkmark   & \checkmark   & \checkmark    \\ \hline
        Initial div factor    & 60           & 140          & 390          & 590          & 170           \\ \hline
        Final div factor      & 5200         & 3400         & 600          & 10           & 5100          \\ \hline
        Warmup period [\%]    & 15           & 33           & 21           & 24           & 4.1           \\ \hline
    \end{tabular}
\end{table}

\begin{figure}[h!]
    \centering
    \includegraphics[width=0.8\linewidth]{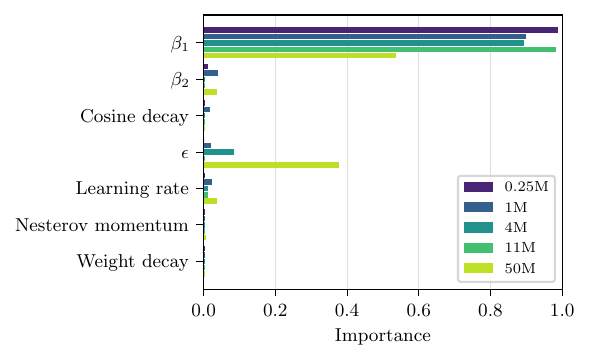}
    \caption{The importance of optimizer settings for transformers depending on model size.}
    \label{fig:importance_transformer}
\end{figure}

\begin{table}[h!]
    \caption{The optimizer settings for \acp{GAN}.}
    \label{tab:optimizer_gan}
    \centering
    \begin{tabular}{|c||c|c|}
        \hline
        \textbf{Parameter}    & \textbf{Discriminator} & \textbf{Generator}  \\
        \hline \hline
        Learning rate         & \num{1.4e-5}           & \num{1e-4}          \\ \hline
        $\beta_1$             & 0.97                   & 0.76                \\ \hline
        $\beta_2$             & 0.96                   & 0.54                \\ \hline
        $\epsilon$            & \num{3.7e-5}           & \num{1.8e-2}        \\ \hline
        Weight decay          & --                     & --                 \\ \hline
        Nesterov momentum     & --                     & \checkmark          \\ \hline
        Cosine decay          & \checkmark             & \checkmark          \\ \hline
        Initial div factor    & 41                     & 260                 \\ \hline
        Final div factor      & 1700                   & 9500                \\ \hline
        Warmup period [\%]    & 43                     & 34                  \\ \hline
    \end{tabular}
\end{table}

\begin{figure}[h!]
    \centering
    \includegraphics[width=0.8\linewidth]{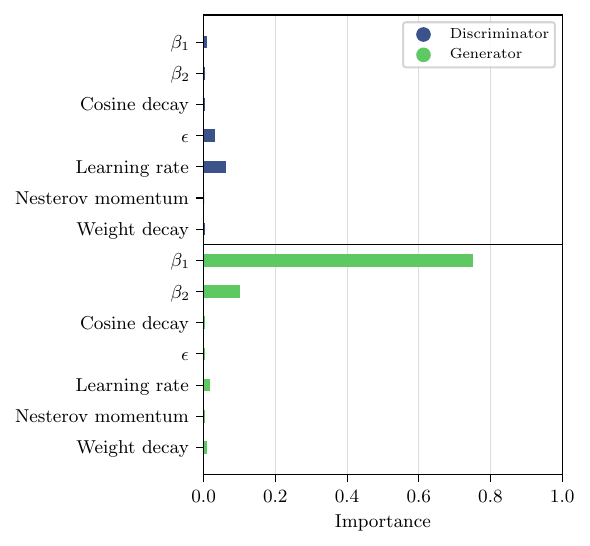}
    \caption{The importance of optimizer settings for \acp{GAN}.}
    \label{fig:importance_gan}
\end{figure}

\begin{table}[h!]
    \caption{The optimizer settings for the diffusion model.}
    \label{tab:optimizer_diffusion}
    \centering
    \begin{tabular}{|c||c|}
        \hline
        \textbf{Parameter}    & \textbf{Diffusion}  \\
        \hline \hline
        Learning rate         & \num{2.8e-5}        \\ \hline
        $\beta_1$             & 0.87                \\ \hline
        $\beta_2$             & 0.82                \\ \hline
        $\epsilon$            & \num{2.5e-5}        \\ \hline
        Weight decay          & \num{5.5e-4}        \\ \hline
        Nesterov momentum     & N/A                 \\ \hline
        Cosine decay          & \checkmark          \\ \hline
        Initial div factor    & N/A                 \\ \hline
        Final div factor      & N/A                 \\ \hline
        Warmup period [\%]    & 10                  \\ \hline
    \end{tabular}
\end{table}

\begin{figure}[h!]
    \centering
    \includegraphics[width=0.8\linewidth]{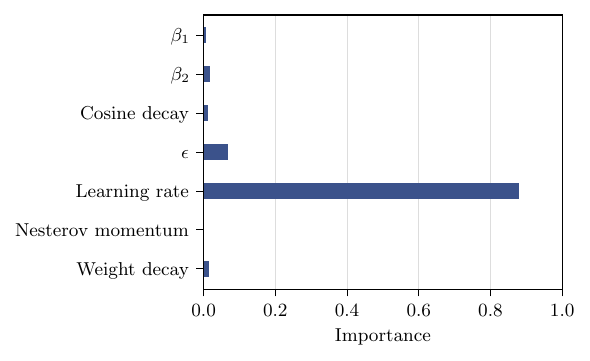}
    \caption{The importance of optimizer settings for the diffusion model.}
    \label{fig:importance_diffusion}
\end{figure}

\end{document}